\documentclass{egpubl}

\usepackage[T1]{fontenc}
\usepackage{dfadobe}  

\biberVersion
\BibtexOrBiblatex
\usepackage[backend=biber,bibstyle=EG,citestyle=alphabetic,backref=true]{biblatex} 
\electronicVersion
\PrintedOrElectronic

\ifpdf \usepackage[pdftex]{graphicx} \pdfcompresslevel=9
\else \usepackage[dvips]{graphicx} \fi

\usepackage{egweblnk}

\usepackage{amsmath}
\usepackage{amssymb}
\usepackage{booktabs} %
\usepackage[capitalize]{cleveref}
\usepackage{enumitem}
\setlist[itemize]{leftmargin=*}
\usepackage{wrapfig}
\usepackage{duckuments}

\usepackage{array}
\usepackage{tabularx}
\usepackage{multirow}
\usepackage{float}

\creflabelformat{equation}{#2\textup{#1}#3}

\floatstyle{ruled}
\newfloat{algorithm}{tbp}{loa}
\providecommand{\algorithmname}{Algorithm}
\floatname{algorithm}{\protect\algorithmname}

\usepackage[noend]{algpseudocode}

\newcommand{\TCOMMENT}[3]{}  %

\newcommand{\rev}[1]{#1}

\global\long\def\var{\operatorname{var}}%
\global\long\def\cov{\operatorname{cov}}%
\global\long\def\corr{\operatorname{corr}}%
\DeclareMathOperator{\sgn}{sgn}

\graphicspath{{./images}}
\usepackage{tikz}
\usepackage[outline]{contour}
\usepackage{ifthen}

\usepackage{xparse} %

\ExplSyntaxOn
\NewDocumentCommand{\aspectratio}{smo}
 {%
  \hbox_set:Nn \l_tmpa_box {\includegraphics{#2}}
  \IfNoValueTF{#3}
   {
    \__student_aspectratio:nn { \box_wd:N \l_tmpa_box } { \box_ht:N \l_tmpa_box }
   }
   {
    \IfBooleanTF{#1}{ \tl_gset:Nx } { \tl_set:Nx } #3
     {
      \__student_aspectratio:nn { \box_wd:N \l_tmpa_box } { \box_ht:N \l_tmpa_box }
     }
   }
 }

\cs_new:Nn \__student_aspectratio:nn
 {
  \fp_eval:n {round( #2 / #1 , 5)}
 }
\ExplSyntaxOff

\newlength{\imgw}
\newlength{\imgh}
\newlength{\offrx}
\newlength{\offry}
\newlength{\offbx}
\newlength{\offby}
\newcommand{\imginset}[8]{%
\aspectratio*{#1}[\imgaspect]
\setlength{\imgh}{\imgaspect\linewidth}
\setlength{\offrx}{#5\linewidth}
\setlength{\offry}{#6\imgh}
\setlength{\offbx}{#7\linewidth}
\setlength{\offby}{#8\imgh}
\begin{tikzpicture}%
\node[anchor=south west,inner sep=0pt,outer sep=0pt] at (0,0) {\includegraphics[width=\linewidth]{#1}};
\draw[orange] (\offrx,\offry) rectangle(\offrx+.125\linewidth,\offry+.125\imgh);
\draw[blue]   (\offbx,\offby) rectangle(\offbx+.125\linewidth,\offby+.125\imgh);
\draw[white](0.0,\imgh) node[below right]{\footnotesize \contour{black}{#2}};
\draw[white](\linewidth,\imgh) node[below left]{\footnotesize \contour{black}{#3}};
\begin{scope}
\ifthenelse{\equal{#4}{bottom}}{%
\clip (0,-0.5\imgh) rectangle(.5\linewidth,0);
\node[anchor=south west,inner sep=0pt,outer sep=0pt] at (-4\offrx,-4\offry-.5\imgh) {\includegraphics[width=4\linewidth]{#1}}; %
}{%
\clip (0,\imgh) rectangle(.5\linewidth,1.5\imgh);
\node[anchor=south west,inner sep=0pt,outer sep=0pt] at (-4\offrx,-4\offry+\imgh) {\includegraphics[width=4\linewidth]{#1}}; %
}
\end{scope}
\begin{scope}
\ifthenelse{\equal{#4}{bottom}}{%
\clip (.5\linewidth,-.5\imgh) rectangle(\linewidth,0);
\node[anchor=south west,inner sep=0pt,outer sep=0pt] at (-4\offbx+.5\linewidth,-4\offby-.5\imgh) {\includegraphics[width=4\linewidth]{#1}}; %
}{%
\clip (.5\linewidth,\imgh) rectangle(\linewidth,1.5\imgh);
\node[anchor=south west,inner sep=0pt,outer sep=0pt] at (-4\offbx+.5\linewidth,-4\offby+\imgh) {\includegraphics[width=4\linewidth]{#1}}; %
}%
\end{scope}
\ifthenelse{\equal{#4}{bottom}}{%
\draw[orange,very thick] (.5pt,-0.5\imgh+.5pt) rectangle(.5\linewidth-.5pt,-.5pt);
\draw[blue,very thick]   (.5pt+.5\linewidth,-0.5\imgh+.5pt) rectangle(\linewidth-.5pt,-.5pt);
}{%
\draw[orange,very thick] (.5pt,\imgh+.5pt) rectangle(.5\linewidth-.5pt,1.5\imgh-.5pt);
\draw[blue,very thick]   (.5pt+.5\linewidth,\imgh+.5pt) rectangle(\linewidth-.5pt,1.5\imgh-.5pt);
}%
\end{tikzpicture}}%

\newcommand{\imgtext}[3]{%
\aspectratio*{#1}[\imgaspect]
\setlength{\imgh}{\imgaspect\linewidth}
\begin{tikzpicture}%
\node[anchor=south west,inner sep=0pt,outer sep=0pt] at (0,0) {\includegraphics[width=\linewidth]{#1}};
\draw[white](0.0,\imgh) node[below right, rectangle, fill=black, text=white, opacity=0.5, text opacity=1]{\small \contour{black}{#2}};
\draw[white](\linewidth,\imgh) node[below left, rectangle, fill=black, text=white, opacity=0.5, text opacity=1]{\small \contour{black}{#3}};
\end{tikzpicture}}%

\newcommand{\imginsetw}[8]{%
\aspectratio*{#1}[\imgaspect]
\setlength{\imgw}{0.66666\linewidth}
\setlength{\imgh}{\imgaspect\imgw}
\setlength{\offrx}{#5\imgw}
\setlength{\offry}{#6\imgh}
\setlength{\offbx}{#7\imgw}
\setlength{\offby}{#8\imgh}
\begin{tikzpicture}%
\node[anchor=south west,inner sep=0pt,outer sep=0pt] at (0,0) {\includegraphics[width=\imgw]{#1}};
\draw[orange] (\offrx,\offry) rectangle(\offrx+.125\imgw,\offry+.125\imgh);
\draw[blue]   (\offbx,\offby) rectangle(\offbx+.125\imgw,\offby+.125\imgh);
\draw[white](0.0,\imgh) node[below right]{\small \contour{black}{#2}};
\draw[white](\imgw,\imgh) node[below left]{\small \contour{black}{#3}};
\begin{scope}
\ifthenelse{\equal{#4}{left}}{%
\clip (-.5\imgw,0) rectangle(0,.5\imgh);
\node[anchor=south west,inner sep=0pt,outer sep=0pt] at (-4\offrx-.5\imgw,-4\offry+.5\imgh) {\includegraphics[width=4\imgw]{#1}}; %
}{%
\clip (\imgw,0) rectangle(1.5\imgw,.5\imgh);
\node[anchor=south west,inner sep=0pt,outer sep=0pt] at (-4\offrx+\imgw,-4\offry+0) {\includegraphics[width=4\imgw]{#1}}; %
}
\end{scope}
\begin{scope}
\ifthenelse{\equal{#4}{left}}{%
\clip (-.5\imgw,.5\imgh) rectangle(0,\imgh);
\node[anchor=south west,inner sep=0pt,outer sep=0pt] at (-4\offbx-.5\imgw,-4\offby+.5\imgh) {\includegraphics[width=4\imgw]{#1}}; %
}{%
\clip (\imgw,.5\imgh) rectangle(1.5\imgw,\imgh);
\node[anchor=south west,inner sep=0pt,outer sep=0pt] at (-4\offbx+\imgw,-4\offby+.5\imgh) {\includegraphics[width=4\imgw]{#1}}; %
}%
\end{scope}
\ifthenelse{\equal{#4}{left}}{%
\draw[orange,very thick] (.5pt-.5\imgw,.5pt) rectangle(-.5pt,.5\imgh-.5pt);
\draw[blue,very thick]   (.5pt-.5\imgw,0.5\imgh+.5pt) rectangle(-.5pt,\imgh-.5pt);
}{%
\draw[orange,very thick] (.5pt+\imgw,.5pt) rectangle(1.5\imgw-.5pt,.5\imgh-.5pt);
\draw[blue,very thick]   (.5pt+\imgw,0.5\imgh+.5pt) rectangle(1.5\imgw-.5pt,\imgh-.5pt);
}%
\end{tikzpicture}}%

\newcommand{\rtext}[1]{\parbox{.8cm}{\raggedleft #1}~}
\newcommand{\ltext}[1]{\raggedright #1}

\title[Regional Adaptive Metropolis Light Transport]%
      {Regional Adaptive Metropolis Light Transport}

\author[H. Otsu et al.]
{
    \parbox{\textwidth}{
        \centering
        H. Otsu$^{1,2}$,
        K. Herveau$^{1}$,
        J. Hanika$^{1}$,
        D. Nowrouzezahrai$^{2}$, and
        C. Dachsbacher$^{1}$
    }
    \\
    \parbox{\textwidth}{
        \centering
        $^1$Karlsruhe Institute of Technology, Germany \\
        $^2$McGill University, Canada
    }
}

\begin{document}

\teaser{
\centering 
\setlength{\tabcolsep}{0.1pt}
\renewcommand{\arraystretch}{.05}
\begin{tabular}{m{0.3cm} *{4}{m{.24\linewidth}}}
    \rotatebox{90}{Fireplace Room}
    &\imgtext{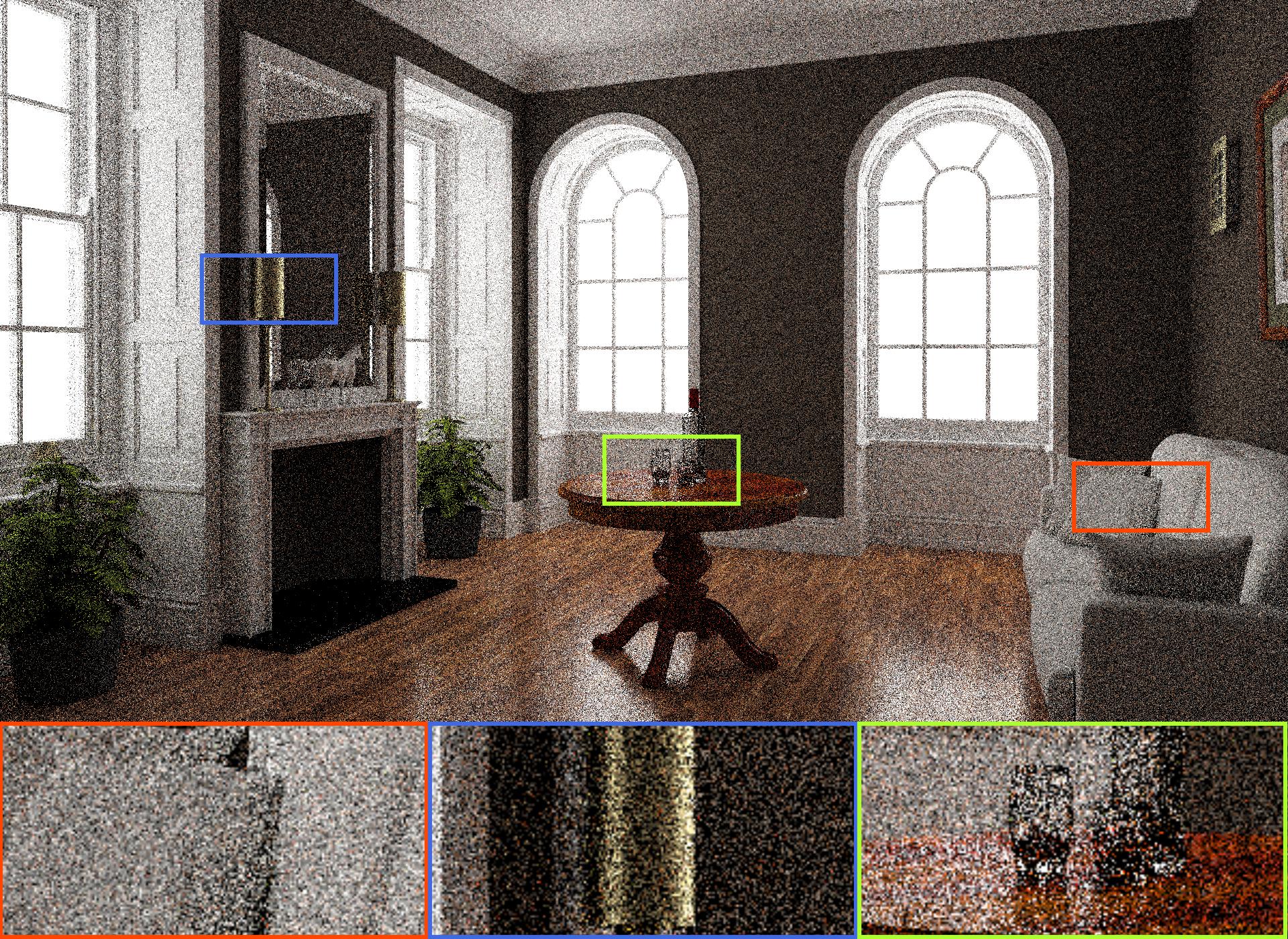}{\ltext{\textsc{Fixed}}}{\rtext{rRMSE 0.6318}}
    &\imgtext{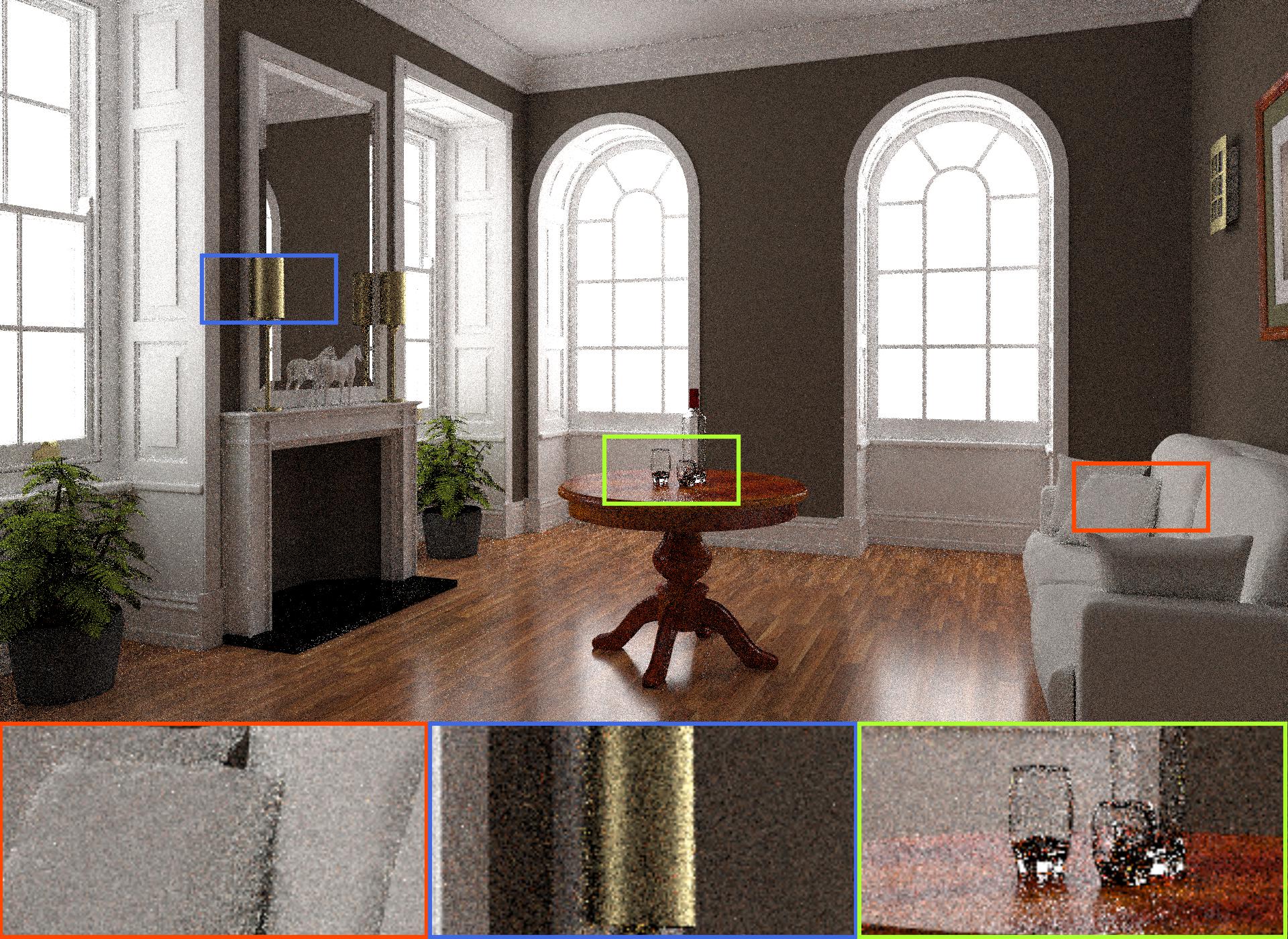}{\ltext{\textsc{Global}}}{\rtext{rRMSE 0.2340}}
    &\imgtext{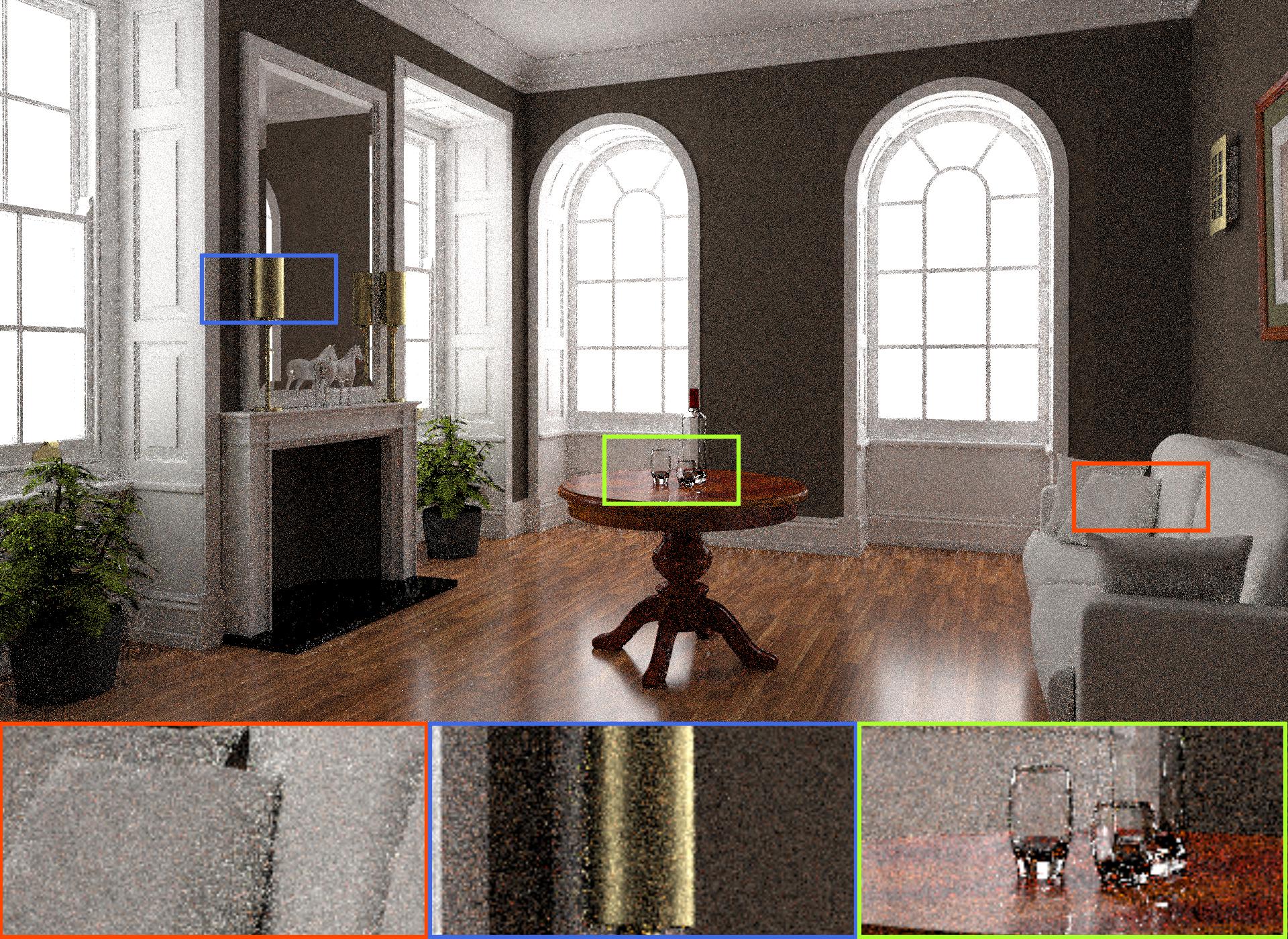}{\ltext{\textsc{RA-Grid}}}{\rtext{rRMSE 0.2501}}
    &\imgtext{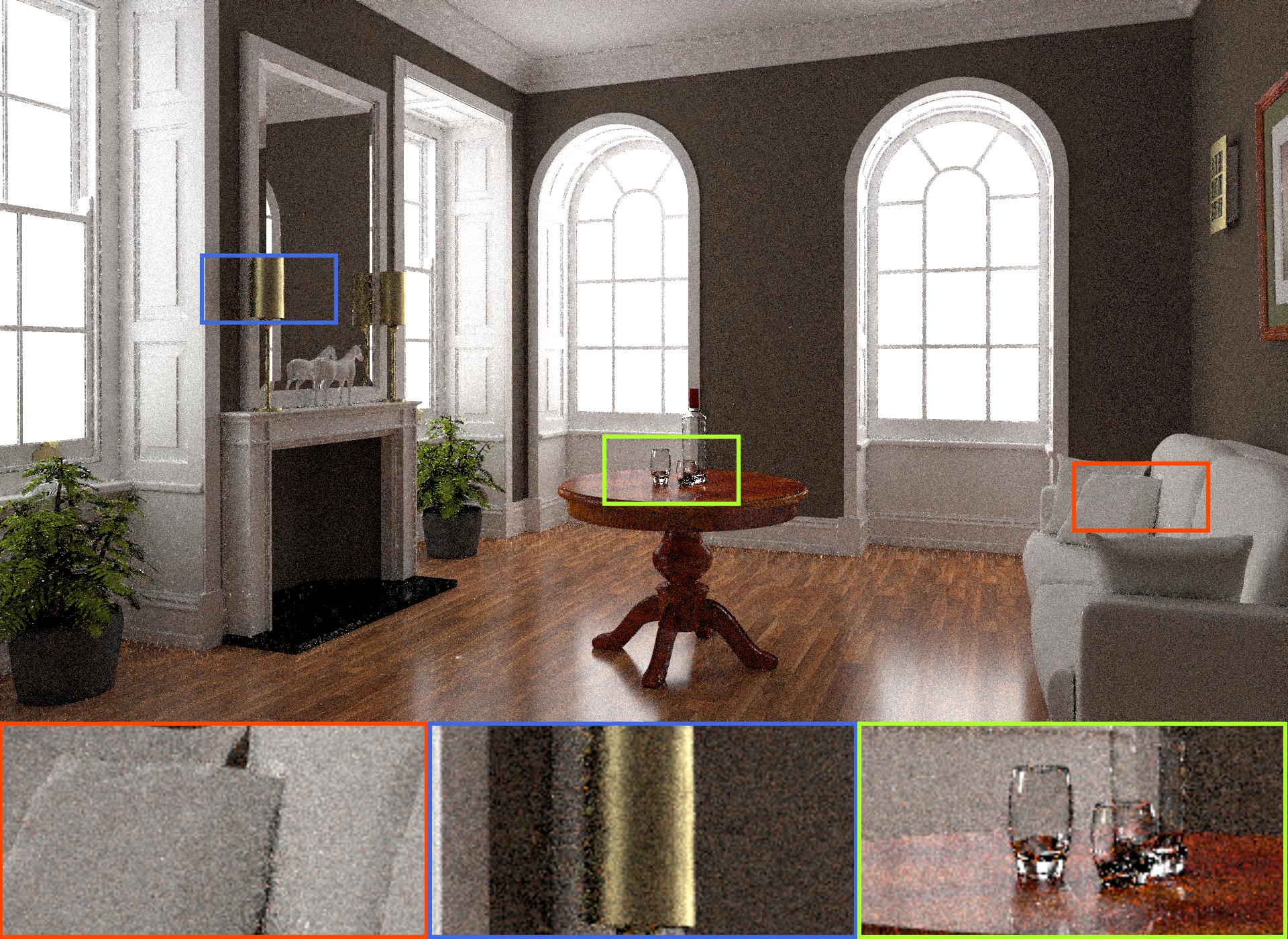}{\ltext{\textsc{RA-Quadtree}}}{\rtext{rRMSE 0.1989}}
  \end{tabular}

\label{fig:rendering_multichain}
\caption{
    Equal-time comparison (10 minutes) of the \emph{Fireplace Room} scene rendered with Metropolis light transport (MLT)~\cite{veach1997mlt} using variants of multi-chain perturbation. 
    In this paper, we develop a regional adaptive path perturbation techniques, which adaptively control the mutation size associated to the current state, based on the information obtained from the previous states.
    Compared to the fixed kernel size (\textsc{Fixed}), the global adaptive appraoch (\textsc{Global}), which assign the single mutation size for the entire path space, can generate better performance.
    But since it is merely adapted to the global average, it tends to miss small details (e.g., reflection on the table). 
    Our regional adaptive path perturbation has a potential to solve this problem. 
    We proposed several partitioning strategies of the path space based on the partitioning of a lower-dimensional canonical space. 
    As well as the naive grid partitioning strategy (\textsc{RA-Grid}), 
    to improve the convergence of the adaptation, we proposed the adaptive partitioning strategy based on quadtree structure (\textsc{RA-Quadtree}).    
}
\label{fig:teaser}
}

\maketitle

\begin{abstract}
    The design of the proposal distributions, and most notably the kernel parameters, are crucial for the performance of Markov chain Monte Carlo (MCMC) rendering.
    A poor selection of parameters can increase the correlation of the Markov chain and result in bad rendering performance.
    We approach this problem by a novel path perturbation strategy for online-learning of state-dependent kernel parameters. We base our approach on the theoretical framework of regional adaptive MCMC which enables the adaptation of parameters depending on the region of the state space which contains the current sample, and on information collected from previous samples. For this, we define a partitioning of the path space on a low-dimensional canonical space to capture the characteristics of paths, with a focus on path segments closer to the sensor. Fast convergence is achieved by adaptive refinement of the partitions.
    Exemplarily, we present two novel regional adaptive path perturbation techniques akin to lens and multi-chain perturbations. Our approach can easily be used on top of existing path space MLT methods to improve rendering efficiency, while being agnostic to the initial choice of kernel parameters.


\end{abstract}

\section{Introduction}

Generating photorealistic synthetic images is one of the original objectives of computer graphics. One important ingredient therefor is the simulation of light transport. 
Markov chain Monte Carlo (MCMC) rendering methods form one class of approaches to this problem. Although it was first introduced for rendering with the original Metropolis light transport (MLT) method~\cite{veach1997mlt} more than two decades ago, research in this field is still active, particularly in the last years.
To sample paths which represent the light transport in a scene, MLT and other MCMC methods generate correlated sequences of samples (paths) according to an arbitrary
target distribution by iteratively mutating the current state to the next state; 
as the latter only depends on the previous state this constitutes a Markov chain.

MCMC rendering is known to be effective in exploring difficult light transport situations, e.g.~challenging visibility configurations.
Its rendering efficiency, however, heavily depends on the design of the proposal distribution, i.e.~how the current state mutates to the next. 
Therefore many approaches leverage domain-specific knowledge about the state space. 
Typically, these proposal distributions are parameterized to provide 
user control for adjustment to a given input scene, which is nearly indispensable for good rendering performance. 
Unfortunately, this is non-trivial and a poor selection of parameters can easily increase the correlation of the Markov chain, leading to drastic efficiency loss. Moreover, these parameters are chosen for individual mutation strategies, but they are \emph{independent of the current state} of the Markov chain. This prohibits fine-grained control of the parameters to adapt to the local structure of the target distribution.

In this paper, we present a novel technique for online learning of a 
\emph{regionally adaptive} proposal distribution. It only requires 
information that is readily available from the execution of the MCMC process itself.
More specifically, we develop a new parametric representation for the proposal
distribution, and a scheme to adaptively update its parameters from the previous states of the Markov chain. 
We base our approach on the well-established theoretical foundation of adaptive MCMC
methods~\cite{roberts2007amcmcergodicity,andrieu2006}, in particular on \emph{regional adaptive} algorithms. 
This theory enables us to update a proposal distribution without spoiling the convergence properties of MCMC, and with only little constraints on the proposal distribution.
Regional adaptive algorithms~\cite{roberts2009amcmcex,andrieu2006} partition the state space in order to adapt to the local structures of the target distribution. 
The parameters associated with each region are adapted independently based on the information collected therein.

Our approach builds on path space MLT and we propose regional adaptive
lens and multi-chain perturbations, extending the original path perturbations~\cite{veach1997mlt}. 
For this, we first define appropriate path space partitioning strategies,
which we obtain by subdividing the lower dimensions of the canonical (random number) space:
For lens perturbations we use the 2D screen space for partitioning, while for the multi-chain perturbation we consider screen space as well as the directions of secondary rays.
To facilitate fast convergence of the parameters to fine-grained partitions,
we develop an adaptive refinement of the partitions based on quadtrees.
Our approach generally provides better rendering performance than previous fixed-sized kernels
or the kernels with global adaptation.
Also, our approach is agnostic to the initial choice of the kernel size.
While we focus on demonstrating its application to lens and multi-chain perturbations, the same idea can be easily combined with arbitrary path perturbation techniques, and thus our approach can be considered as a plug-in method on the top of existing MCMC renderers.

\vspace{2mm}
\noindent
In summary, our contributions are: 
\begin{itemize}
\item a novel MCMC rendering technique utilizing state-dependent mutations based
on regional adaptive MCMC.
\item regional adaptive path perturbation techniques extending the original lens and
caustic perturbations.
\item adaptive path space refinement strategies for better adaptation of parameters.
\end{itemize}

\section{Related Work}

This paper applies adaptive Markov chain Monte Carlo. Here we focus on the most closely related previous work, and refer the reader to PBRT \cite{PBRT3e} for an extensive introduction to rendering.
In \Cref{sec:background} we will establish context and notation.

\subsection{MCMC Rendering}
Hoberock and Hart~\cite{hoberock2010} have shown how to adapt the target function for MLT using
a screen space data structure. Gruson et al.~\cite{gruson2016spartial} take this idea a step
further and use bidirectional light transport and a 3D indexing scheme. Both
approaches guide MCMC by changing the target function. In contrast, we adapt the
mutation step size.

\subsection{Adaptive MCMC}
There are only few works on \emph{adaptive} MCMC in the rendering community.

Zsolnai and Szirmay-Kalos~\cite{zsolnai13mlt} propose a framework to adjust the step size of 
small-step mutations in the primary sample space context. Since this is done
globally, it cannot adapt to different features in the image. We will compare
against such global approaches in our results section.

Li et al.~\cite{li2015h2lt} propose an MCMC method which adapts to the local structure of
the integrand in the context of Hamiltonian Monte Carlo. This method depends on
gradients as well as on the Hessian. This limits the generality and can lead to
issues with finely displaced geometry where local derivatives can be
misleading.

The statistics literature has a larger body of adaptive MCMC papers. We will detail the
most related ones in \cref{sec:regional-adaptive}.
In short, the problem in applying them to rendering is that our state space has a large dimensionality.
It is not easy to see how to best partition it for regional adaptivity. In particular,
the primary sample space makes the task intractable because of the high number of dimensions.
We thus use the path space and propose a way to adaptively subdivide it.

\subsection{Path Guiding}

Another family of path sampling algorithms achieves adaptive
sampling by analyzing all previously traced paths and explicitly constructing
Monte Carlo sampling densities from them.
These \emph{path guiding} approaches, e.g.~\cite{vorba2014pathguiding,mueller2017pathguiding,vorba2019pgp}, depend on specialized acceleration structures to construct
and sample the densities. It can be difficult to balance prior and
learned density, and the implementation effort to arrive at efficient
data structures for guiding is significant. In our approach, we depend on a similar
data structure to store the regionally adaptive step size for our random walks,
but the data load is comparatively light: we only store a scalar to
indicate the isotropic step size for our mutation strategy.

\section{Background}
\label{sec:background}

\subsection{Path Integral}

Light transport simulation estimates solutions to the path integral~\cite{veach1998rmc},
where the intensity of the $j$-th pixel can be written as
\begin{equation}
I_{j}=\int_{\mathcal{P}}h_{j}\left(\bar{x}\right)f\left(\bar{x}\right)d\mu\left(\bar{x}\right).\label{eq:path_integral}
\end{equation}
Here $\mathcal{P}=\cup_{k=2}^{\infty}\mathcal{P}_{k}$ is the \emph{path space} and 
$\mathcal{P}_{k}$ is the set of paths with length $k-1$. 
An element (path) is denoted as $\bar{x}=\mathbf{x}_{1}\mathbf{x}_{2}\cdots\mathbf{x}_{k}\in \mathcal{P}_{k}$, where $\text{\ensuremath{\mathbf{x}_{1}}}$ is the vertex on the
sensor of the camera, and $\mathbf{x}_{k}$ the vertex on a light.
Note that we use s slightly different convention (initial path vertex $\mathbf{x}_{1}$ on the sensor), compared to Veach and Guibas~\cite{veach1997mlt}.

The integrand consists of the reconstruction filter $h_{j}$ and the measurement contribution function $f_{j}$ with respect to the product area measure $\mu$. 
MCMC rendering estimates all pixel intensities simultaneously, essentially ``splatting'' the path contributions to the screen. 
In case of tristimulus rendering, the measurement contribution function is a vector-valued function.

\subsection{Metropolis Light Transport}

Metropolis light transport (MLT) is the first application of MCMC
to rendering~\cite{veach1997mlt}. It uses the Metropolis-Hasting (MH)
algorithm~\cite{metropolis1953,hastings1970monte} to 
generate a sequence of samples following a \emph{target distribution} with (possibly unnormalized) density function $\pi$; the sequence is a Markov chain as every new sample depends on the previous one only.

The state space of MLT's Markov chain is the path space $\mathcal{P}$ and the target
distribution $\pi$ is a scalar contribution function and defined as $\pi\left(\bar{x}\right):=l\left(f\left(\bar{x}\right)\right)$,
where $l$ is a luminance function. 
The samples are eventually distributed according to the normalized scalar contribution function
$\pi/b$, where $b=\int_{\mathcal{P}}\pi\left(\bar{x}\right)d\mu\left(\bar{x}\right)$
is a normalization factor. The factor is typically estimated with independent
MC sampling techniques in a preprocess.

\paragraph*{Metropolis-Hasting Update.}

Given a current path $\bar{x}_{i}$, the MH algorithm first proposes
a tentative path $\bar{y}$ using a \emph{transition kernel} 
$T$ with $\bar{y}\sim T\left(\cdot\mid\bar{x}\right)$. The
proposal $\bar{y}$ is \emph{accepted} as the next state with a certain \emph{acceptance
probability}. If \emph{rejected}, the current path $\bar{x}$ remains
also as the next state: 
\[
\bar{x}_{i+1}=\begin{cases}
\bar{y} & \text{with probability \ensuremath{a\left(\bar{y}\mid\bar{x}_{i}\right)}}\\
\bar{x}_{i} & \text{otherwise}
\end{cases}
\]
where $a\left(\bar{y}\mid\bar{x}\right)$ is 
\[
a\left(\bar{y}\mid\bar{x}\right)=\min\left(1,\frac{\pi\left(\bar{y}\right)T\left(\bar{x}\mid\bar{y}\right)}{\pi\left(\bar{x}\right)T\left(\bar{y}\mid\bar{x}\right)}\right).
\]

Given an initial state $\bar{x}_{1}$ this process generates a Markov
chain $\left\{ \bar{x}_{i}\right\} _{1,\dots,N}$ where $n$ is the number
of mutations. Using this Markov chain we can estimate the path integral:
\[
I_{j}\approx\hat{I}_{j}=\frac{1}{N}\sum_{i=1}^{N}\frac{f\left(\bar{x}_{i}\right)}{\pi\left(\bar{x}_{i}\right)/b}=\frac{b}{N}\sum_{i=1}^{N}\frac{f\left(\bar{x}_{i}\right)}{\pi\left(\bar{x}_{i}\right)}.
\]

\paragraph*{Combining Multiple Mutation Strategies.}
For better overall efficiency, MLT often combines multiple
mutation strategies. We denote the set of transition kernels for $M$ different strategies as $\left\{ T_{j}\right\} _{j=1,\dots,M}$.
The heuristically defined discrete density $s\left(j\mid\bar{x}\right)$ is used to choose a
strategy $j$ for a given path $\bar{x}$ according to its \emph{suitability}:
$j\sim s\left(\cdot\mid\bar{x}\right)$. The acceptance probability of the tentative path $\bar{y}\sim T_{j}\left(\cdot\mid\bar{x}\right)$ then becomes:
\[
a\left(\bar{y}\mid\bar{x}\right)=\min\left(1,\frac{\pi\left(\bar{y}\right)T_{j}\left(\bar{x}\mid\bar{y}\right)s\left(j\mid\bar{y}\right)}{\pi\left(\bar{x}\right)T_{j}\left(\bar{y}\mid\bar{x}\right)s\left(j\mid\bar{x}\right)}\right).
\]

\subsection{Path Perturbations}

The mutation strategies in path space MLT form two categories:
bidirectional mutations and path perturbations. The former are required for the global exploration of the path space 
and thus to guarantee the ergodicity of the Markov chain (the prerequisite for MCMC). 
Path perturbations, on the other hand, are responsible for local exploration: They are designed to be efficient for light transport in specific parts of the path space and
enable MLT to achieve an overall better performance.
Path perturbations are typically not ergodic by themselves and are thus combined with bidirectional mutations.

Our work introduces regional adaptive path perturbations based on Veach's perturbations, which in turn are inspired by bidirectional path sampling techniques. However, instead of regenerating eye and light subpaths as in bidirectional path tracing~\cite{veach94bidirectional}, perturbations are designed to keep a part of the path structure.

\paragraph*{Path Proposal.}

Proposing a path $\bar{x}'$ based on a path $\bar{x}=\mathbf{x}_{1}\cdots\mathbf{x}_{k}$ generally consists of the following three steps:
\begin{enumerate}
\item Deterministically select a connecting edge which splits
the path into the eye subpath $\bar{y}=\mathbf{y}_{1}\cdots\mathbf{y}_{s}$
and the light subpath $\bar{z}=\mathbf{z}_{t}\cdots\mathbf{z}_{1}$, with $\bar{x}=\bar{y}\bar{z}$.
\item Create tentative new subpaths (for one or both subpaths): $\bar{y}'\sim T_{E}\left(\cdot\mid\bar{y}\right)$ and $\bar{z}'\sim T_{L}\left(\cdot\mid\bar{z}\right)$.
Note that subpaths can remain unchanged when $T_{E}$ or $T_{L}$ are delta distributions.
\item The new proposed path is obtained by reconnecting the two perturbed
subpaths: $\bar{x}'=\bar{y}'\bar{z}'$.
\end{enumerate}
Note that in contrast to bidirectional mutations the connecting edge is deterministic. This simplifies the computation of the transition probability to the product of the two subpath's transition probabilities:
\[
T\left(\bar{x}'\mid\bar{x}\right)=T_{E}\left(\bar{y}'\mid\bar{y}\right)T_{L}\left(\bar{z}'\mid\bar{z}\right).
\]

The concrete path perturbations are described by providing the distributions for the each step.
In the following, we will summarize lens and multi-chain perturbations~\cite{veach1997mlt}.

\begin{figure}
\begin{centering}
\def\svgwidth{.9\linewidth}
\hspace{10pt}
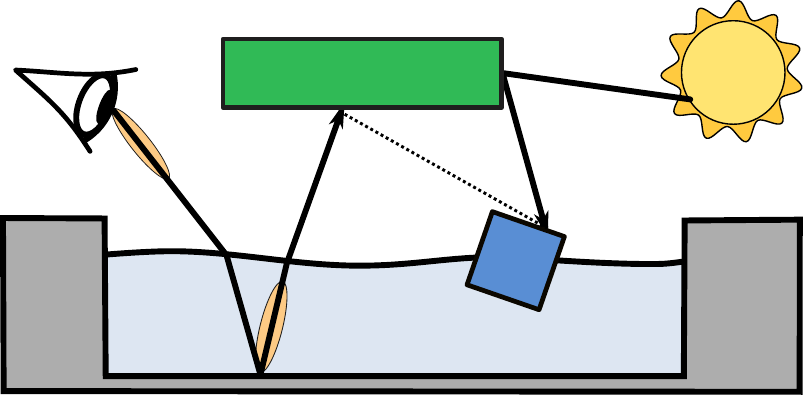
\end{centering}
\caption{\label{fig:path_perturbations}Path perturbations based on bidirectional
path sampling (lens and multi-chain perturbations). The proposal process
first splits the current path into the eye and light subpaths. Each
subpath is perturbed by their respective transition kernels. The final proposal
is generated by reconnecting the perturbed subpaths.
The lens perturbation is moving the outgoing direction at $\boldsymbol{y}_1$, indicated
with an orange lobe, and the multi-chain perturbation additionally mutates the
outgoing direction at $\boldsymbol{y}_3$. The path is propagated deterministically
through the specular interactions at $\boldsymbol{y}_2$ and $\boldsymbol{y}_4$, respectively.}
\end{figure}

\paragraph*{Lens Perturbation.}

Lens perturbations determine two path vertex indices $\left(s,t\right)$ such that either
$\mathbf{y}_{s}$ is a point on the light, or $\mathbf{y}_{s}$ and
$\mathbf{z}_{t}$ are both non-specular vertices. If no such $\left(s,t\right)$
exists, lens perturbation cannot be applied to the path.
This insight can be used to modify the aforementioned suitability function and 
by this perform early rejection, which is beneficial to avoid costly operations involved in proposing tentative paths.

With lens perturbations, light subpaths $\bar{z}$ remain unchanged while the eye subpath $\bar{y}$ is perturbed by mutating the primary ray direction $\omega_{\mathbf{y}_{1}\to\mathbf{y}_{2}}$. 
As all other vertices except for the endpoint $\mathbf{y}_{s}$ are specular
vertices, they can be deterministically computed by ray tracing. 
If a non-specular surface is hit while doing so, the proposal is immediately rejected.

\paragraph*{Multi-chain Perturbation.}

The multi-chain perturbation (\cref{fig:path_perturbations}) is
an extension of the lens perturbation. The eye subpath is chosen such
that $\bar{y}$ contains multiple consecutive specular chains surrounded
by non-specular vertices.
The condition for the endpoint of the subpath is the same as for the lens perturbation: either $\mathbf{y}_{s}$ is a point
on the light, or $\mathbf{y}_{s}$ and $\mathbf{z}_{t}$ are both
non-specular vertices.
Let $c_1,c_2,\dots, c_l$ be the indices of non-specular vertices along the eye subpath (which are not necessarily adjacent).
We consider the point on the camera is always non-specular, i.e.~$c_1 = 1$.
{For each $i\in [1..l]$, similarly to lens perturbations, the multi-chain perturbation
perturbs the ray direction $\omega_{\mathbf{x}_{c_i}\to\mathbf{x}_{c_i+1}}$},
and traces the rays until it hits a non-specular vertex. The
vertex becomes a proposal for \rev{$\mathbf{x}_{c_{i+1}}$}. 
If inconsistencies, such as different path length or non-matching vertex types are found, the proposal is immediately rejected. The light subpath $\bar{z}$ remains the same.

\paragraph*{Perturbing Ray Directions.}

To perturb a ray direction, we first represent the direction in
spherical coordinates $\left(\theta,\phi\right)$ in the coordinate frame around the current direction $\omega$.
We only consider isotropic mutation kernels around the primary ray, and thus 
the azimuthal angle $\phi$ is perturbed by taking a new sample
from the uniform distribution: $\phi'\sim\mathcal{U}_{\left[0,2\pi\right)}\left(\cdot\right)$.
On the other hand, the polar angle $\theta$ is taken from the 
distribution
concentrated around the origin: $\theta'\sim p_{\theta}\left(\cdot\mid\sigma\right)$,
where $\sigma$ is the parameter to control the mutation size.

The original path perturbations by Veach and Guibas use the truncated reciprocal
distribution parameterized by valid range of distribution. In this
paper, for simpler parameterization, we use the truncated normal distribution
defined on the range $\theta\in\left[-\pi,\pi\right]$. Note that
the truncation is introduced to avoid infinite domain of the normal
distribution, which necessitates the evaluation of complicated marginal
densities.

\section{Regional Adaptive MCMC}
\label{sec:regional-adaptive}

\subsection{Motivation}

The performance of MCMC rendering heavily depends on the design of
proposal distributions and the choice of the parameters for the
mutation strategies. A poor selection of parameters can easily
result in bad performance. Hand-tuning parameters, however,
is often challenging for users as it requires deep understanding 
of the mutation strategies, especially when multiple strategies are combined.

The motivation of adaptive MCMC is to determine parameters automatically, solely based on the information obtained from the sampling process itself. Its theoretical and methodological aspects have been studied in computational statistics, however, it has only rarely been applied to MCMC rendering, e.g.~for photon tracing~\cite{hachisuka2011vis}. Our aim is to leverage the concept of adaptive MCMC for path space MLT.

\subsection{Optimal Scaling}
\label{subsec:optimal_scaling}

The correlation of a Markov chain is a quality measure for a MCMC process. 
Several equivalent measures exist, e.g.~autocorrelation time or effective sample size (Appendix~\ref{sec:appendix_correlation_and_convergence}).
If the correlation is large (the effective sample size is small), the variance converges more slowly.
In the context of the MH algorithm, the correlation depends on the design of the underlying transition kernel, which underpins the importance of {its design}.

The problem of choosing an appropriate parameter to control the
kernel size is known as the \emph{optimal scaling problem}. Robert et
al.~\cite{roberts1997} derived the optimal scale for a variant
of the MH algorithm called the \emph{random walk Metropolis} (RWM) algorithm,
{where the offset from the current state is taken from the normal distribution independently for each dimension.}
It makes the assumption that the target distribution
can be represented by {a product of smooth functions, each depending on only one dimension of the state.}
{Based on that, they determine the optimal scaling parameter that yields the minimum asymptotic variance.
The optimal parameter leads to an acceptance probability of 0.234, the well-known constant optimal acceptance probability.}
This result, however,
{is based on an asymptotic analysis}
and poses a strong assumption on the shape of the target distribution.
{Thus in practice the optimal parameter can deviate from this result~\cite{roberts2001}.}

\subsection{Adaptive MCMC\label{subsec:adaptive_mcmc}}

Adaptive MCMC attempts to achieve better performance by automatically
\emph{learning} the parameters of the proposal distribution from 
run-time information. In this adaptive setting the states not only depend
on the previous state and thus the generated sequence of samples is no
longer a Markov chain. Therefore additional constraints are
necessary to guarantee convergence.

Roberts and Rosenthal~\cite{roberts2007amcmcergodicity} proved
the ergodicity of adaptive MCMC (under certain conditions for transition
kernel discussed below). Let the transition kernel be $P_{\Gamma}$ where $\Gamma\in\mathcal{Y}$
is the parameter of the kernel and $\mathcal{Y}$ is the set of all
combinations of parameters. It is used to generate a chain $X_{0},X_{1},\dots$
in the space $\mathcal{X}$, where $X_{n+1}\sim P_{\Gamma_{n}}\left(X_{n},\cdot\right)$,
and $\Gamma_{n}$ is chosen based on the past output. According to
their results, the asymptotic convergence can be proven
(see
Roberts and Rosenthal~\cite{roberts2007amcmcergodicity} for details), assuming
(a) the \emph{diminishing adaptation} condition:
\[
\lim_{n\to\infty}\sup_{x\in\mathcal{X}}\left\Vert P_{\Gamma_{n+1}}\left(x,\cdot\right)-P_{\Gamma_{n}}\left(x,\cdot\right)\right\Vert =0\quad\text{in probability}
\]
and (b) the \emph{containment} condition:
\[
\forall\epsilon>0,\left\{ M_{\epsilon}\left(X_{n},\Gamma_{n}\right)\right\} _{n=0}^{\infty}\quad\text{is bounded in probability},
\]
where $\left\Vert \cdots\right\Vert $ is the total variation distance
and 
\[
M_{\epsilon}\left(x,\gamma\right)=\inf\left\{ n\geq1\mid\left\Vert P_{\gamma}^{n}\left(x,\cdot\right)-\pi\left(\cdot\right)\right\Vert \leq\epsilon\right\} .
\]
In practice, the condition (b) can be
largely ignored~\cite{brooks2011handbook}. They proved that condition
(b) holds when $\mathcal{X}\times\mathcal{Y}$ is finite, or compact
in some topology in which the transition kernels have continuous density.
Specifically, it holds for RWM under general conditions~\cite{bai2011}.
On the other hand, condition (a) depends on the design of the
adaptive algorithm and therefore has to be handled carefully. Intuitively,
condition (a) can be described as the ``amount'' of adaptation
that has to diminish as $n\to\infty$. For instance, if the parameter
is estimated by an empirical average (e.g., empirical covariance matrix),
the condition is automatically satisfied for $n\to\infty$.

If the parameter $\Gamma_{n}$ converges to a fixed value, then condition (a) is automatically satisfied. But we note that the convergence of $\Gamma_{n}$ is merely a sufficient condition.
Condition (a) may still be satisfied even if $\Gamma_{n}$ does not converge~\cite{roberts2009amcmcex}.

\paragraph*{{Design} Criteria for Adaptation.}

Adaptive MCMC techniques can be categorized according to their adaptation strategy~\cite{andrieu2001}:
\begin{itemize}
\item \emph{Moment matching.} This family of techniques attempts to match
the moments of the proposal distribution to those of the target distribution, e.g.~the approach by Haario et al.~\cite{haario2001} which we describe below. 
The motivation is to determine a parametric approximation of
the proposal distribution. 
\item \emph{Minimizing asymptotic variance.} These techniques indirectly or directly
attempt to optimize for the asymptotic variance (Appendix~\ref{sec:appendix_correlation_and_convergence}, \cref{eq:variance}). Often results of the aforementioned optimal scaling are used, e.g.~to control the parameters of the proposal distribution so that the
estimated acceptance probability gets closer to 0.234.
\end{itemize}
Andrieu et al.~\cite{andrieu2001,andrieu2008adaptivemcmc}
formulated adaptive MCMC as a stochastic optimization problem; this discussion is outside the scope of this paper and we refer to their works.

\paragraph*{Adaptive Metropolis.}

Haario et al.~\cite{haario2001} developed an \emph{adaptive Metropolis} (AM) algorithm.
Their approach is based on the observation that the optimal covariance
matrix of the normal proposal distribution is $\left(2.38^{2}/d\right)\Sigma$
where $\Sigma$ is the covariance matrix of the target distribution.
The core idea of the approach is simple: Instead of using the unknown
covariance matrix of the target distribution, the approach uses
the \emph{estimate} of the target covariance matrix.

\paragraph*{Adaptive Scaling.}

Andrieu and Thoms~\cite{andrieu2008adaptivemcmc} pointed out that directly using the constant scaling factor $2.38^{2}/d$ often yields
bad results. They introduce an adaptive scaling parameter multiplied to the estimated covariance matrix so that the empirical acceptance ratio is close to the target value (e.g., 0.234).
{In our method, we apply a simplified version of this scheme where the covariance matrix is fixed to the identity matrix. We detail it further in \cref{sec:ra_path}.
}

\subsection{Regional Adaptive MCMC}

The adaptive Metropolis algorithm~\cite{haario2001} aims to optimize a single global covariance
matrix. This is not effective once a target distribution becomes more complex, e.g.~when is contains multiple modes with support of different sizes.
\begin{wrapfigure}{r}{.35\linewidth}
\begin{centering}
\def\svgwidth{.9\linewidth}
\begingroup%
  \makeatletter%
  \providecommand\color[2][]{%
    \errmessage{(Inkscape) Color is used for the text in Inkscape, but the package 'color.sty' is not loaded}%
    \renewcommand\color[2][]{}%
  }%
  \providecommand\transparent[1]{%
    \errmessage{(Inkscape) Transparency is used (non-zero) for the text in Inkscape, but the package 'transparent.sty' is not loaded}%
    \renewcommand\transparent[1]{}%
  }%
  \providecommand\rotatebox[2]{#2}%
  \newcommand*\fsize{\dimexpr\f@size pt\relax}%
  \newcommand*\lineheight[1]{\fontsize{\fsize}{#1\fsize}\selectfont}%
  \ifx\svgwidth\undefined%
    \setlength{\unitlength}{316.78385096bp}%
    \ifx\svgscale\undefined%
      \relax%
    \else%
      \setlength{\unitlength}{\unitlength * \real{\svgscale}}%
    \fi%
  \else%
    \setlength{\unitlength}{\svgwidth}%
  \fi%
  \global\let\svgwidth\undefined%
  \global\let\svgscale\undefined%
  \makeatother%
  \begin{picture}(1,0.7138427)%
    \lineheight{1}%
    \setlength\tabcolsep{0pt}%
    \put(0,0){\includegraphics[width=\unitlength,page=1]{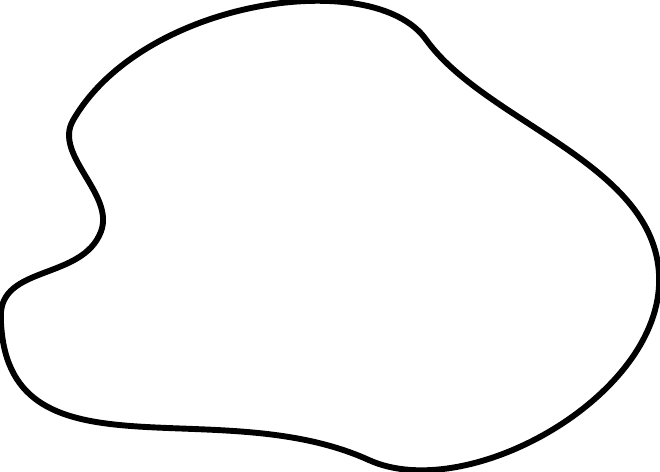}}%
    \put(0.78845353,0.62117991){\makebox(0,0)[lt]{\lineheight{1.25}\smash{\begin{tabular}[t]{l}$\Omega$\end{tabular}}}}%
    \put(0,0){\includegraphics[width=\unitlength,page=2]{partition.pdf}}%
    \put(0.31926076,0.57118551){\makebox(0,0)[lt]{\lineheight{1.25}\smash{\begin{tabular}[t]{l}$\Omega_1$\end{tabular}}}}%
    \put(0.14971626,0.18721116){\makebox(0,0)[lt]{\lineheight{1.25}\smash{\begin{tabular}[t]{l}$\Omega_2$\end{tabular}}}}%
    \put(0.46218532,0.36266457){\makebox(0,0)[lt]{\lineheight{1.25}\smash{\begin{tabular}[t]{l}$\Omega_3$\end{tabular}}}}%
    \put(0.5313103,0.15679757){\makebox(0,0)[lt]{\lineheight{1.25}\smash{\begin{tabular}[t]{l}$\cdots$\end{tabular}}}}%
  \end{picture}%
\endgroup%

\end{centering}
\label{fig:regional_adaptive_mcmc}
\end{wrapfigure}
A better adaptation can be achieved when the mutation kernel, or its scaling parameters, depend on the current state of the chain. These so-called \emph{state-dependent} parameters are used in the regional adaptive Metropolis algorithm~\cite{roberts2009amcmcex}, which therefore partitions the state space $\Omega$ into a finite number of disjoint regions
$\Omega_{1},\Omega_{2},\dots,\Omega_{m}$, with $\Omega=\cup_{i=1}^{m}\Omega_{i}$
and for all $i$ and $j$, $\Omega_{i}\cap\Omega_{j}=\varnothing$.
For each partition, different (sets of) parameters are used and each proposal distribution is only updated using states visiting the respective partition. For example, it is possible to optimize the parameters for $\Omega_{i}$ according to empirical acceptance probabilities from jumps off of this partition, and thus get an estimate close to the target acceptance probability.

\section{Regional Adaptive Path Perturbations}
\label{sec:ra_path}

\begin{algorithm}
    \caption{\label{alg:ramlt_base}High-level overview of the sample generation
    by regional adaptive path perturbation.}
    
    \small
    
    \begin{algorithmic}[1]
    
    \State $\bar{x}_{1}\gets$ Generate initial path
    
    \For{$k\gets\text{1 to \ensuremath{m}}$}
    
        \State $\lambda_{k}\gets$ Initial scaling parameter
    
        \State $i_{k}\gets0$ \Comment{Number of visits to $k$-th region}
    
        \State $A_{k}\gets0$ \Comment{Accumulated acceptance probability}
    
        \State $n_{k}\gets1$ \Comment{Number of parameter updates}
    
    \EndFor
    
    \For{$i\gets\text{1 to \ensuremath{M}}$}
    
    \State $k\gets$ Find $k\in\left[1..m\right]$ such that $\bar{x}\in\mathcal{P}_{k}$
    
    \State $\sigma_{k}\gets\sqrt{\exp\left(\lambda_{k}\right)}$
    
    \State $\bar{y}_{i}\sim T_{i}\left(\cdot\mid\bar{u}_{i},\sigma_{k}\right)$
    \Comment{\cref{eq:regional_proposal_comp}}
    
    \State $A_{k}\gets A_{k}+\alpha(\bar{x}_{i},\bar{y}_{i})$
    
    \State $i_{k}\gets i_{k}+1$
    
    \If{$i_{k}>L$}
    
        \State $\hat{\alpha}_{k}\gets A_{k}/L$ \Comment{\cref{eq:est_exp_acc}}
    
        \State $\gamma\gets\min\left(\gamma_{\mathrm{max}},\gamma_{\mathrm{scale}} n_{k}^{-1/2}\right)$
    \Comment{\cref{eq:step_size_n}}
    
        \State $\lambda_{k}\gets\lambda_{k}+\gamma\cdot\sgn{(\hat{\alpha}_{k}-\bar{\alpha}^{*})}$
    \Comment{\cref{eq:est_exp_acc}}
    
        \State $i_{k},A_{k}\gets0$
    
        \State $n_{k}\gets n_{k}+1$
    
    \EndIf
    
    \State$\bar{x}_{i+1}\gets$ MH update with $\bar{x}_{i}$ and $\bar{y}_{i}$
    
    \EndFor
    
    \end{algorithmic}
\end{algorithm}

\subsection{Overview}

Our goal is to adapt the idea of regional adaptive MCMC to improve the performance
of path space MLT. In this section, we will introduce two \emph{regional
adaptive path perturbations}, which are regional adaptive extensions
of the corresponding path perturbations. Specifically, we will develop
a regional adaptive variant of the lens and multi-chain perturbation.

In order to apply regional adaptation, we need to partition the
state space. Each region in this partition will contain their own set of adaptive parameters.
The proposal distribution of the adaptive path perturbation
then depends on the parameters stored in the partition,
using only the information obtained from the region containing the current state.

Specifically, let $\left\{ \mathcal{P}_{k}\right\} _{k=1,\dots,m}$
denote a partition of the path space, where $m$ is the number of
regions in the partition. By definition, the union of all regions
is the entire path space: $\mathcal{P}=\cup_{k=1}^{m}\mathcal{P}_{k}$,
and the regions are pairwise disjoint: for all $k$ and $l$ such
that $k\neq l$, $\mathcal{P}_{k}\cap\mathcal{P}_{l}=\varnothing$.

\paragraph*{Algorithm.}

\cref{alg:ramlt_base} describes the high level overview
of the sample generation algorithm using our regional adaptive
path perturbation. A practical implementation will also select from multiple
mutation strategies but this step is omitted for brevity.

This algorithm generates a sequence of samples $\bar{x}_{1},\bar{x}_{2},\cdots,\bar{x}_{M}$
using a regionally adaptive proposal. For each step, given the current
path $\bar{x}_{i}$, a corresponding region $k\in\left[1..m\right]$
is selected (line~8). Since $\left\{ \mathcal{P}_{k}\right\} $ is
a partition, there always exists a unique $k\in\left[1..m\right]$ such
that $\bar{x}_{i}\in\mathcal{P}_{k}$. The update of the scaling parameter
$\lambda_{i}$ occurs once in $L$ iterations. Here, the number of
visits $i_{k}$ is managed independently for each region. If the counter
goes beyond $L$, the update process is executed (line~13). The expected
acceptance ratio is accumulated as $A_{k}$, which is used to estimate
$\bar{\alpha}_{k}$ (line~14). The update of the scaling parameter
is done according to the difference between the estimated acceptance
probability $\hat{\alpha}_{k}$ and the target acceptance probability
$\bar{\alpha}^{*}$ (line~16). The amount of adjustment of $\lambda_{k}$
is controlled by the variable $\gamma$ (line~15), which diminishes
with the number of updates. The details of the equations in each
step of the algorithm are explained in this section.

\subsection{State-Dependent Proposal\label{subsec:state_dependent_proposal}}

Let $\bar{\sigma}$ be a vector of parameters to control the mutation size
of a path perturbation. For the lens perturbation, $\bar{\sigma}:=\left(\sigma_{1}\right)$,
where $\sigma_{1}$ controls the perturbation of the primary ray direction.
For the multi-chain perturbation, $\bar{\sigma}:=\left(\sigma_{1},\sigma_{2}\right)$,
where $\sigma_{1}$ controls the perturbation of the primary ray direction
and $\sigma_{2}$ controls the perturbation of the ray direction originating
from the first non-specular vertex to the next vertex. We
denote the proposal distribution of the path perturbations by $T\left(\cdot\mid\bar{x},\bar{\sigma}\right)$,
where $\bar{x}\in\mathcal{P}$ is the current state.

The assignment of the parameters for each region can be expressed
by the state-dependent parameters defined by

\[
\bar{\sigma}\left(\bar{x}\right)=\sum_{k=1}^{m}\mathbf{1}_{\mathcal{P}_{k}}\left(\bar{x}\right)\bar{\sigma}_{k},
\]
where $\mathbf{1}_{\mathcal{P}_{k}}\left(\bar{x}\right)$ is the indicator
function and $\bar{\sigma}_{k}$ is the parameter vector for the $k$-th region.
We can express the same distribution in the form of the weighted mixture
of the per-region proposal kernels:
\begin{align}
T\left(\cdot\mid\bar{x}\right) & =T\left(\cdot\mid\bar{x},\bar{\sigma}\left(\bar{x}\right)\right)\nonumber \\
 & =\sum_{k=1}^{m}\mathbf{1}_{\mathcal{P}_{k}}\left(\bar{x}\right)T\left(\cdot\mid\bar{x},\bar{\sigma}_{k}\right)=\sum_{k=1}^{m}\mathbf{1}_{\mathcal{P}_{k}}\left(\bar{x}\right)T_{k}\left(\cdot\mid\bar{x}\right),\label{eq:regional_proposal_mixture}
\end{align}
where 
\begin{equation}
T_{k}\left(\cdot\mid\bar{x}\right):=T\left(\cdot\mid\bar{x},\bar{\sigma}_{k}\right).\label{eq:regional_proposal_comp}
\end{equation}
This formulation implies that we select the parameters according to
the location of the current state $\bar{x}$. If the current state
is in the region $\mathcal{P}_{k}$, the parameters $\bar{\sigma}_{k}$
are used for the perturbation.

\subsection{Adapting Parameters\label{subsec:adapting_parameters}}

\paragraph*{Criterion for Adaptation.}

We adopt the criteria used for the improved version of the AM algorithm
using an adaptive scaling parameter~\cite{andrieu2008adaptivemcmc}.
For each region, the expected acceptance probability $\bar{a}$ \emph{within
the region} is estimated, so that it is close to the target value
$\bar{\alpha}^{*}$. This can for instance be configured as
 $\bar{\alpha}^{*}=0.234$. We note that the adaptation can
be driven only from the information obtained within a single region.
This is important for an efficient implementation, since it
does not require global synchronization to update the parameters.

This property can be justified by the fact that the difference between
the expected acceptance probability $\bar{\alpha}$ and the target
acceptance probability $\bar{\alpha}^{*}$ can be expressed by 
\[
\bar{\alpha}-\bar{\alpha}^{*}=\sum_{k=1}^{m}b_{k}\left(\mathbb{E}_{\pi\otimes T_{k}}\left[\alpha\left(\bar{x},\bar{y}\right)\mid\bar{x}\in\mathcal{P}_{k}\right]-\bar{\alpha}^{*}\right),
\]
where $\bar{x}\in\mathcal{P}$ is the current state, $b_{k}$ is a
constant only dependent on $k$-th region, and $\mathbb{E}_{\pi\otimes T_{k}}\left[\alpha\left(\bar{x},\bar{y}\right)\mid\bar{x}\in\mathcal{P}_{k}\right]$
is the conditional expectation of the acceptance probability given
the current state $\bar{x}$ is in the $k$-th region. The derivation
of this equation is found in Appendix~\ref{sec:appendix_region_independence}.

\paragraph*{Estimating Expected Acceptance Probability.}

The adaptation of the parameters is driven by the estimation of the
acceptance probability. As we discussed above, the estimation can
be executed independently based on the region the current state
lies in. Let $\bar{x}_{1},\bar{x}_{2},\dots,\bar{x}_{M}$ be a sequence
of states created by the MCMC process and its corresponding proposals $\bar{y}_{1},\bar{y}_{2},\dots,\bar{y}_{M}$.
We denote a sub-sequence $\left\{ \bar{x}_{1}^{k},\bar{x}_{2}^{k},\dots,\bar{x}_{M_{k}}^{k}\right\} \subseteq\left\{ \bar{x}_{1},\bar{x}_{2},\dots,\bar{x}_{M}\right\} $
where the states visit the region $\mathcal{P}_{k}$ (that is, for
all $l\in\left[1..M_{k}\right]$, $\bar{x}_{l}^{k}\in\mathcal{P}_{k}$).
Note that the probability that the state $\bar{x}\in\mathcal{P}_{k}$
can be estimated by simply counting the number of visits to the region.
Therefore, the estimate $\hat{\alpha}_{k}$ of $\bar{\alpha}_{k}$
is obtained by
\begin{equation}
\bar{\alpha}_{k}\approx\hat{\alpha}_{k}:=\frac{1}{M}\frac{\sum_{i=1}^{M}\alpha\left(\bar{x}_{i},\bar{y}_{i}\right)\mathbf{1}_{\mathcal{P}_{k}}\left(\bar{x}_{i}\right)}{M_{k}/M}=\frac{1}{M_{k}}\sum_{l=1}^{M_{k}}\alpha\left(\bar{x}_{l}^{k},\bar{y}_{l}^{k}\right).\label{eq:est_exp_acc}
\end{equation}

\paragraph*{Updating the Scaling Parameter.}

Instead of updating $\sigma$ directly to control the mutation size, we use a
logarithmic mapping $\lambda=\log\sigma^{2}$, similar to the one used in Andrieu and Thoms~\cite{andrieu2008adaptivemcmc}. The $k$-th region maintains an estimate of
$\lambda_{k}$ and obtains the corresponding
$\sigma_{k}=\sqrt{\exp\left(\lambda_{k}\right)}$.

To update the scaling parameter $\lambda_{k}$, we adopt a simple
update scheme used in literature~\cite{andrieu2008adaptivemcmc,roberts2009amcmcex}
based on the observation that the expected acceptance ratio changes monotonically
with the parameter $\lambda_{k}$. Let $j$ be the iteration
count of the parameter update (which is not necessarily same as number
of mutations so far). Then we denote the parameter $\lambda_{k}$
in the $j$-th iteration by $\lambda_{k}^{j}$, and the update formula
is given by
\begin{equation}
\lambda_{k}^{j+1}=\lambda_{k}^{j}+\gamma_{j+1}\cdot\sgn{\left(\hat{\alpha}_{k}-\bar{\alpha}^{*}\right)}.\label{eq:update_lambda}
\end{equation}
The equation can be understood by the following intuition: if the
acceptance ratio is smaller than $\bar{\alpha}^{*}$, the kernel size
should be bigger ($\lambda_{k}$ should be bigger), if the acceptance
ratio is bigger than $\bar{\alpha}^{*}$, the kernel size should be
smaller ($\lambda_{k}$ should be smaller).

The parameter $\gamma_{j+1}$ controls the amount of change of the
parameter $\lambda_{k}$ in iteration $j+1$. We may consider this parameter
as the \emph{step size}. We configured the parameter according to
the suggestion by Roberts and Rosenthal~\cite{roberts2009amcmcex}:
\begin{equation}
\gamma_{j}:=\min\left(\gamma_{\mathrm{max}},\gamma_{\mathrm{scale}} \cdot j^{-1/2}\right),\label{eq:step_size_n}
\end{equation}
where $\gamma_{\mathrm{max}}$ is a value to prevent large change of the parameter,
especially in the early state of the adaptation.
$\gamma_{\mathrm{scale}}$ controls the speed at which the adjustment of the parameter vanishes.
This step is necessary to meet the diminishing adaptation condition
(\cref{subsec:adaptive_mcmc}).

\subsection{Path Space Partition\label{subsec:path_space_partition}}

\begin{figure}
\begin{centering}
\def\svgwidth{.9\linewidth}
\quad
\begingroup%
  \makeatletter%
  \providecommand\color[2][]{%
    \errmessage{(Inkscape) Color is used for the text in Inkscape, but the package 'color.sty' is not loaded}%
    \renewcommand\color[2][]{}%
  }%
  \providecommand\transparent[1]{%
    \errmessage{(Inkscape) Transparency is used (non-zero) for the text in Inkscape, but the package 'transparent.sty' is not loaded}%
    \renewcommand\transparent[1]{}%
  }%
  \providecommand\rotatebox[2]{#2}%
  \newcommand*\fsize{\dimexpr\f@size pt\relax}%
  \newcommand*\lineheight[1]{\fontsize{\fsize}{#1\fsize}\selectfont}%
  \ifx\svgwidth\undefined%
    \setlength{\unitlength}{385.23248147bp}%
    \ifx\svgscale\undefined%
      \relax%
    \else%
      \setlength{\unitlength}{\unitlength * \real{\svgscale}}%
    \fi%
  \else%
    \setlength{\unitlength}{\svgwidth}%
  \fi%
  \global\let\svgwidth\undefined%
  \global\let\svgscale\undefined%
  \makeatother%
  \begin{picture}(1,0.49175754)%
    \lineheight{1}%
    \setlength\tabcolsep{0pt}%
    \put(0,0){\includegraphics[width=\unitlength,page=1]{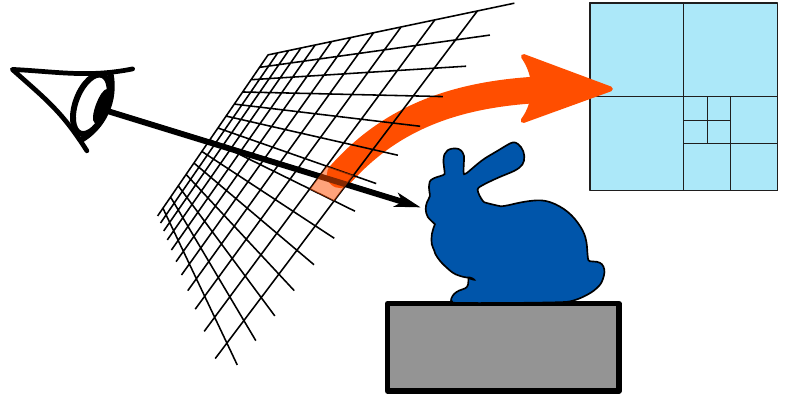}}%
    \put(0.34586621,0.44914092){\makebox(0,0)[lt]{\lineheight{1.25}\smash{\begin{tabular}[t]{l}$\mathcal{S}$\end{tabular}}}}%
    \put(0.76444455,0.42967216){\makebox(0,0)[lt]{\lineheight{1.25}\smash{\begin{tabular}[t]{l}$\mathcal{T}$\end{tabular}}}}%
  \end{picture}%
\endgroup%

\end{centering}
\caption{\label{fig:path_space_partition}Path space partitioning for lens and
multi-chain perturbations. For the lens perturbation, we use a 2d partition
in screen space. For the multi-chain perturbation, we use a 4d partition
in screen space and the ray direction from the first non-specular vertex
from the camera.}
\end{figure}

Since regional adaptation depends on a partition of the state space,
it is inherently limited by the curse of dimensionality. Naively increasing
the number of regions does not result in better performance, since
by increasing the regions, the probability to visit a region decreases.
This means that there will be only few updates of the parameters in each region,
leading to bad adaptation and thus suboptimal performance.

Therefore it is important to choose a partition which is moderately
small, but still expressive enough to capture the details of the
state space. In this paper, we use two low-dimensional subspace partitions
tailored for the lens and the multi-chain perturbation, respectively (\cref{fig:path_space_partition}).
In the next section, we will discuss these is detail.

\paragraph*{2d Partition for the Lens Perturbation.}

For lens perturbation, we use (canonical) screen space, where a point
on the space corresponds to the primary ray direction, and vice versa.
We define a 2d partition $\left\{ \mathcal{S}_{k}\right\} _{k=1,\dots,m}$
of $\left[0,1\right]^{2}$. Then the path space partition can be defined by $\left\{ P_{k}\right\} _{k=1,\dots,m}$ , where
\[
\mathcal{P}_{k}=\left\{ \mathbf{x}_{1}\mathbf{x}_{2}\cdots\in\mathcal{P}\mid\text{\ensuremath{\text{rp}\left(\omega_{\mathbf{x}_{1}\to\mathbf{x}_{2}}\right)\in\mathcal{S}_{k}}}\right\} ,
\]
and $\text{rp}\left(\omega\right)$ is the raster position of 
to the direction $\omega$.

\paragraph*{4d Partition for the Multi-chain Perturbation.}

Let $\mathbf{x}_{c}$ denote the first non-specular vertex in the
current path. For the multi-chain perturbation, the partition is defined
on the space $\left[0,1\right]^{4}$, where the first two dimensions
are in screen space, and the last two dimensions signify the ray direction
originated from $\mathbf{x}_{c}$. We define a 4d partition
$\left\{ \mathcal{S}_{l}\times\mathcal{T}_{j}\right\} _{l,j}$
of $\left[0,1\right]^{4}$, where $\mathcal{S}_{l}\subseteq\left[0,1\right]^{2}$
and $\mathcal{T}_{j}\subseteq\left[0,1\right]^{2}$. Then the path
space partition can be obtained by
\begin{multline}
\mathcal{P}_{k}=\{\mathbf{x}_{1}\mathbf{x}_{2}\cdots\mathbf{x}_{c}\mathbf{x}_{c+1}\cdots\in\mathcal{P}\mid\\
\ensuremath{\text{rp}\left(\omega_{\mathbf{x}_{1}\to\mathbf{x}_{2}}\right)\in\mathcal{S}_{l}}\land\ensuremath{\text{cy}\left(\omega_{\mathbf{x}_{c}\to\mathbf{x}_{c+1}}\right)\in\mathcal{T}_{j}}\},
\end{multline}
where $\text{cy}\left(\omega\right)$ is
the direction $\omega$ in 2d cylindrical coordinates.

\section{Adaptive Partitioning}
\label{sec:adaptive_partitioning}

\begin{figure}
    \begin{centering}
        \def\svgwidth{\linewidth}
        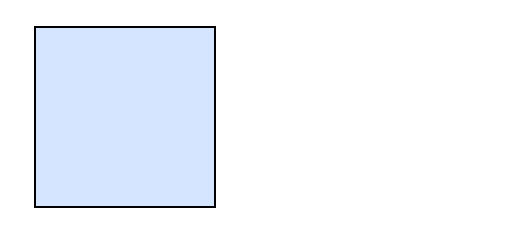
    \end{centering}
    \vspace*{-1em}
    \caption{\label{fig:grid_and_quadtree_partitions}
    Grid and quadtree partitioning
    applied to the 2d canonical space $\mathcal{S}=\left[0,1\right]^{2}$.
    The grid partition split the space into the region with uniform sizes.
    On the other hand, quadtree partition slipts the space into the regions with different sizes.}
\end{figure}

In this section, we will introduce the partitioning strategies for
the subspaces discussed in \cref{subsec:path_space_partition}.
First, we will focus on the partitioning of the 2d canonical space $\mathcal{S}:=\left[0,1\right]^{2}$,
which is used for the lens perturbation. Next, we will
extend the discussion to 4d for the multi-chain perturbation in
\cref{subsec:4d_partition}.

\subsection{2d Partition\label{subsec:2d_partition}}

\paragraph*{Grid Partition.}

The obvious approach to define a partition is using a uniform grid (\cref{fig:grid_and_quadtree_partitions} left),
where each region $\mathcal{S}_{k}$ corresponds to a cell in the $N\times N$ grid.

\paragraph*{Quadtree Partition.}

\begin{figure}[t]
    \centering
    \scriptsize
    \begin{tabular}{
        @{}
        >{\centering\arraybackslash}p{.2\linewidth}
        @{\hspace{.005\linewidth}}
        >{\centering\arraybackslash}p{.2\linewidth}
        @{\hspace{.005\linewidth}}
        >{\centering\arraybackslash}p{.2\linewidth}
        @{\hspace{.005\linewidth}}
        >{\centering\arraybackslash}p{.2\linewidth}
        @{\hspace{.005\linewidth}}
        >{\centering\arraybackslash}p{.2\linewidth}
        @{}
    }
        Error / Grid size &
        $N=1$ &
        $N=5$ &
        $N=15$ &
        $N=19$ \\
        \includegraphics[width=\linewidth]{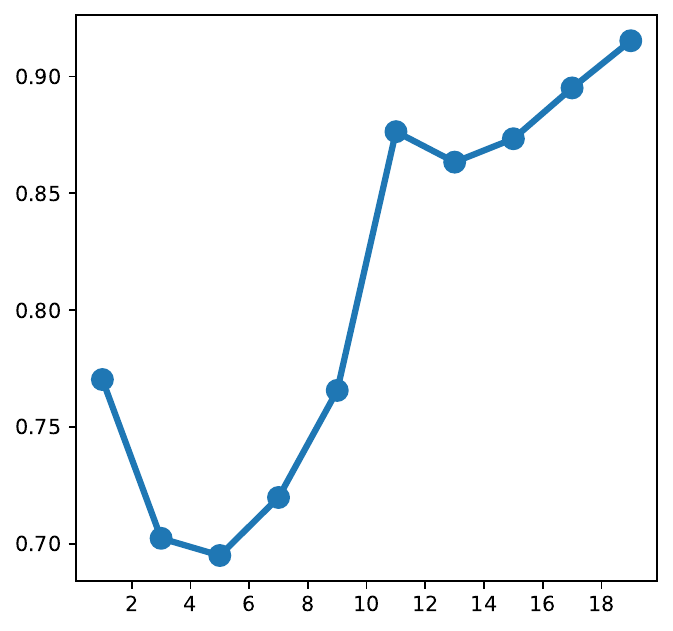} &
        \includegraphics[width=\linewidth]{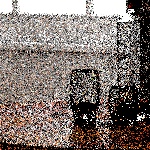} &
        \includegraphics[width=\linewidth]{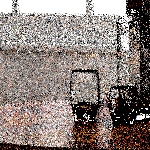} &
        \includegraphics[width=\linewidth]{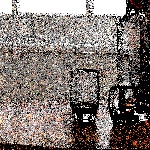} &
        \includegraphics[width=\linewidth]{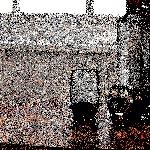} \\
        &
        rRMSE 0.770 &
        rRMSE 0.695 &
        rRMSE 0.873 &
        rRMSE 0.915 \\
    \end{tabular}
    \caption{The plot on the left shows the error with different grid sizes used for partitioning for the regional adaptive multi-chain perturbation (Fireplace Room scene).
    {We use the 4d partitioning with the same parameter $N$ for both top and bottom level grids corresponding to primary and secondary ray directions.}
    We can observe that results with too small as well as too large grid sizes are suboptimal.
    }
    \label{fig:grid_size}
\end{figure}

Although the grid partitioning is simple to implement and mostly effective
for low-dimensional cases, there are two issues. First, the selection
of the grid size $N$ is not obvious. As we discussed in \cref{subsec:path_space_partition},
the performance of the Markov chain vitally depends on the resolution, but a larger $N$ is not
always better {(\cref{fig:grid_size})}. Second, a uniform grid does not adapt to
the variation of the state space.

To resolve these problems, we propose a \emph{quadtree partition
}(\cref{fig:grid_and_quadtree_partitions} right), where the
regions are the leaf nodes of the quadtree. This structure can better adapt
to local properties of the state space without increasing the overall
number of regions, thus keeping the number of samples to estimate regional parameters high.

\subsection{Refining Quadtree Partition\label{subsec:refining_quadtree_partition}}

\begin{figure}
    \begin{centering}
        \def\svgwidth{\linewidth}
        \begingroup%
  \makeatletter%
  \providecommand\color[2][]{%
    \errmessage{(Inkscape) Color is used for the text in Inkscape, but the package 'color.sty' is not loaded}%
    \renewcommand\color[2][]{}%
  }%
  \providecommand\transparent[1]{%
    \errmessage{(Inkscape) Transparency is used (non-zero) for the text in Inkscape, but the package 'transparent.sty' is not loaded}%
    \renewcommand\transparent[1]{}%
  }%
  \providecommand\rotatebox[2]{#2}%
  \newcommand*\fsize{\dimexpr\f@size pt\relax}%
  \newcommand*\lineheight[1]{\fontsize{\fsize}{#1\fsize}\selectfont}%
  \ifx\svgwidth\undefined%
    \setlength{\unitlength}{245.07087527bp}%
    \ifx\svgscale\undefined%
      \relax%
    \else%
      \setlength{\unitlength}{\unitlength * \real{\svgscale}}%
    \fi%
  \else%
    \setlength{\unitlength}{\svgwidth}%
  \fi%
  \global\let\svgwidth\undefined%
  \global\let\svgscale\undefined%
  \makeatother%
  \begin{picture}(1,0.47423209)%
    \lineheight{1}%
    \setlength\tabcolsep{0pt}%
    \put(0,0){\includegraphics[width=\unitlength,page=1]{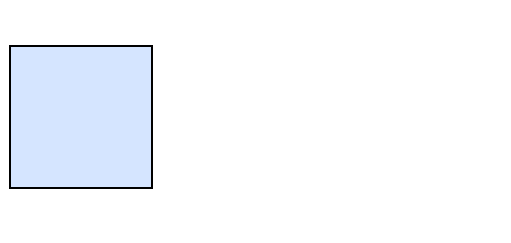}}%
    \put(0.02567461,0.04827584){\color[rgb]{0,0,0}\makebox(0,0)[lt]{\lineheight{1.25}\smash{\begin{tabular}[t]{l}Current partition\end{tabular}}}}%
    \put(0,0){\includegraphics[width=\unitlength,page=2]{fig_quadtree_refinement.pdf}}%
    \put(0.71125471,0.04795705){\color[rgb]{0,0,0}\makebox(0,0)[lt]{\lineheight{1.25}\smash{\begin{tabular}[t]{l}Refined partition\end{tabular}}}}%
    \put(0,0){\includegraphics[width=\unitlength,page=3]{fig_quadtree_refinement.pdf}}%
  \end{picture}%
\endgroup%

    \par\end{centering}
    \caption{\label{fig:quadtree_refinement}Refinement of quadtree partition.
    For each leaf node, it determines whether it is split or not according
    to the criterion based on the number of visit of the region.}
\end{figure}

Here we will introduce a strategy to refine the quadtree partition
adaptively (\cref{fig:quadtree_refinement}).
Initially, the quadtree only contains one node (the root node), which
represents all of $\mathcal{S}$.

The refinement process is executed once in $M_{\text{refine}}$ mutations,
which is selected as a user parameter. While sampling, the number
of visits to each region is recorded, which is later used as a criterion
for the refinement. On adaptation, we traverse the leaf nodes of the
tree and determine if we split the node or not. We adopted a simple
counting-based criterion similar to practical path guiding~\cite{mueller2017pathguiding}.
If the number of visits to the region is beyond a given threshold $M_{\text{split}}$,
then the node is split into four child nodes. Upon the creation of
the child node, the adaptation parameters are inherited:
The scaling parameter is copied from the parent node
as it is. The number of adaptive updates of the child node is set
to $n_{k}/4$, where $n_{k}$ is the number of updates of the parent
node.

The splitting is executed only once for each leaf node {in one refinement process}, and the
generated child nodes are not targeted for the further split operations.
After the split operation, the area of the region of one leaf is reduced by a factor
of 4. This means the region will receive fewer samples
in subsequent iterations and is thus less likely to be split again.

\subsection{4d Partition\label{subsec:4d_partition}}

\begin{figure}
    \begin{centering}
        \def\svgwidth{\linewidth}
        \begingroup%
  \makeatletter%
  \providecommand\color[2][]{%
    \errmessage{(Inkscape) Color is used for the text in Inkscape, but the package 'color.sty' is not loaded}%
    \renewcommand\color[2][]{}%
  }%
  \providecommand\transparent[1]{%
    \errmessage{(Inkscape) Transparency is used (non-zero) for the text in Inkscape, but the package 'transparent.sty' is not loaded}%
    \renewcommand\transparent[1]{}%
  }%
  \providecommand\rotatebox[2]{#2}%
  \newcommand*\fsize{\dimexpr\f@size pt\relax}%
  \newcommand*\lineheight[1]{\fontsize{\fsize}{#1\fsize}\selectfont}%
  \ifx\svgwidth\undefined%
    \setlength{\unitlength}{245.07087527bp}%
    \ifx\svgscale\undefined%
      \relax%
    \else%
      \setlength{\unitlength}{\unitlength * \real{\svgscale}}%
    \fi%
  \else%
    \setlength{\unitlength}{\svgwidth}%
  \fi%
  \global\let\svgwidth\undefined%
  \global\let\svgscale\undefined%
  \makeatother%
  \begin{picture}(1,0.47423209)%
    \lineheight{1}%
    \setlength\tabcolsep{0pt}%
    \put(0.14426274,0.01384703){\color[rgb]{0,0,0}\makebox(0,0)[lt]{\lineheight{1.25}\smash{\begin{tabular}[t]{l}Top-level grid\end{tabular}}}}%
    \put(0,0){\includegraphics[width=\unitlength,page=1]{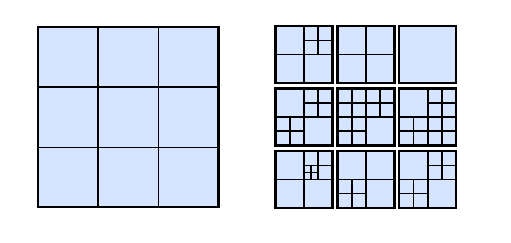}}%
    \put(0.53322989,0.01317918){\color[rgb]{0,0,0}\makebox(0,0)[lt]{\lineheight{1.25}\smash{\begin{tabular}[t]{l}Bottom-level quadtree\end{tabular}}}}%
  \end{picture}%
\endgroup%

    \par\end{centering}
    \caption{\label{fig:grid_quadtree_4d_partition}4d partition using a
    2d grid and a 2d quadtree. The grid is used for screen space, and the
    quadtree for the outgoing direction.}
\end{figure}

For the 4d partition of the multi-chain perturbations,
we use both a grid and a quadtree (\cref{fig:grid_quadtree_4d_partition}).
The first two dimensions in screen space use
the grid. And each grid cell holds a quadtree representing the other two
dimensions corresponding to the outgoing direction at the first non-specular vertex.
The refinement of the quadtree is executed the same as before
\cref{subsec:refining_quadtree_partition}. Note that we opted for a simpler combination of grid and quadtree, as we did not observe artifacts at image tile (grid cell) boundaries with multi-chain perturbations. Moreover, nested quadtrees would require a more complicated splitting heuristic.

\section{Implementation}

\paragraph*{Parallel Chains and Synchronization.}

Following the standard practice of the parallelization of MCMC rendering, 
the parallelization is achieved by maintaining multiple chains for each thread. 
Thus the statistics needed for adaptation can be accessed from multiple threads, requiring synchronization on update. 
However, it is inefficient to use a mutex lock for the every invocation of the adaptive update.
Instead, we use atomic counters to maintain the number of visits $i_k$ and the accumulated acceptance probability $A_k$, and introduce a mutex lock only when the scaling parameter is actually updated (if-clause from line~13 in Algorithm~\ref{alg:ramlt_base}).
This reduces the number of possible locks by the factor of $L$.

\paragraph*{Explicit Synchronization for Refinement of Partitions.}
The refinement process of the quadtree partition must be synchronized,
since it is involved in the operation that changes the structure of the quadtree.
To support this, we applied a small modification to the top-level loop of the path space MLT implementation, which explicitly synchronizes the parallel process for a given number of mutations.
It gives a chance for the adaptive mutation strategy to update the underlying structure.

\begin{figure*}[t]
    \centering 
    \setlength{\tabcolsep}{0.1pt}
    \renewcommand{\arraystretch}{.05}
    \begin{tabular}{m{0.3cm} *{4}{m{.24\linewidth}}}
\rotatebox{90}{Necklace}
&\imgtext{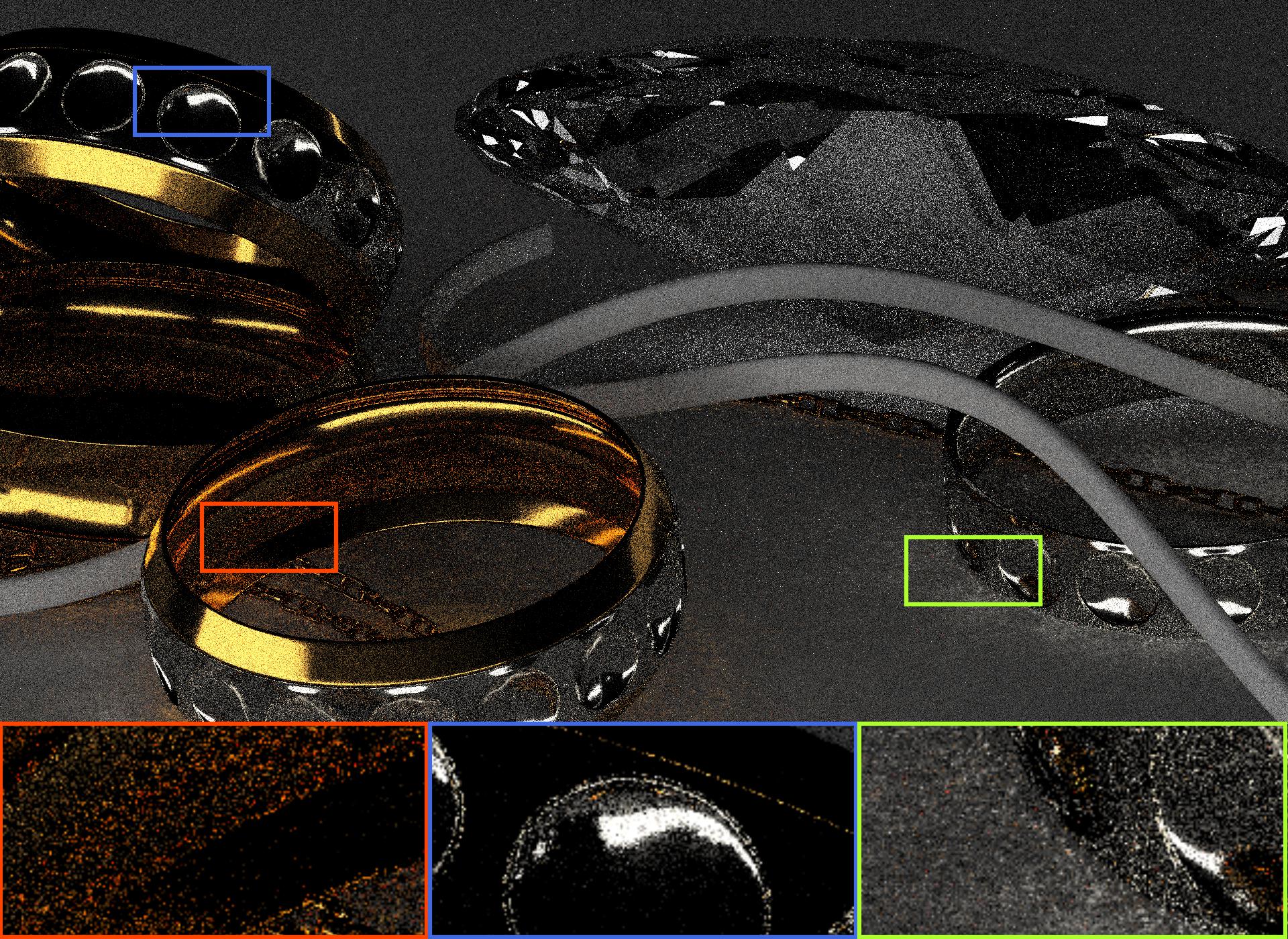}{\ltext{\textsc{Fixed}}}{\rtext{rRMSE 2.1921}}
&\imgtext{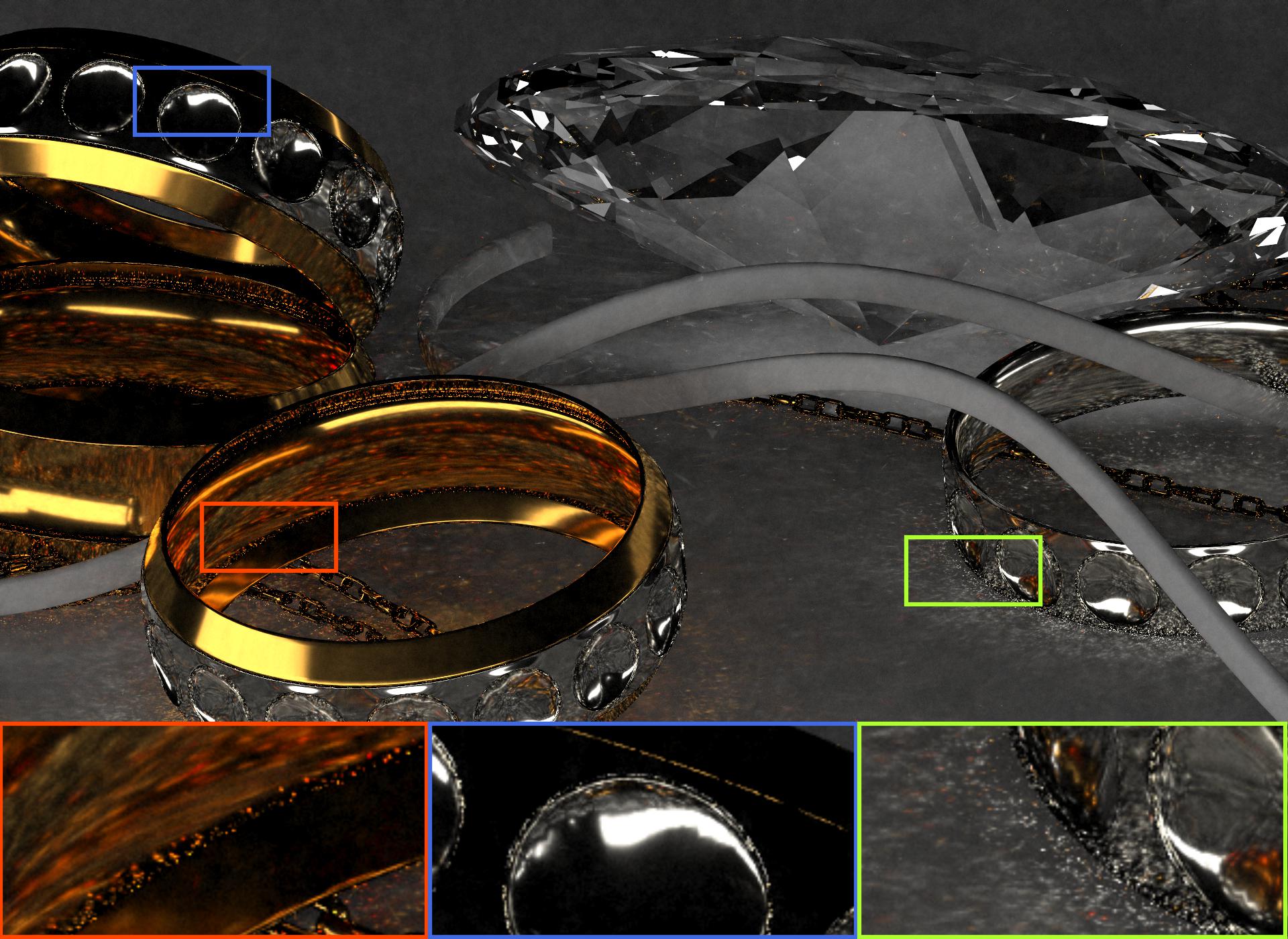}{\ltext{\textsc{Global}}}{\rtext{rRMSE 0.8403}}
&\imgtext{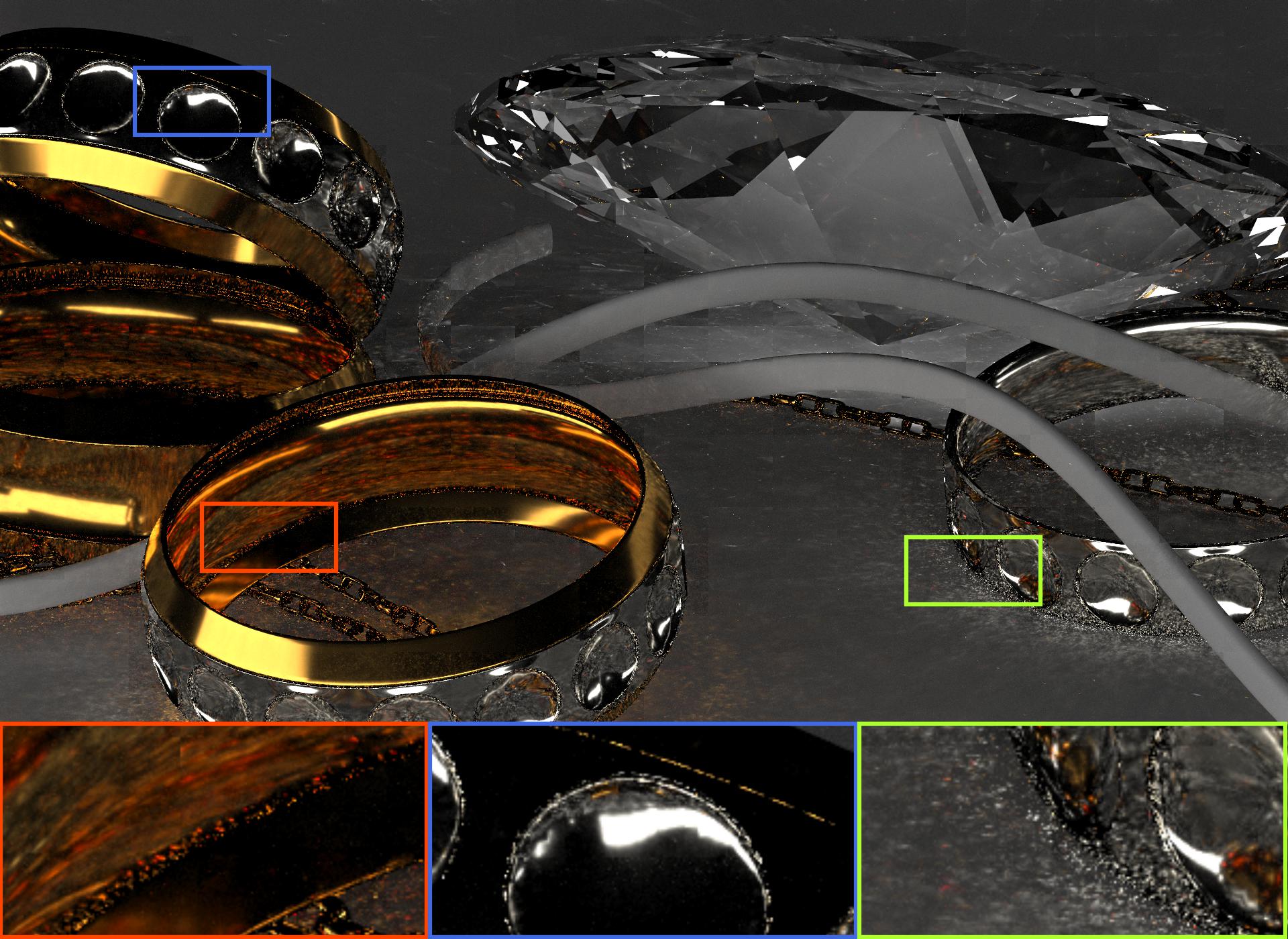}{\ltext{\textsc{RA-Grid}}}{\rtext{rRMSE 0.8805}}
&\imgtext{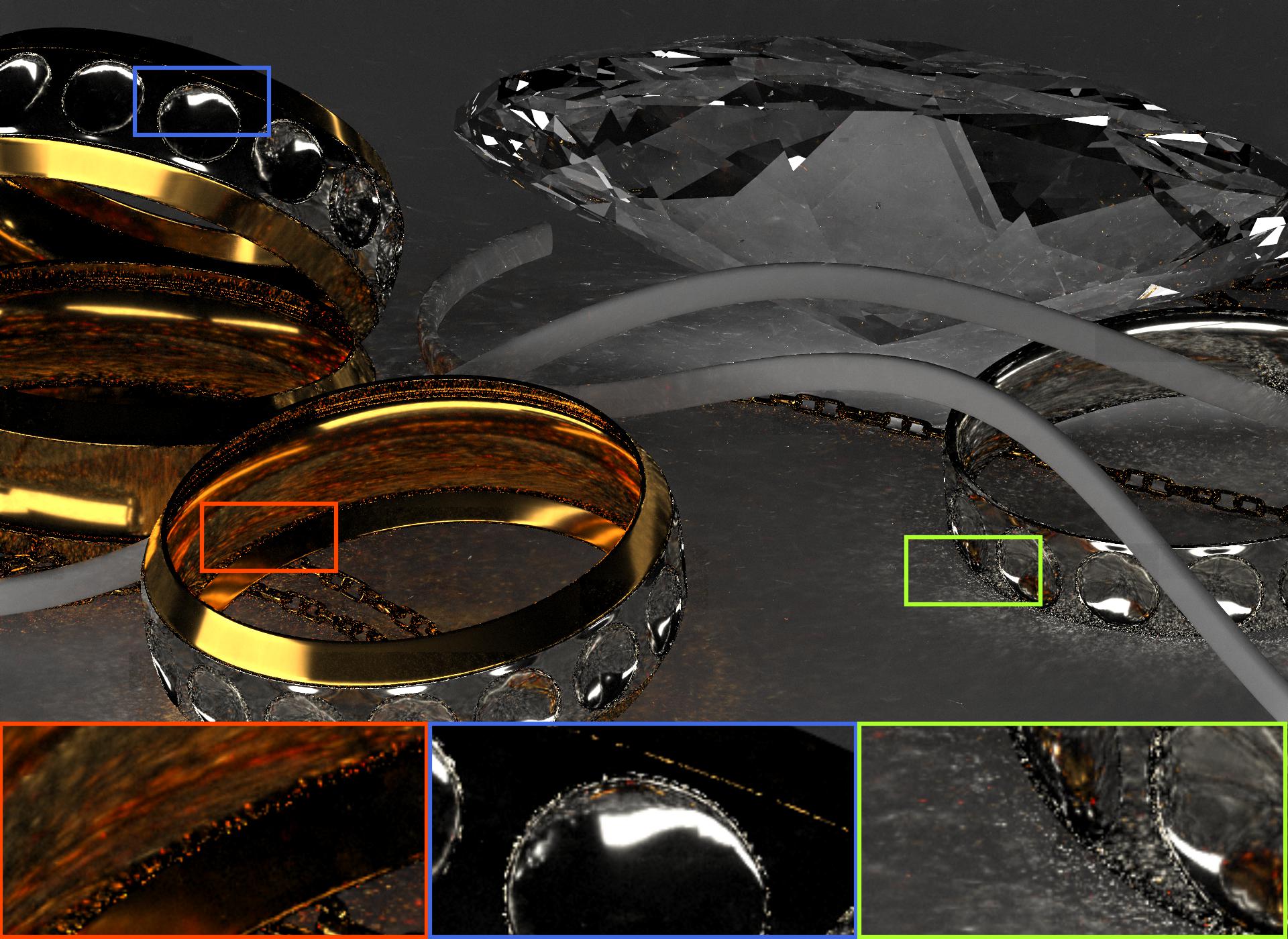}{\ltext{\textsc{RA-Quadtree}}}{\rtext{rRMSE 0.9537}}
\\
\rotatebox{90}{Living Room}
&\imgtext{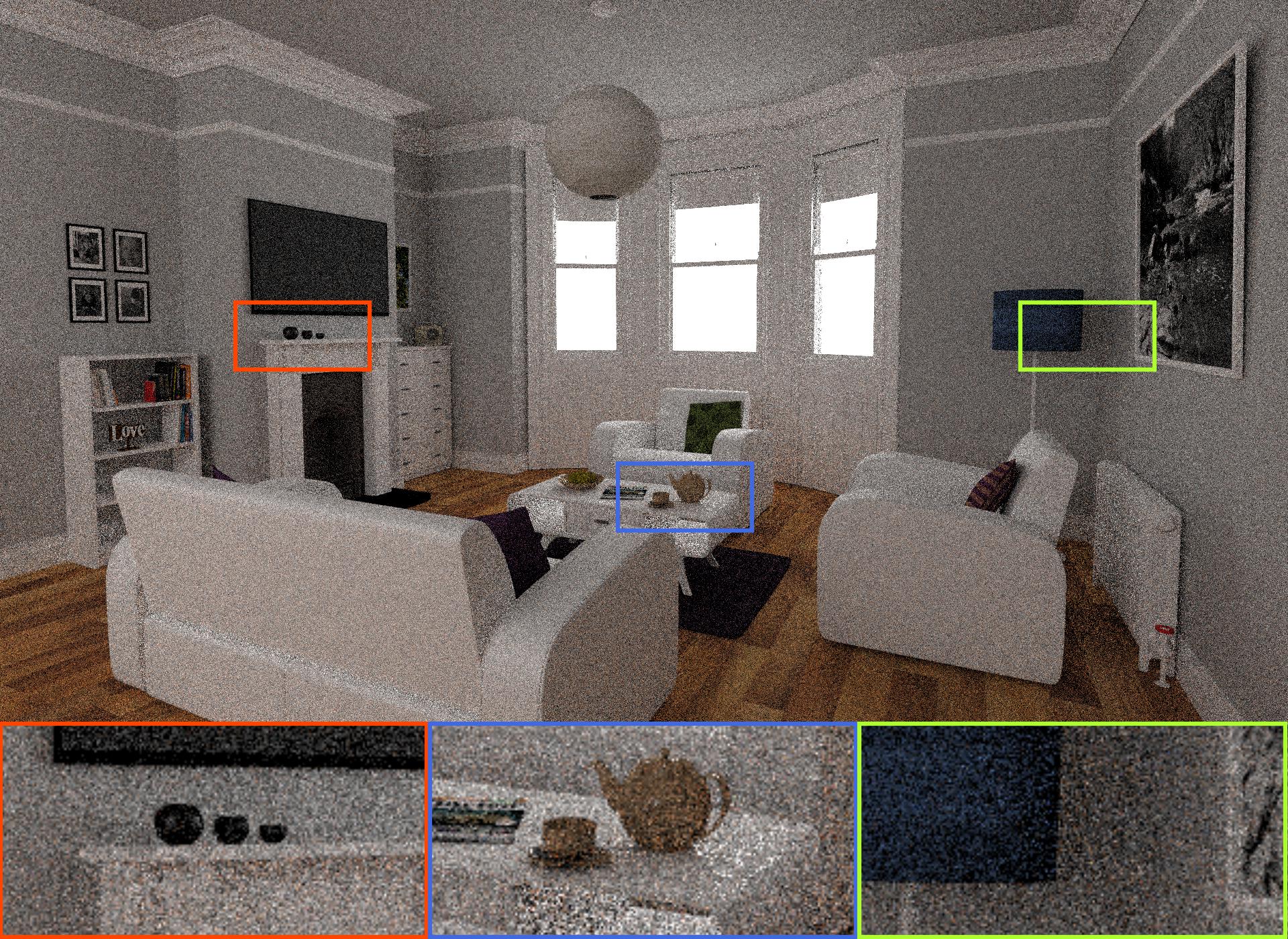}{\ltext{\textsc{Fixed}}}{\rtext{rRMSE 0.4933}}
&\imgtext{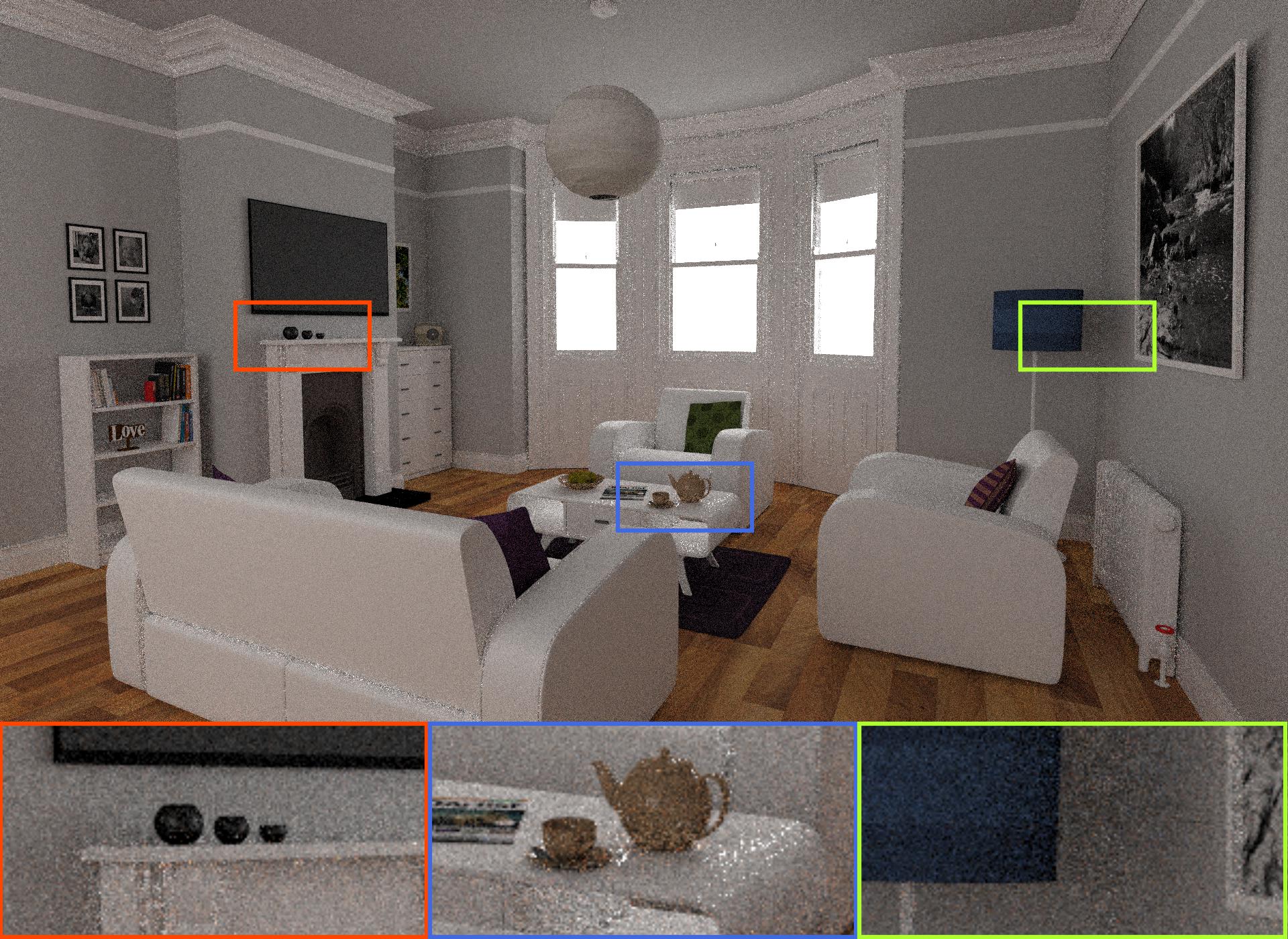}{\ltext{\textsc{Global}}}{\rtext{rRMSE 0.2314}}
&\imgtext{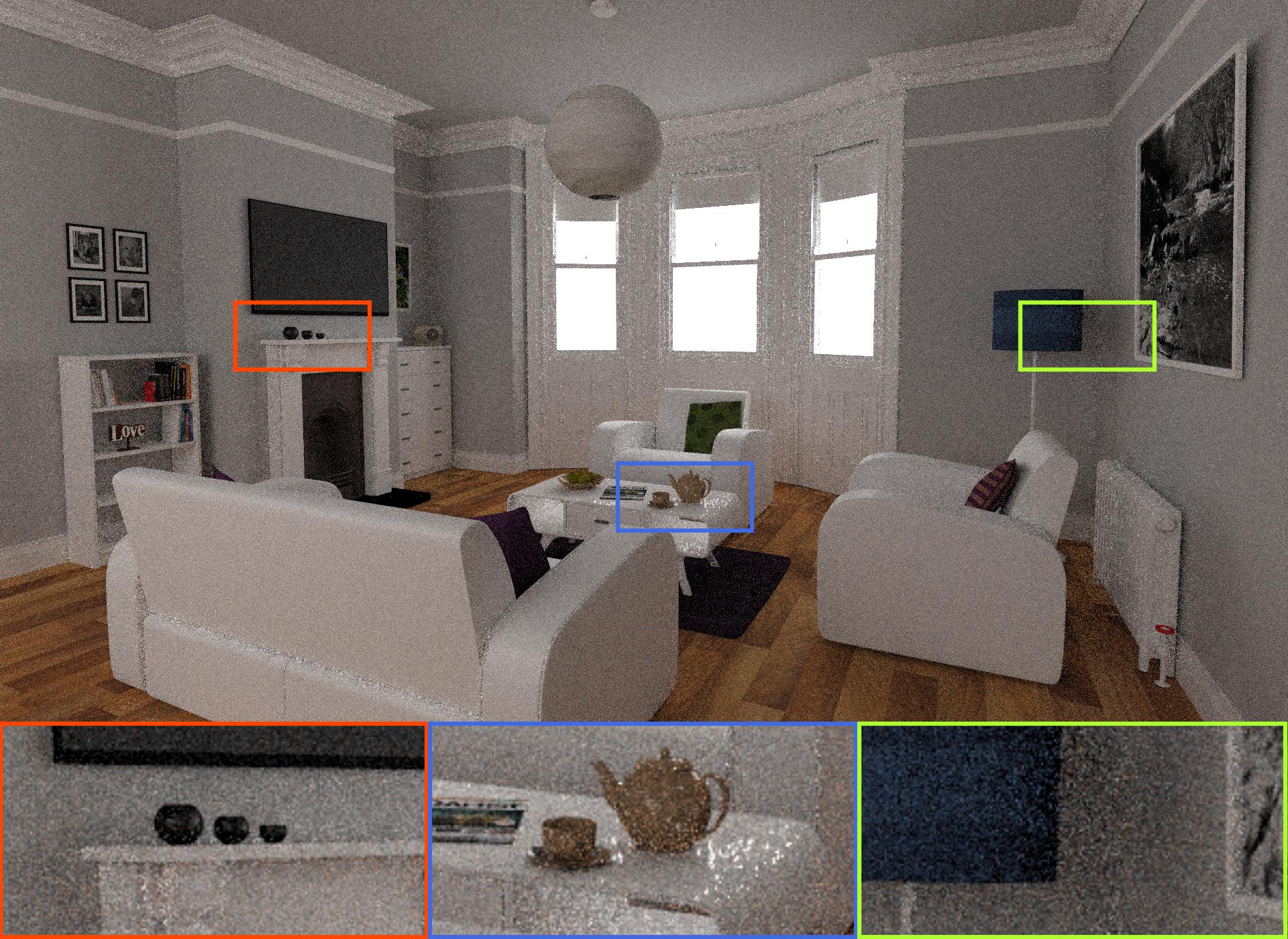}{\ltext{\textsc{RA-Grid}}}{\rtext{rRMSE 0.2375}}
&\imgtext{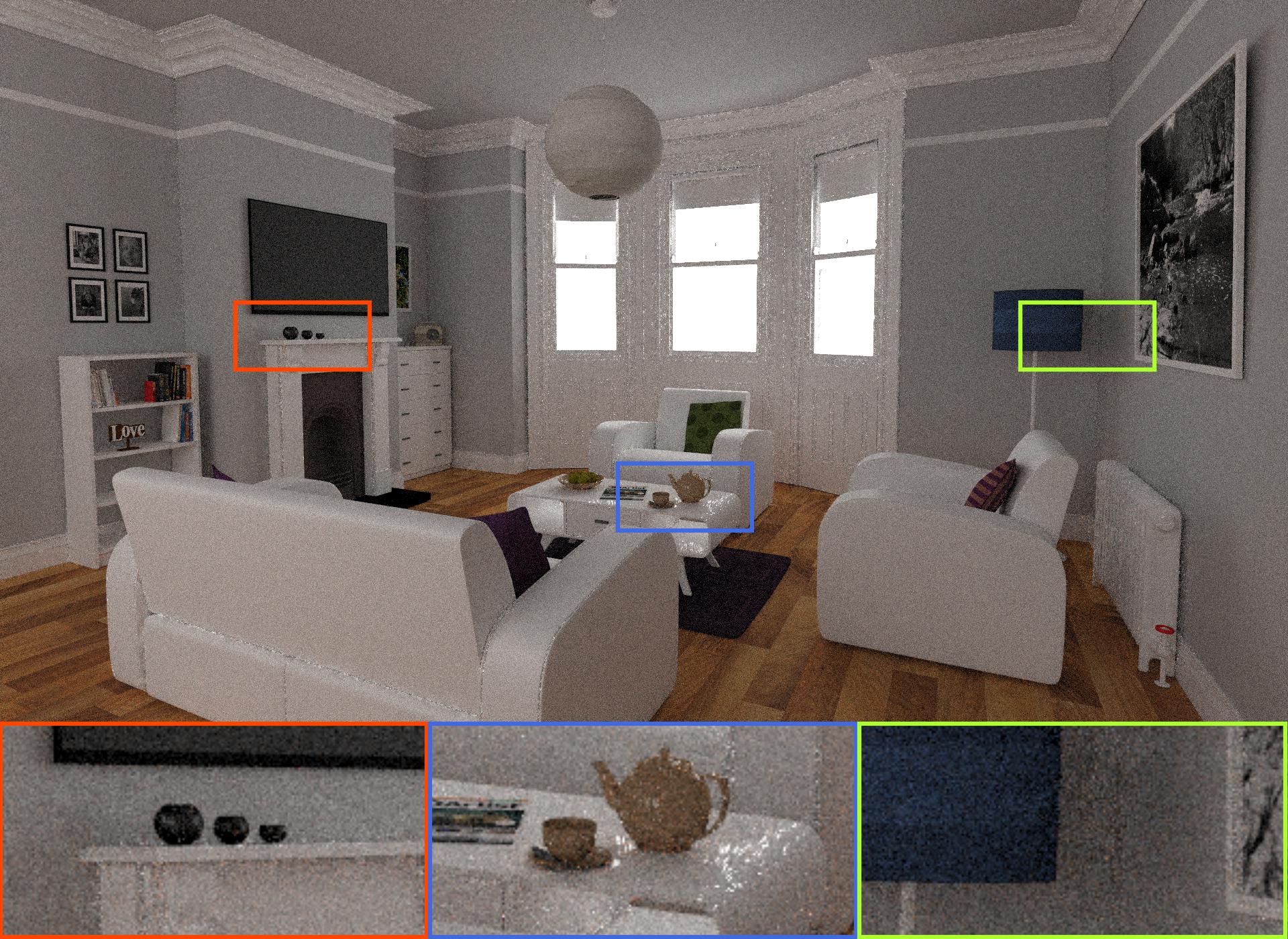}{\ltext{\textsc{RA-Quadtree}}}{\rtext{rRMSE 0.2098}}
\\
\rotatebox{90}{Ajar Door}
&\imgtext{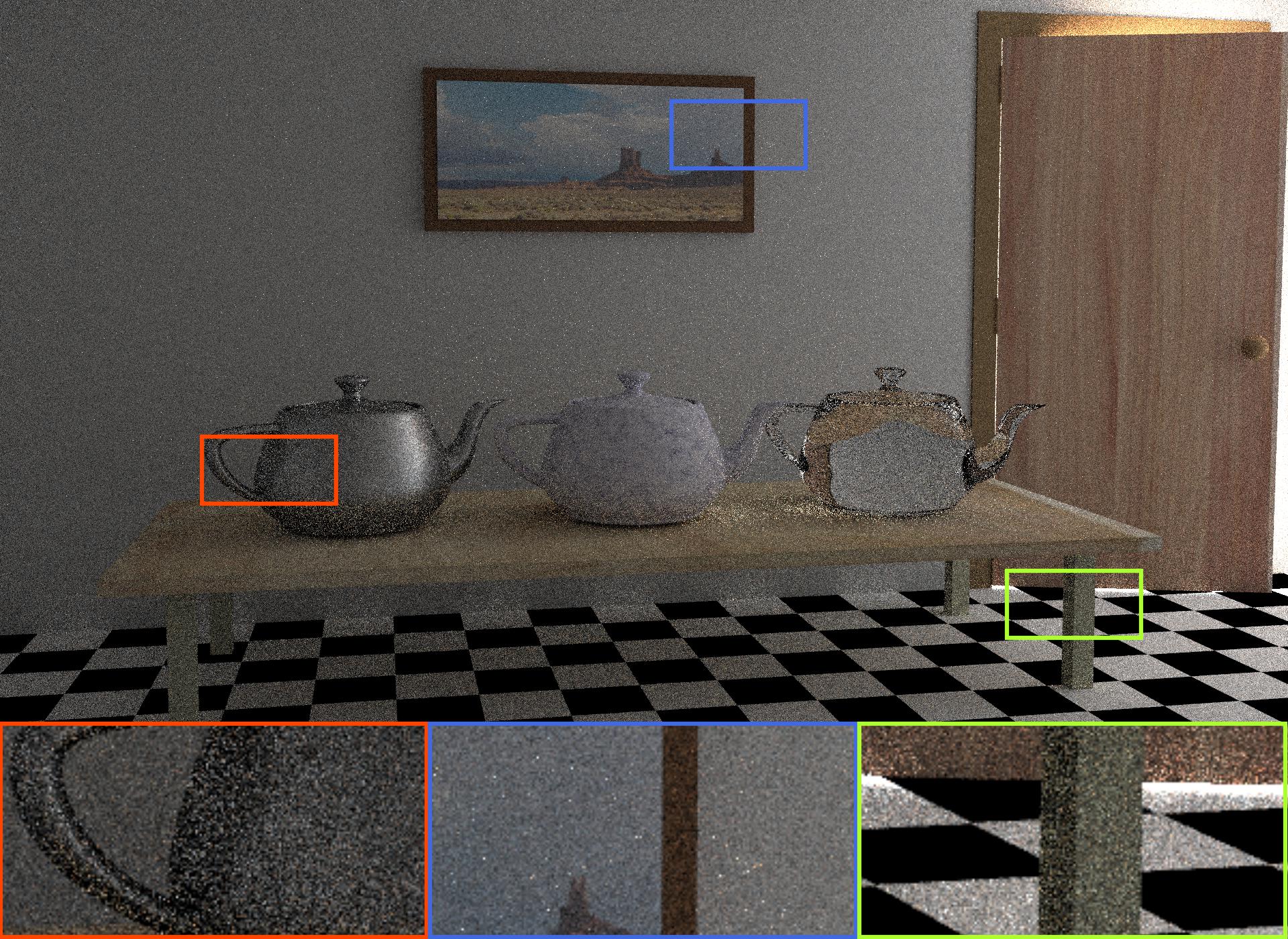}{\ltext{\textsc{Fixed}}}{\rtext{rRMSE 0.9388}}
&\imgtext{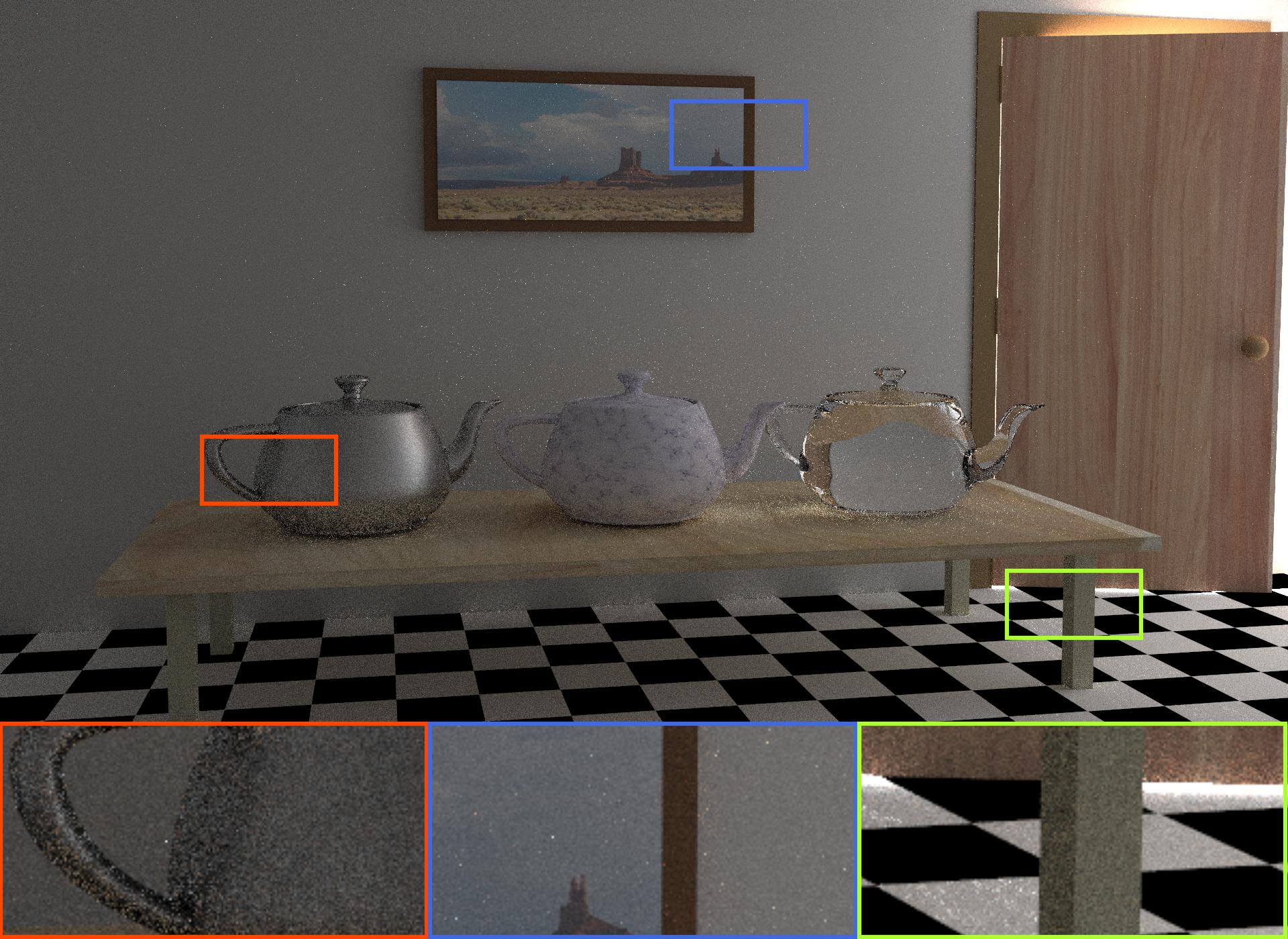}{\ltext{\textsc{Global}}}{\rtext{rRMSE 0.4232}}
&\imgtext{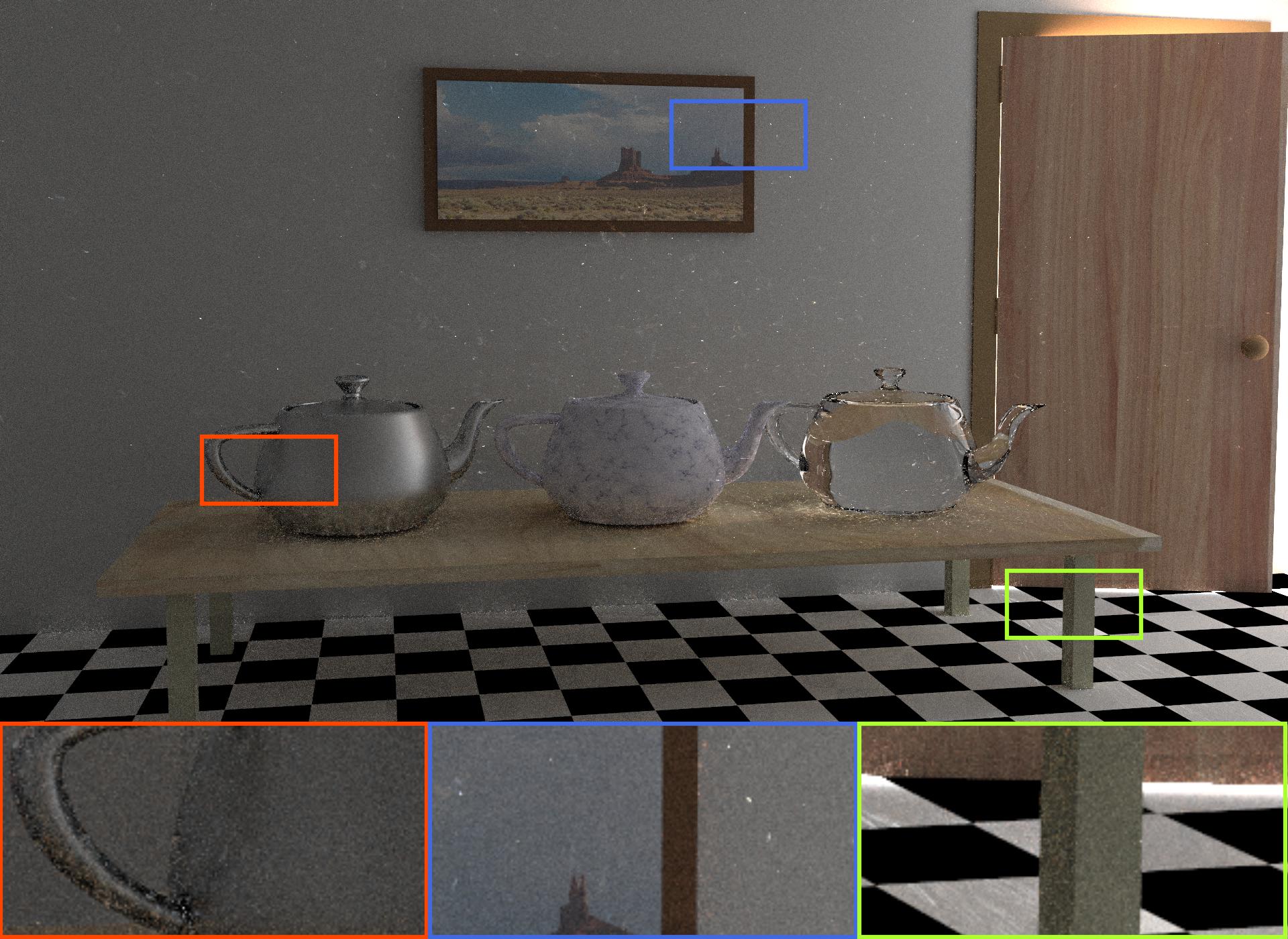}{\ltext{\textsc{RA-Grid}}}{\rtext{rRMSE 0.3650}}
&\imgtext{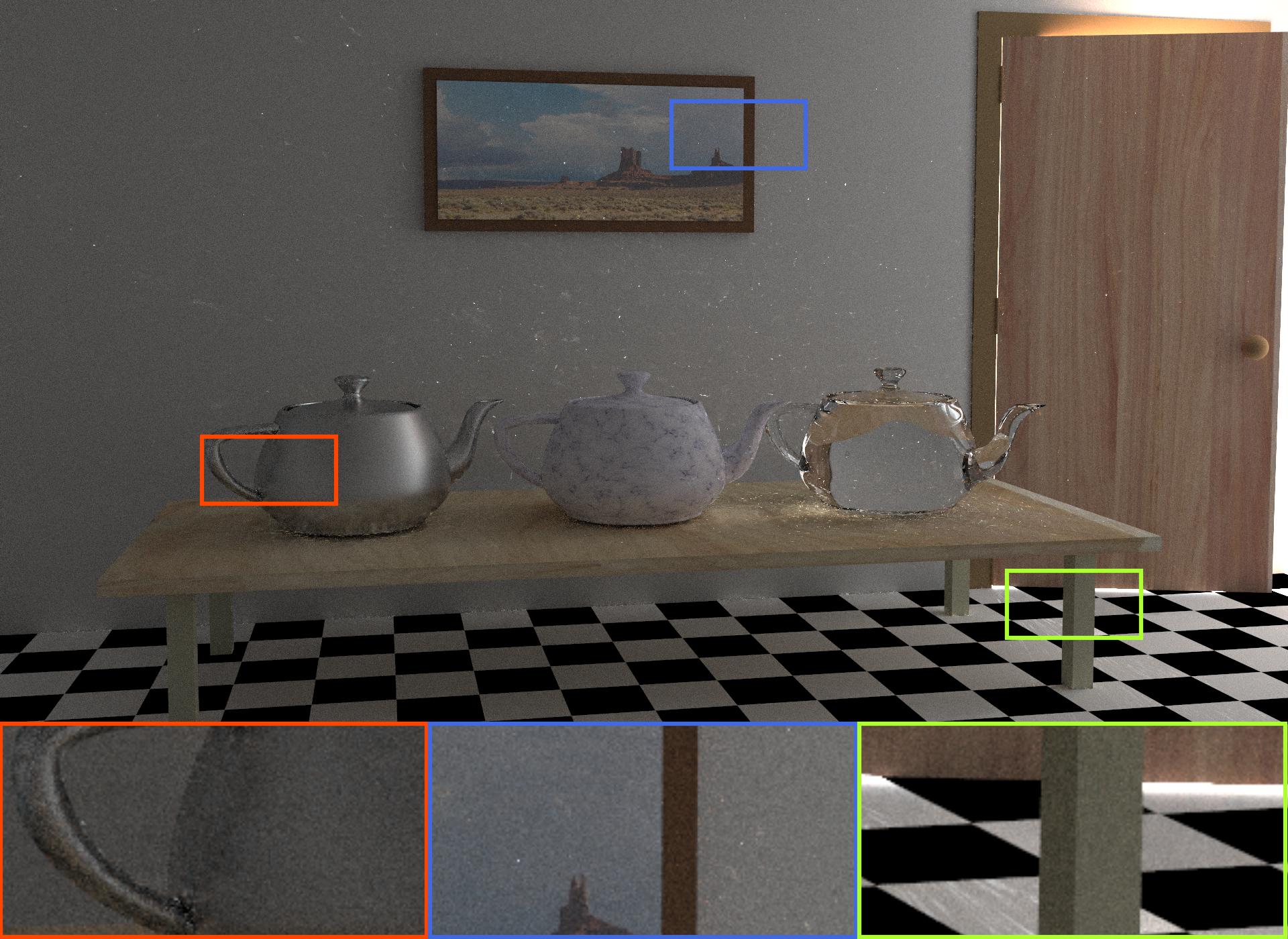}{\ltext{\textsc{RA-Quadtree}}}{\rtext{rRMSE 0.3176}}
\\
\end{tabular}

    \caption{Equal-time comparisons of the four variants of multi-chain perturbations (\textsc{Fixed}, \textsc{Global}, \textsc{RA-Grid}, \textsc{RA-Quadtree}) rendered with the three scenes (Necklace, Living Room, Ajar Door).}
    \label{fig:rendering_multichain}
\end{figure*}

\begin{figure*}[t]
    \centering 
    \setlength{\tabcolsep}{0.3pt}
    \renewcommand{\arraystretch}{.1}
    \begin{tabular}{m{0.3cm} *{4}{m{.24\linewidth}}}
\rotatebox{90}{Fireplace Room}
&\imgtext{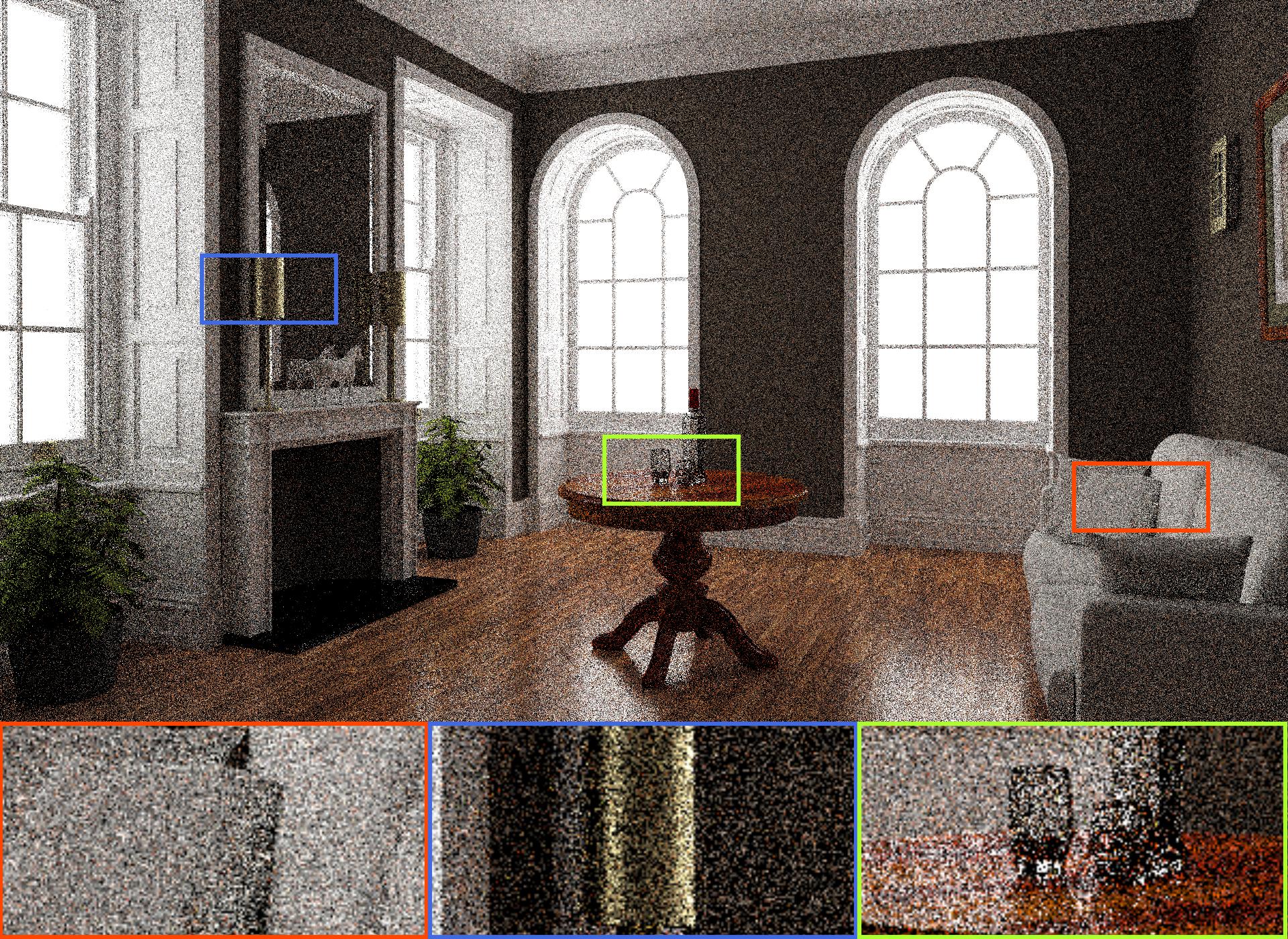}{\ltext{\textsc{Fixed}}}{\rtext{rRMSE 0.6329}}
&\imgtext{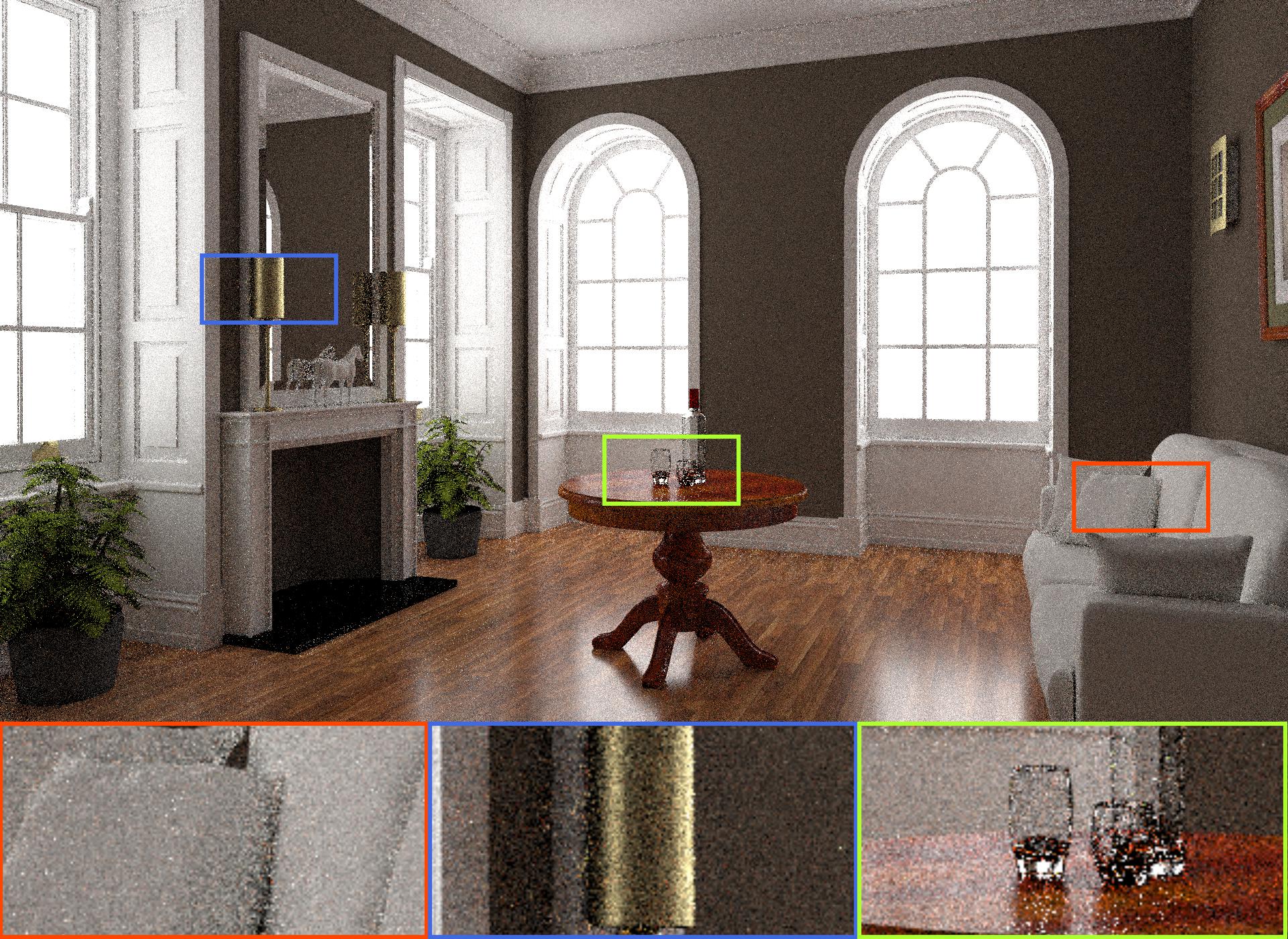}{\ltext{\textsc{Global}}}{\rtext{rRMSE 0.2458}}
&\imgtext{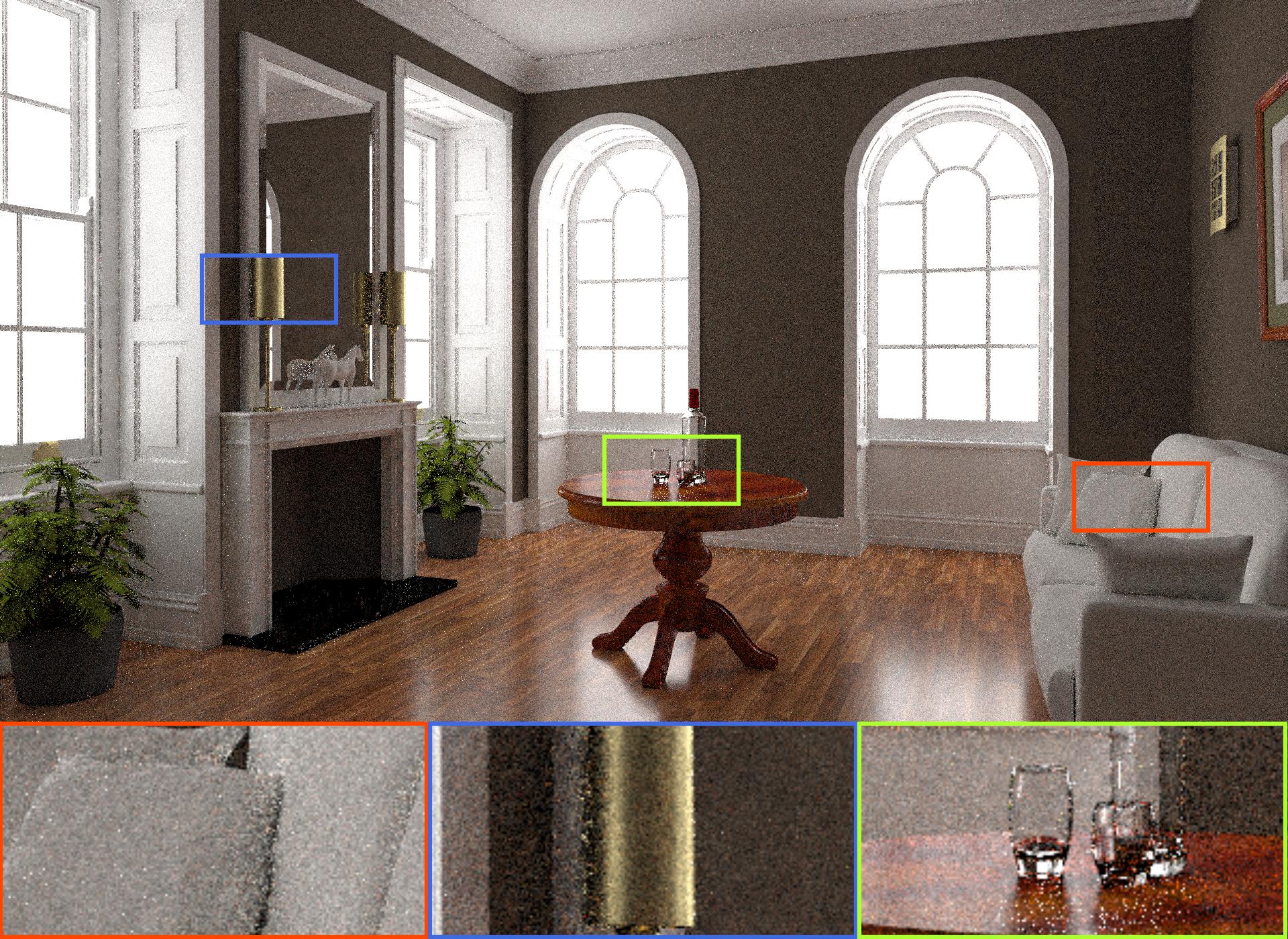}{\ltext{\textsc{RA-Grid}}}{\rtext{rRMSE 0.2179}}
&\imgtext{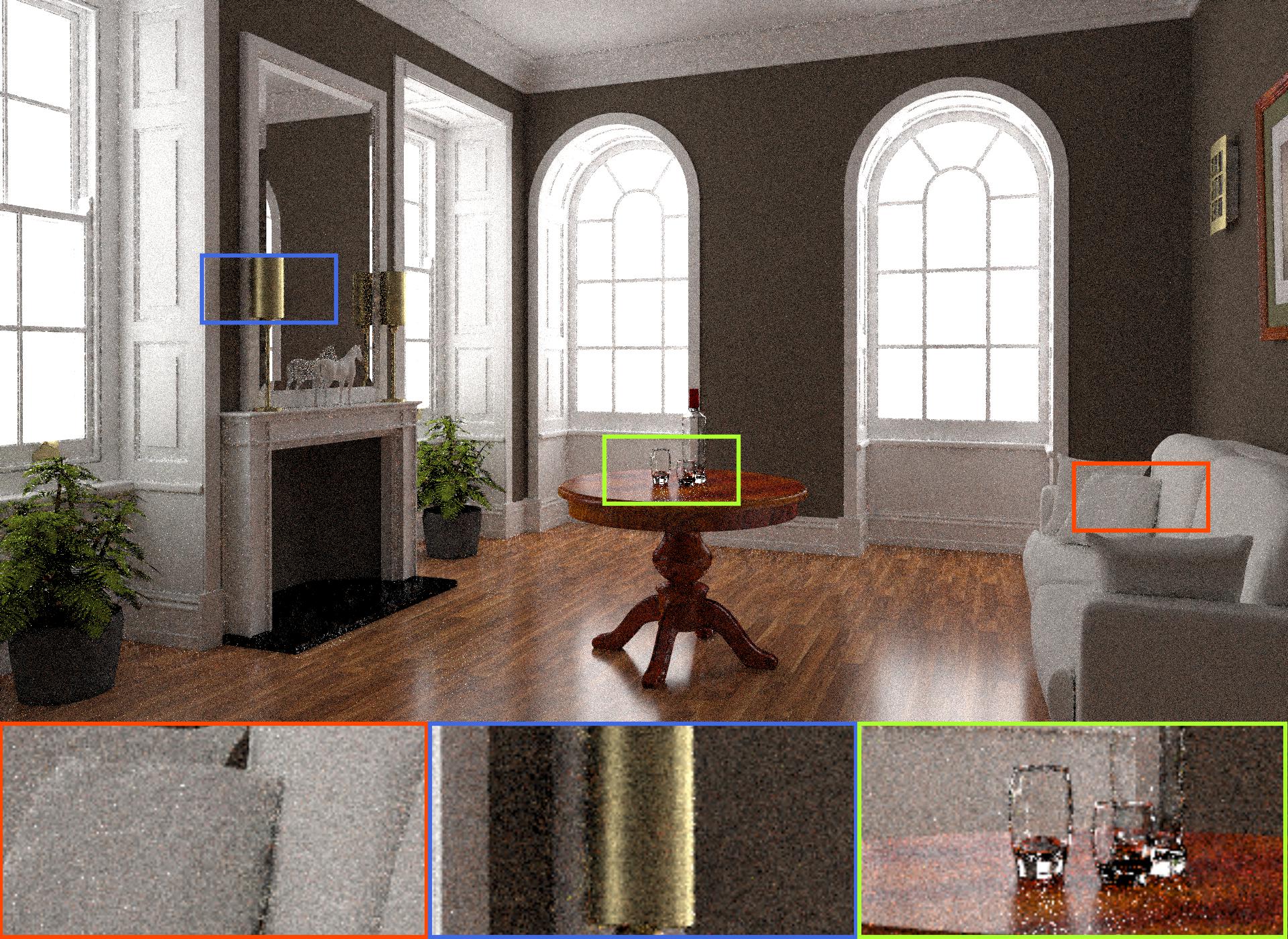}{\ltext{\textsc{RA-Quadtree}}}{\rtext{rRMSE 0.2142}}
\\
\rotatebox{90}{Necklace}
&\imgtext{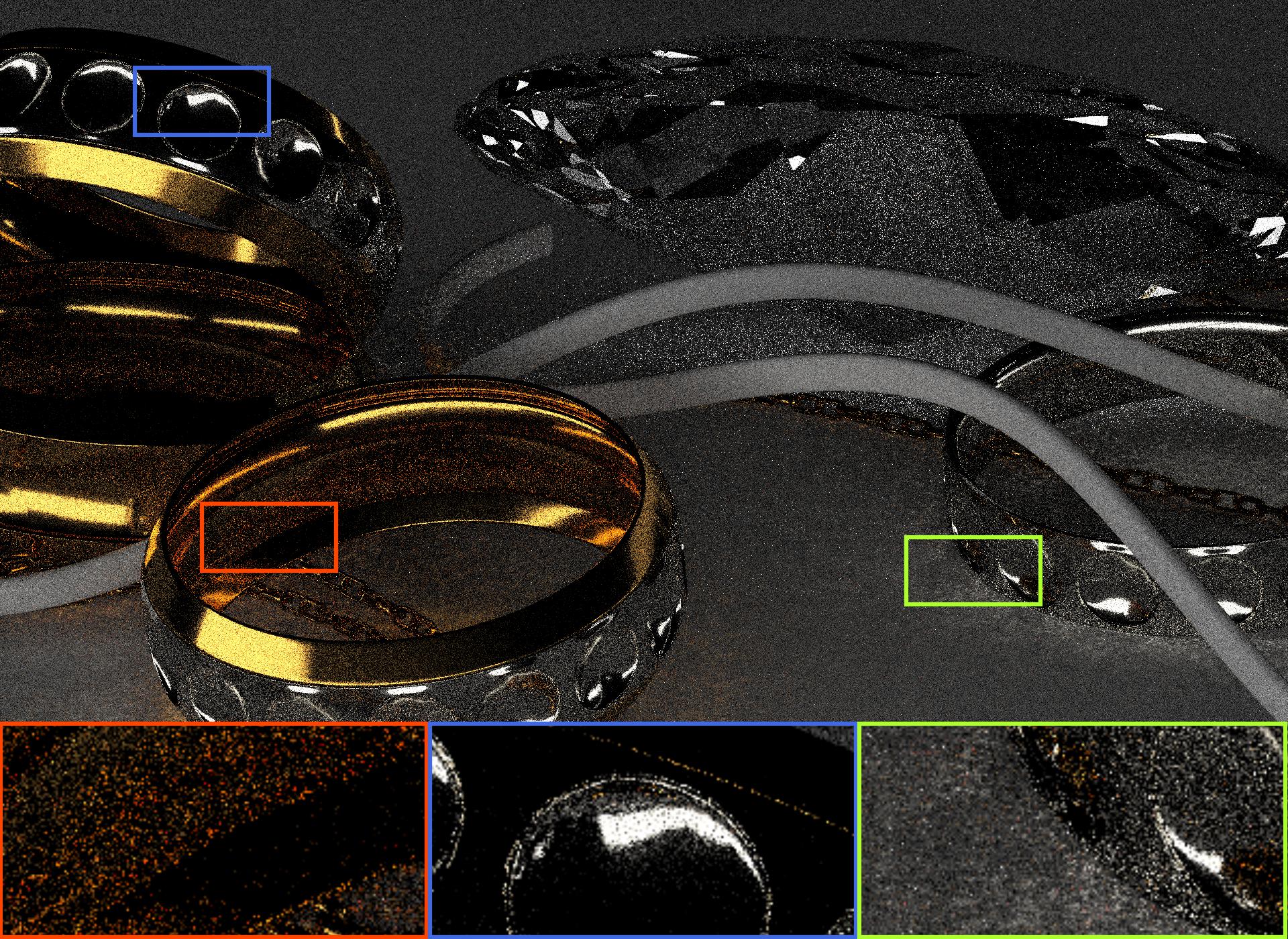}{\ltext{\textsc{Fixed}}}{\rtext{rRMSE 2.5103}}
&\imgtext{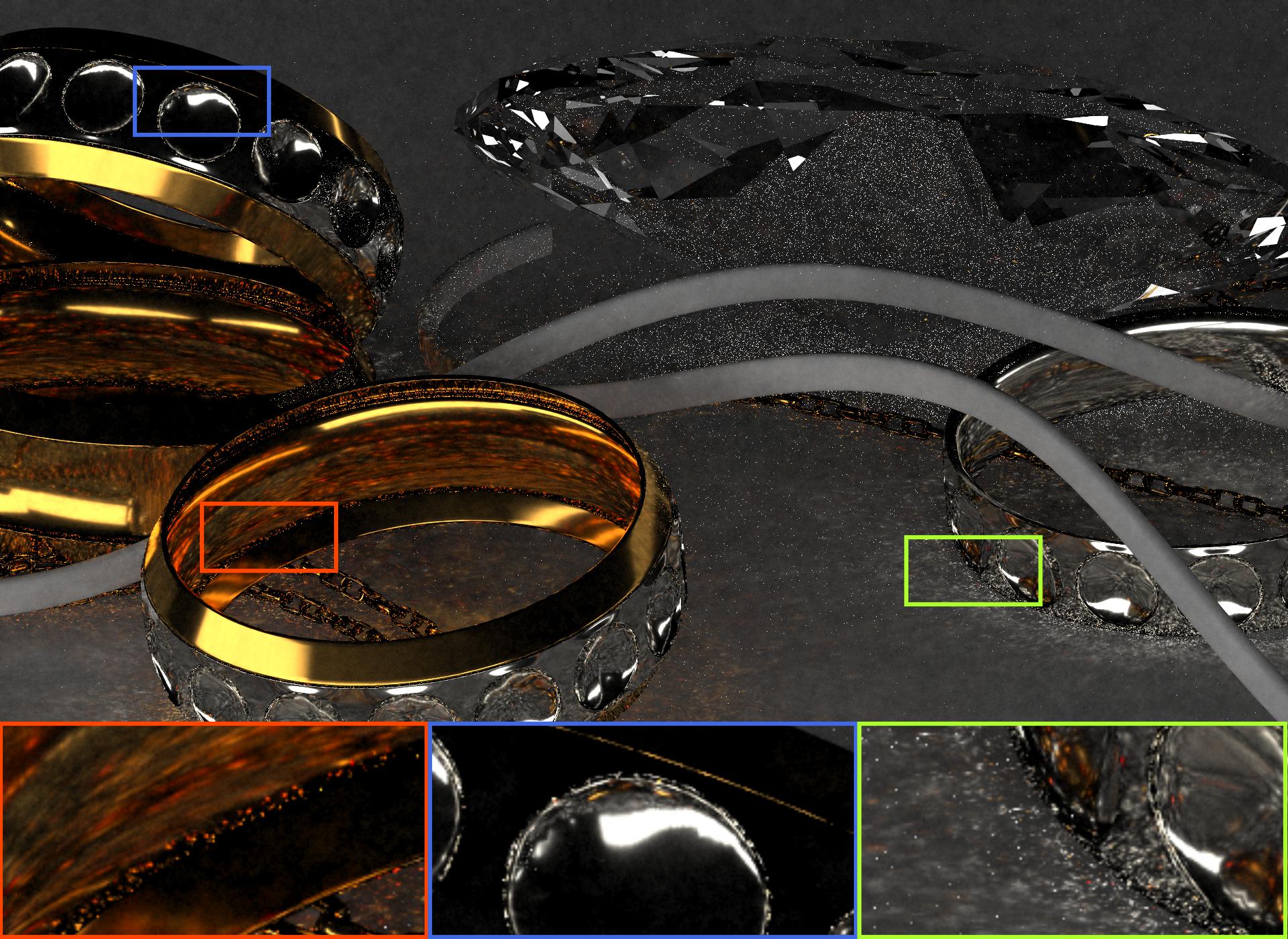}{\ltext{\textsc{Global}}}{\rtext{rRMSE 2.6005}}
&\imgtext{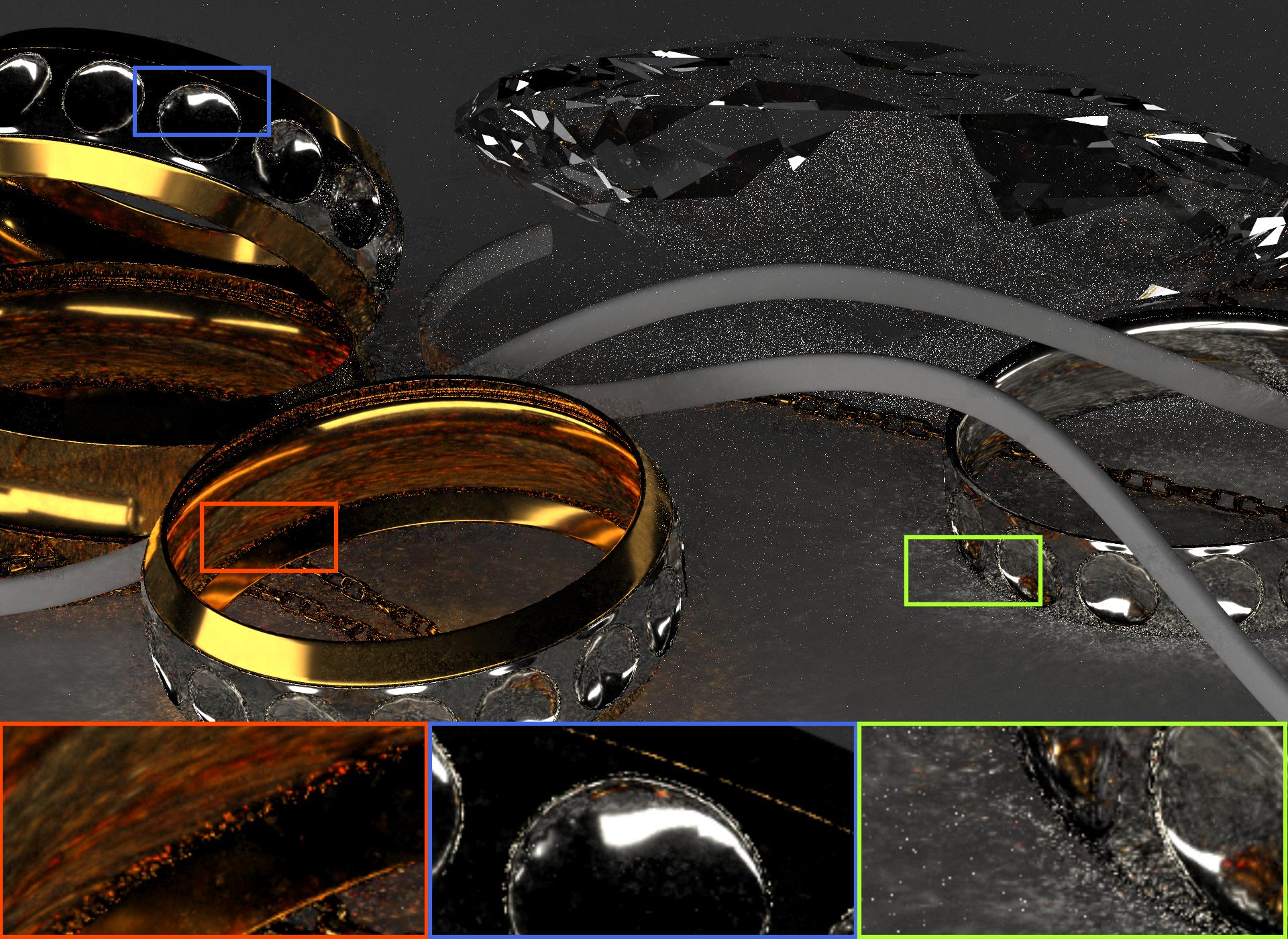}{\ltext{\textsc{RA-Grid}}}{\rtext{rRMSE 2.0825}}
&\imgtext{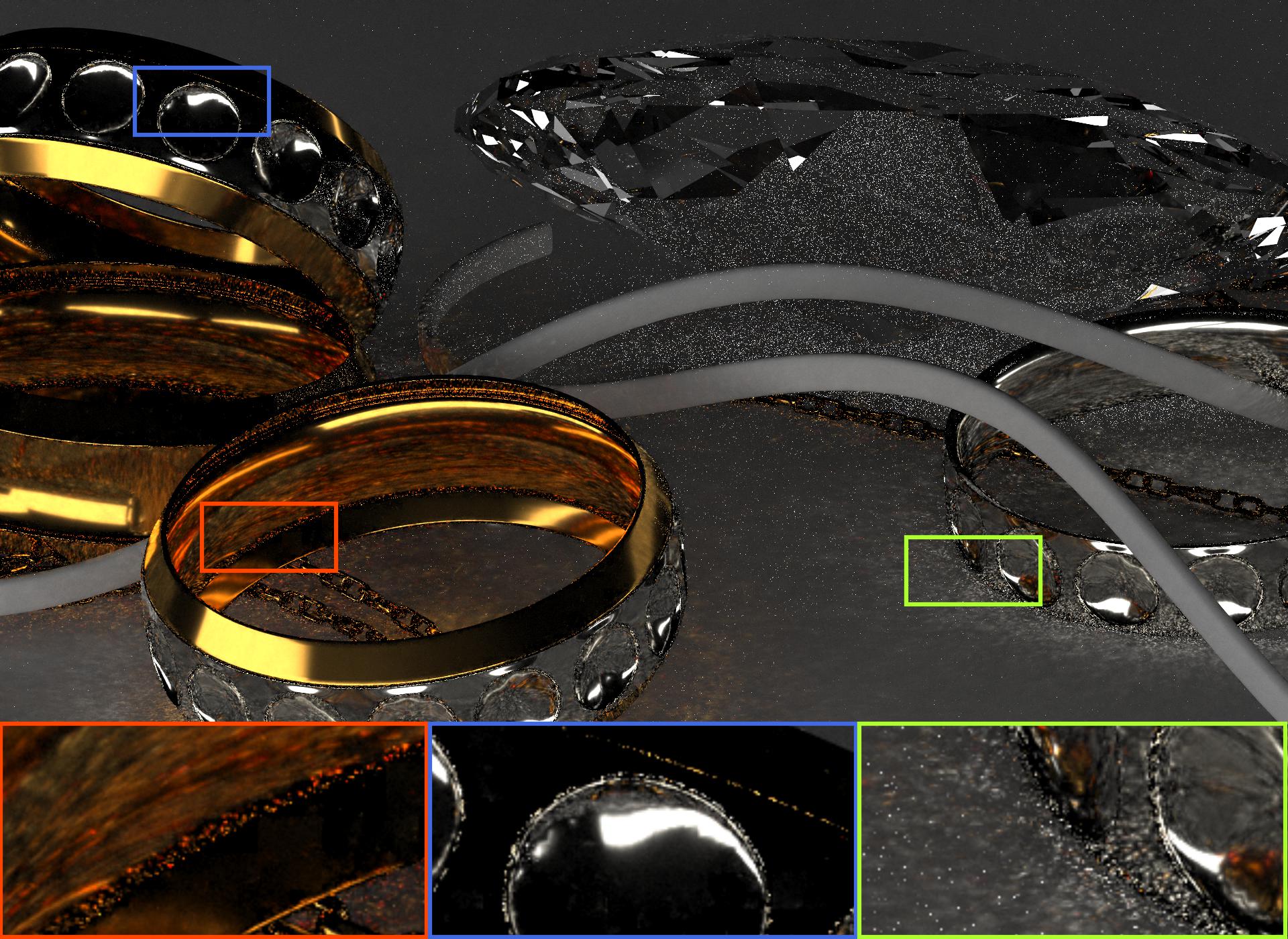}{\ltext{\textsc{RA-Quadtree}}}{\rtext{rRMSE 2.1058}}
\\
\rotatebox{90}{Living Room}
&\imgtext{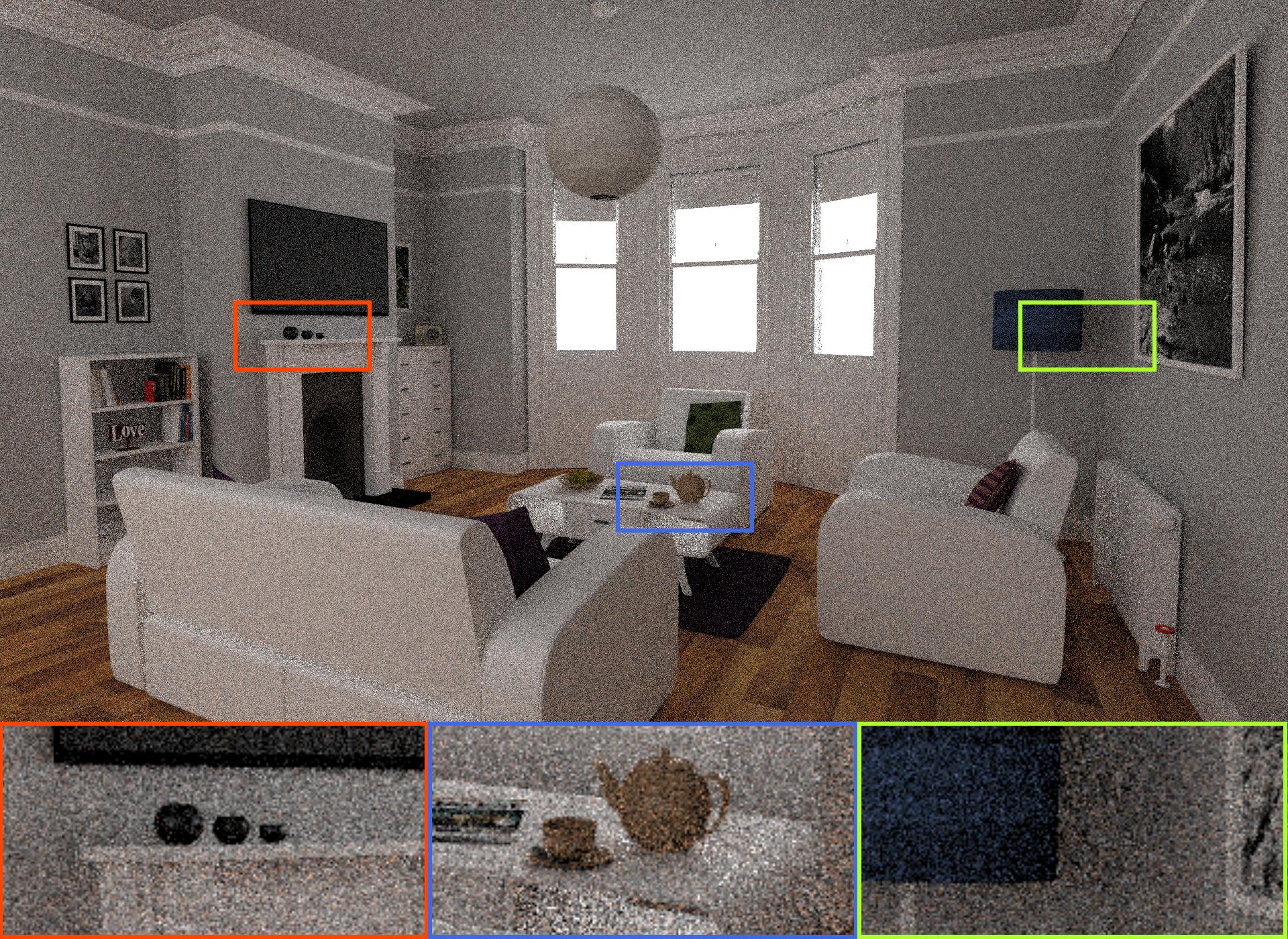}{\ltext{\textsc{Fixed}}}{\rtext{rRMSE 0.4993}}
&\imgtext{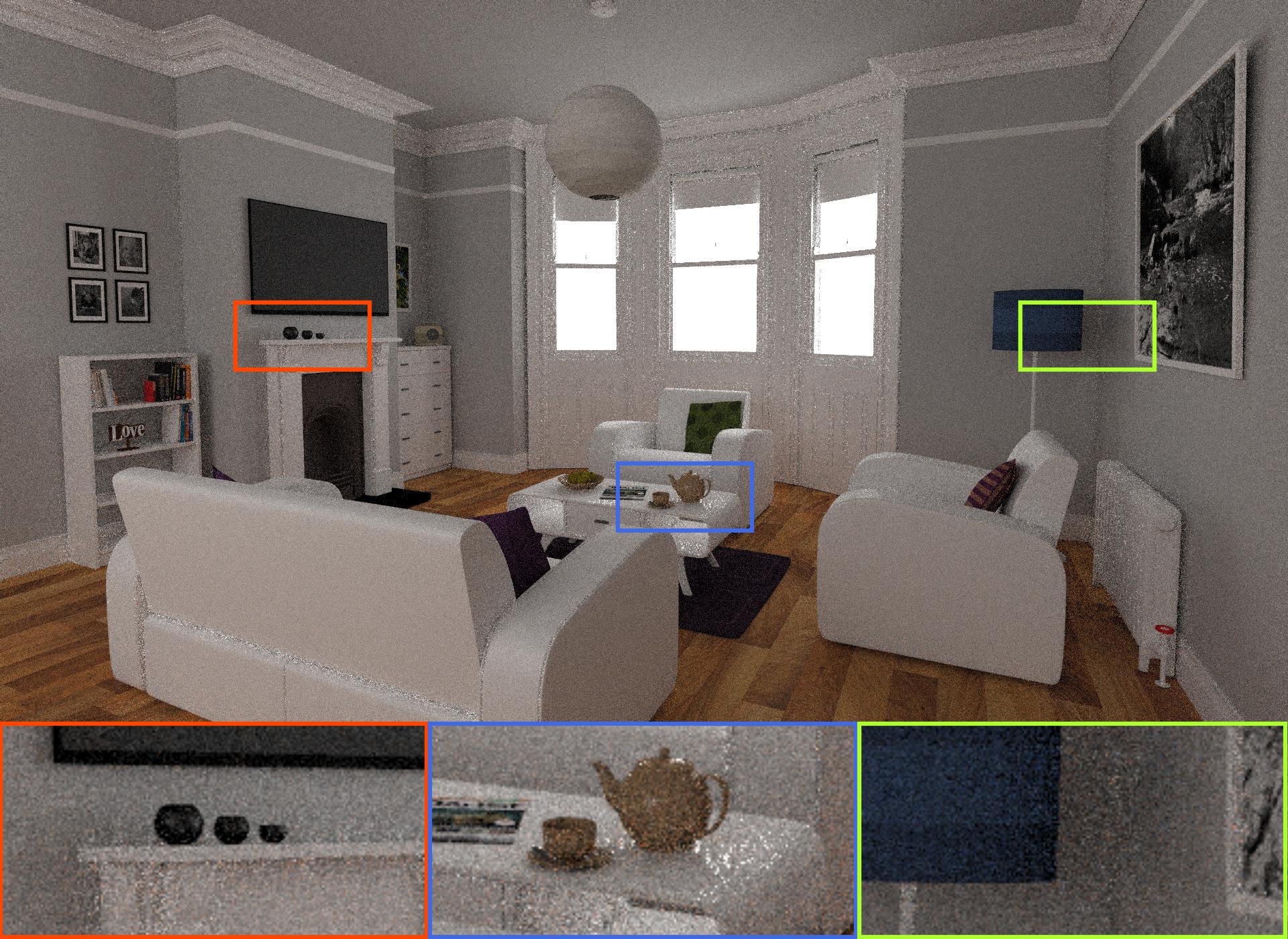}{\ltext{\textsc{Global}}}{\rtext{rRMSE 0.2315}}
&\imgtext{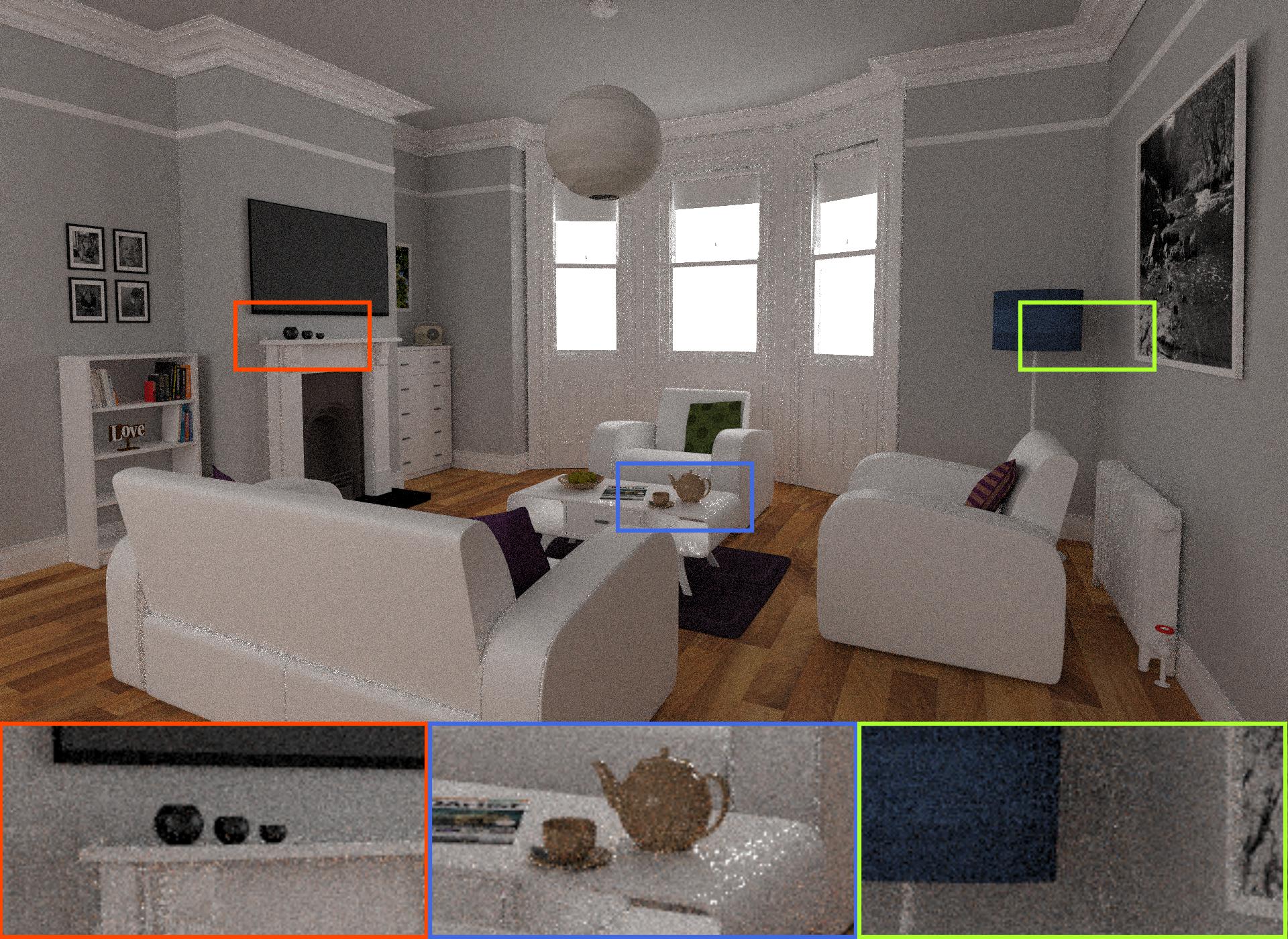}{\ltext{\textsc{RA-Grid}}}{\rtext{rRMSE 0.2209}}
&\imgtext{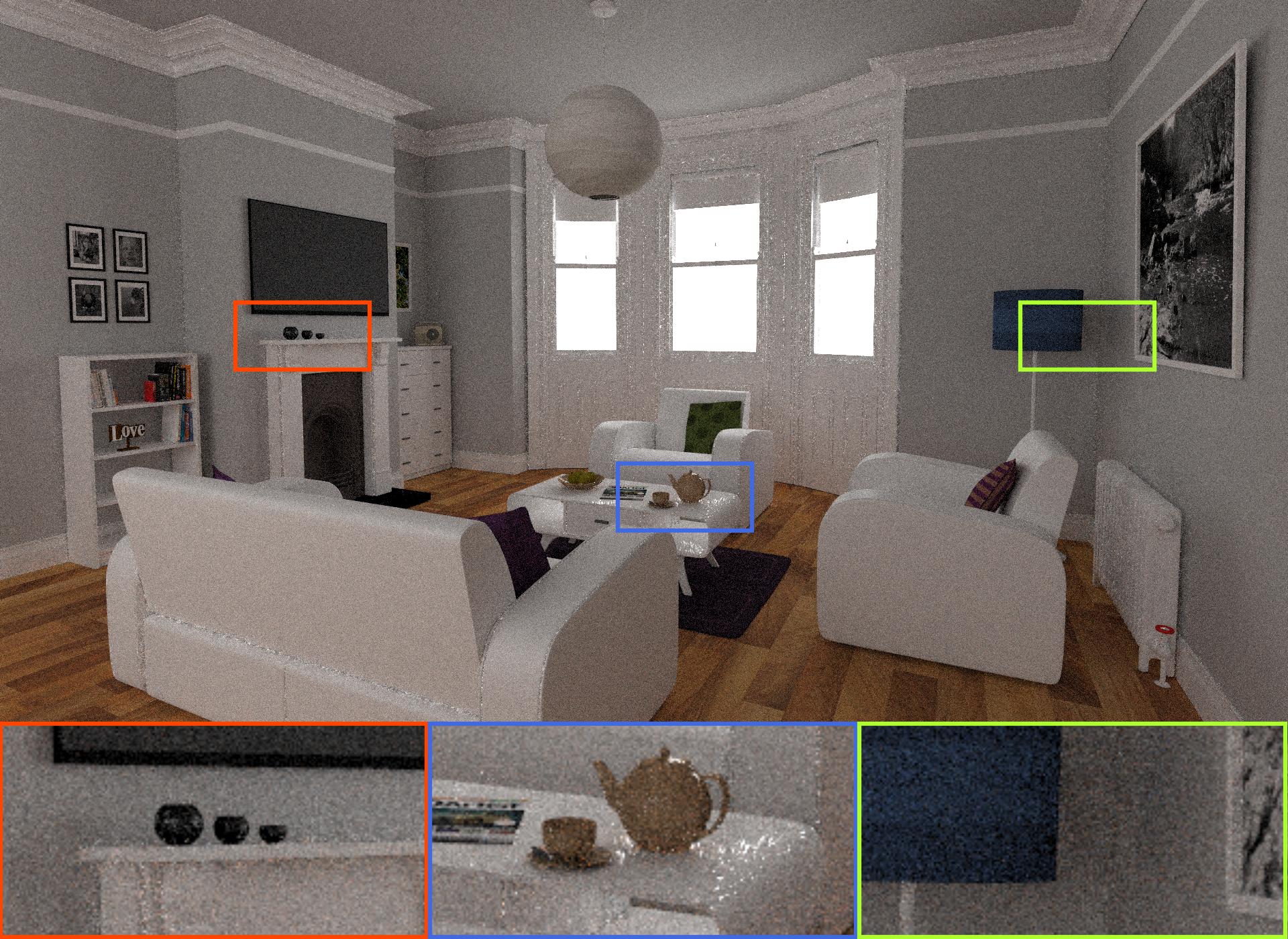}{\ltext{\textsc{RA-Quadtree}}}{\rtext{rRMSE 0.2173}}
\\
\rotatebox{90}{Ajar Door}
&\imgtext{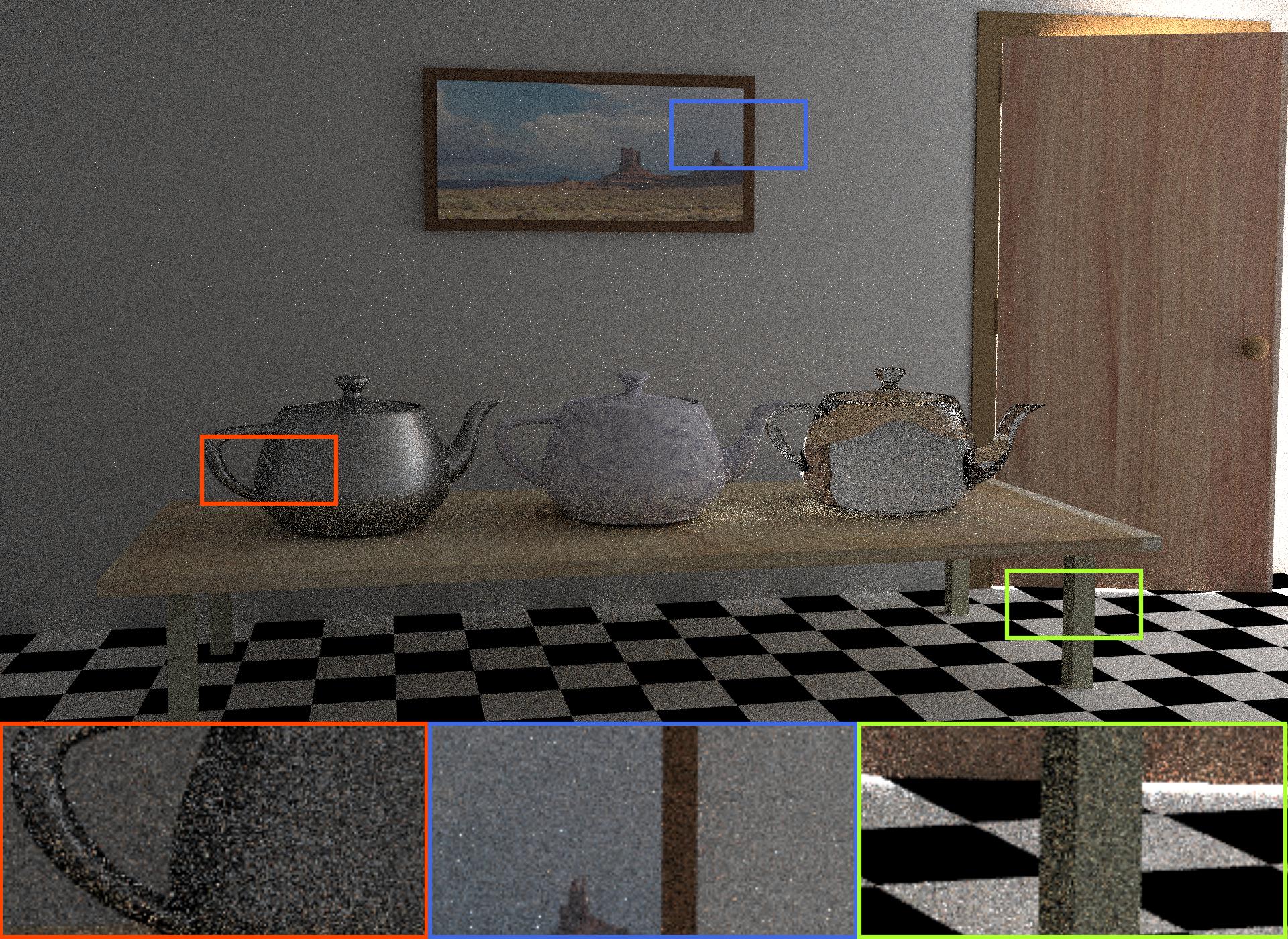}{\ltext{\textsc{Fixed}}}{\rtext{rRMSE 0.8745}}
&\imgtext{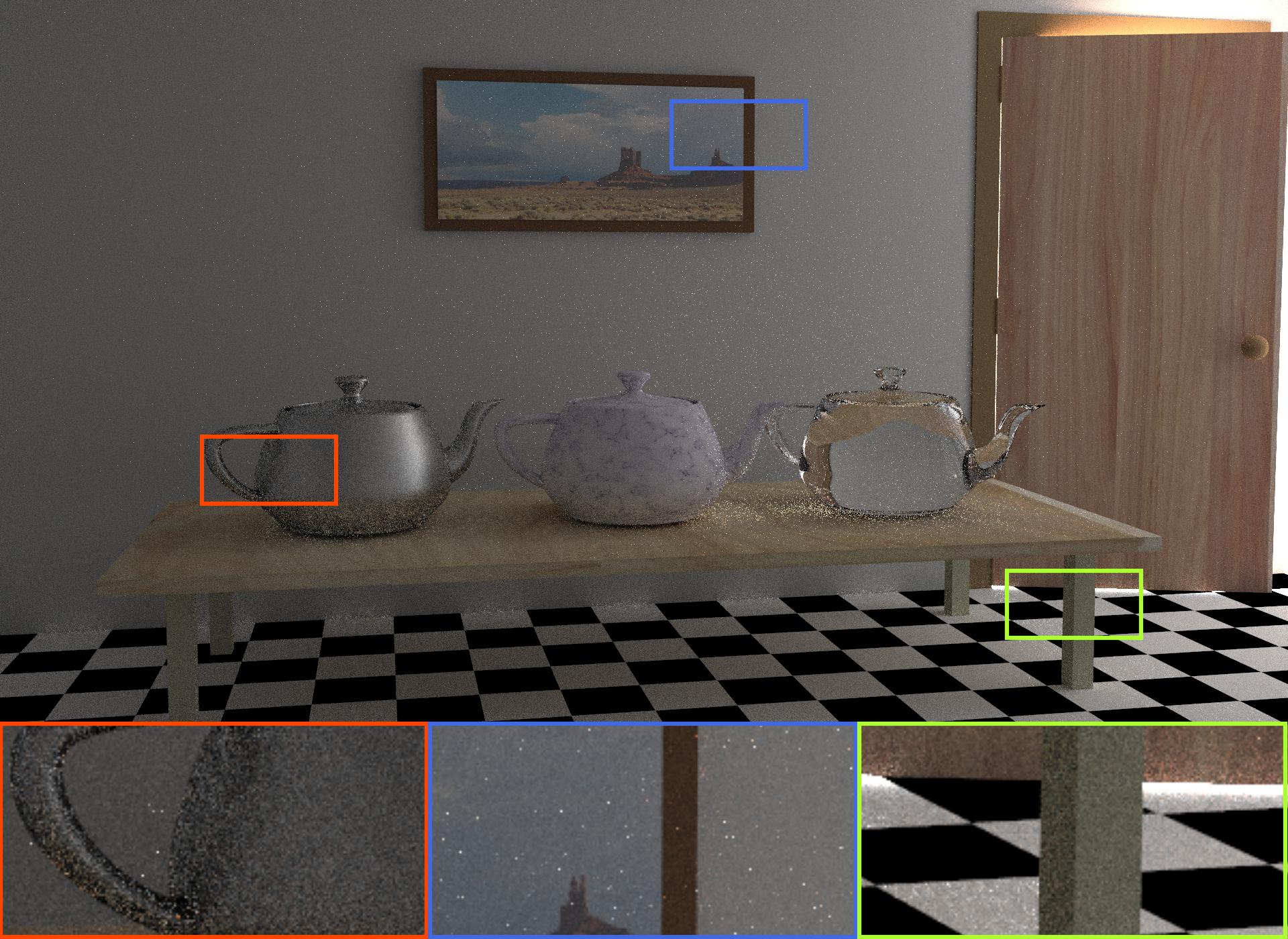}{\ltext{\textsc{Global}}}{\rtext{rRMSE 0.5408}}
&\imgtext{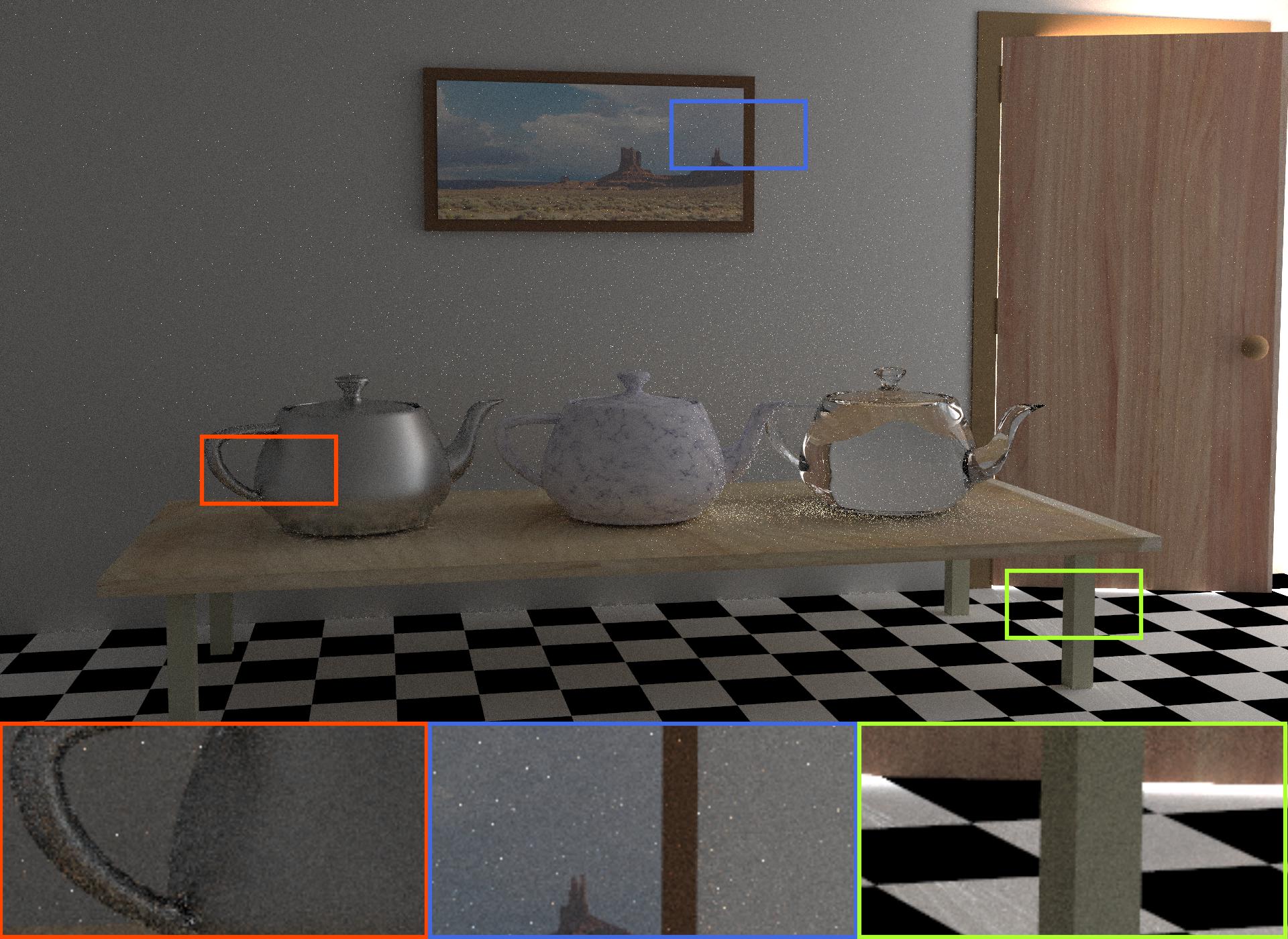}{\ltext{\textsc{RA-Grid}}}{\rtext{rRMSE 0.4612}}
&\imgtext{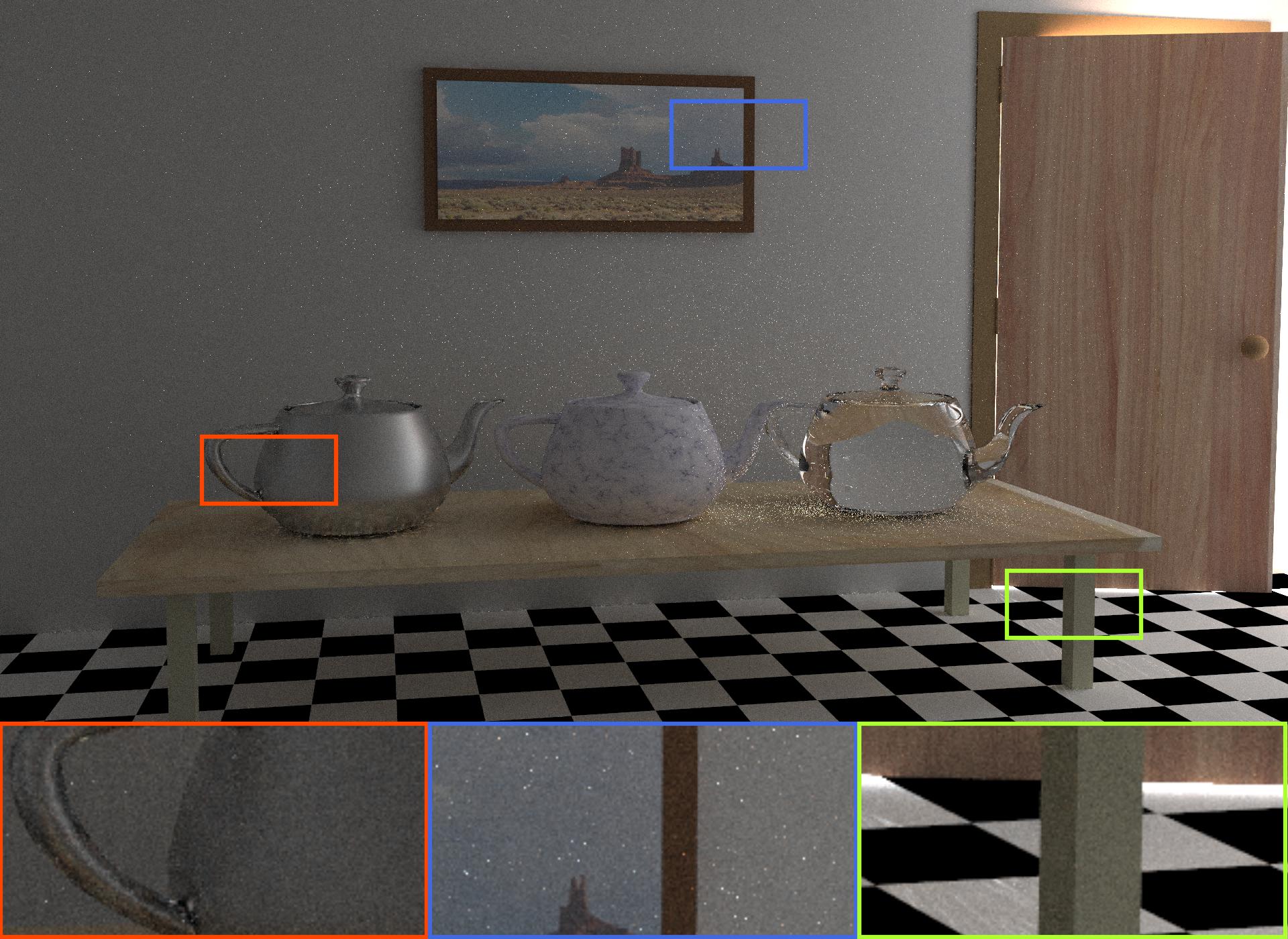}{\ltext{\textsc{RA-Quadtree}}}{\rtext{rRMSE 0.4706}}
\\
\end{tabular}

    \caption{Equal-time comparisons of the four variants of lens perturbations (\textsc{Fixed}, \textsc{Global}, \textsc{RA-Grid}, \textsc{RA-Quadtree}) rendered with the four scenes (Fireplace Room, Necklace, Living Room, Ajar Door).}
        \label{fig:rendering_lens}
\end{figure*}

\section{Results}

\subsection{Setup}

We implemented our proposed approaches in an open source renderer which implements path space MLT~\cite{veach1997mlt}.
All measurements are conducted on a machine with an Intel Xeon E5-2698 v3 CPU at 2.3 GHz using 32 threads. 
The comparisons are equal-time renderings of 10 minutes, except for the \emph{Necklace} scene (20 minutes).
The reference images were rendered with bidirectional path tracing for at least one day using a machine with 8 Intel Xeon E5-8867 v3 CPUs at 2.5 GHz using 256 threads. 
The maximum number of path vertices is configured to 20.

\paragraph*{Approaches.}
We compared several variants of lens and multi-chain perturbation including our regional adaptive extensions. 
For each perturbation technique, we configured four variants:
\begin{enumerate}
\item Fixed kernel size (\textsc{Fixed}): The same kernel size is used throughout the process. This is equivalent to using the original path perturbations.
\item Global adaptation (\textsc{Global}): A single global parameter is updated throughout the adaptation process. This is equivalent to considering a partition having a single region covering the entire path space. This approach is included as a baseline for adaptive techniques.
\item Regional adaptive with grids (\textsc{RA-Grid}): The proposed regional adaptive path perturbations using a regular grid partitioning.
\item Regional adaptive with quadtrees (\textsc{RA-Quadtree}): The proposed regional adaptive path perturbations using an adaptive quadtree partitioning (\cref{sec:adaptive_partitioning}).
\end{enumerate}
{To guarantee global exploration, we always combine the path perturbation (\textsc{Fixed}, \textsc{Global}, \textsc{RA-Grid}, \textsc{RA-Quadtree}) with a bidirectional mutation using a probability of $0.5$ for both. If the path perturbation cannot mutate the path, the bidirectional mutation is used.}

\paragraph*{Parameters.}

For a faithful comparison, we use the same initial scaling parameters as the \textsc{Fixed} approach for all adaptive techniques. The initial scaling is set to $\lambda = 1$.
For adaptation, we set the parameters introduced in \cref{sec:ra_path}
to $L=10$, $\gamma_{\mathrm{max}}=1$, and $\gamma_{\mathrm{scale}}=5$.
To avoid numerical instabilities when the scaling parameter becomes too small, the minimum scaling parameter is limited to -30.
For grid partitions, the top- and bottom-level grid size is configured by $N_{\mathrm{top}}=20$ and $N_{\mathrm{bottom}}=50$. The same top-level grid size is used for both 2d and 4d partitions, and also used for the top-level grid of the 4d partition.
For quadtree partitions, we use a split threshold of $M_\mathrm{split}=5000$. The refinement process is executed once in $M_\mathrm{refine}=10^7$ mutations.

We found that the target acceptance ratio $\bar{\alpha}$ should be higher than the theoretical optimal value of 0.234 by Roberts et al.~\cite{roberts2001}.
We used $\bar{\alpha}=0.5$ by default, except for the \emph{Necklace} scene where we used $\bar{\alpha}=0.8$.
We will discuss this later in \cref{sec:discussion}.

\subsection{Equal-Time Comparisons}

\paragraph*{Multi-chain Perturbation.}
\cref{fig:teaser} shows the equal-time comparison of four variants of multi-chain perturbations for the \emph{Fireplace Room} scene, which is an indoor scene that is illuminated by area lights attached to the windows.
The relative root mean square error (rRMSE) is used to compare the image to the reference.
The rendering quality of the fixed kernel (\textsc{Fixed}) is the worst. 
Since the mutation size is configured to be large, the proposed path tends to be rejected, {resulting} in higher correlations.
The global adaptation (\textsc{Global}) generally performs better. However, as it can only adapt to the global average, it often misses the exploration of the small details (e.g., the reflection on the table, {green} inset).
The regional adaptation with grid partitioning (\textsc{RA-Grid}) partially performs better (e.g., red inset), but in large parts of scene it is worse than global adaptation; the rRMSE values shows this as well.
This behavior can be explained by the premature adaptation.
The adaptation is triggered by the number of visits to a region. Now if the regions are separated into many disjoint regions from the beginning on, it is less likely that {a path} visits a region. 
Less-visited nodes then undergo fewer adaptive updates. 
Quadtree partitioning (\textsc{RA-Quadtree}) solves this problem. 
The quadtree starts from a single region and gradually splits the regions.
When a node is split, the information is inherited from the parent node. 
In other words, the child nodes can continue adaptation based on the value that the parent had.

\cref{fig:rendering_multichain} shows the comparisons for three scenes (\emph{Necklace}, \emph{Living Room}, \emph{Ajar Door}):  
The \emph{Necklace} scene is characterized by a mixture of difficult light transport phenomena involving glossy materials as well as relatively simple diffuse materials. The \textsc{Global}-strategy leads to clearly visible correlation artifacts on the diffuse surfaces in this scene. 
For glossy surfaces, a small kernel size should be preferred. For diffuse surfaces, a larger kernel size is better. The global adaptation only maintains a single parameter and is thus not able to adapt to both at the same time. 
\textsc{RA-Grid} could assign two different modes to the surfaces with two different materials, but box-like artifacts appears on the boundary of grid cells.
Note how the \textsc{RA-Quadtree} performs well in this case.

The \emph{Living Room} scene has similar characteristics as the \emph{Fireplace Room} scene. We can observe the improvement of \rev{\textsc{RA-Quadtree}} over two other two adaptive approaches.

The \emph{Ajar Door} scene is characterized by difficult indirect light transport. The light source lies in the back room, shining through the door. The objects on the table have different material properties. We can observe that \textsc{RA-Quadtree} outperforms the other strategies, in particular visible at the geometric edges or boundaries between two objects with different materials.

\begin{figure}[t]
    \centering
    \scriptsize
    \begin{tabular}{
        @{}
        >{\centering\arraybackslash}p{.02\linewidth}
        @{}
        >{\centering\arraybackslash}p{.23\linewidth}
        @{}
        >{\centering\arraybackslash}p{.23\linewidth}
        @{}
        >{\centering\arraybackslash}p{.23\linewidth}
        @{}
        >{\centering\arraybackslash}p{.23\linewidth}
        @{}
    }
        &
        Fireplace Room &
        Necklace &
        Living Room &
        Ajar Door
        \\
        \multirow{1}{*}[1.3cm]{\rotatebox[origin=c]{90}{Multi-chain}} &
        \includegraphics[width=\linewidth]{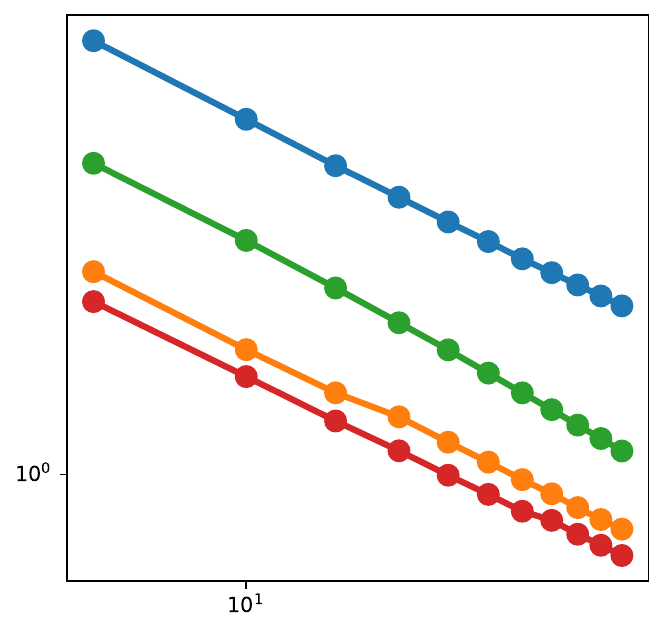} &
        \includegraphics[width=\linewidth]{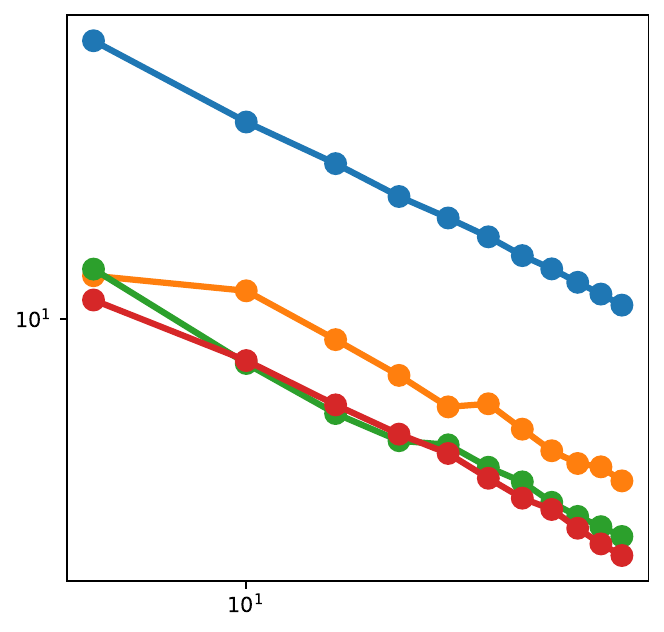} &
        \includegraphics[width=\linewidth]{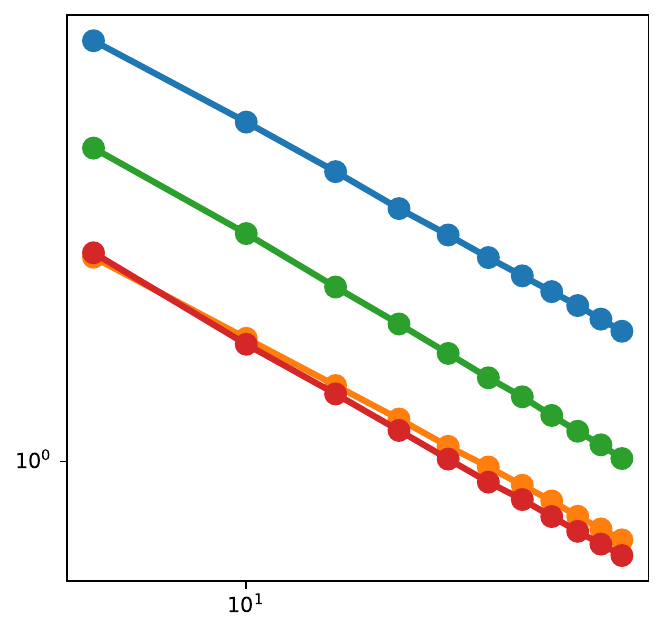} &
        \includegraphics[width=\linewidth]{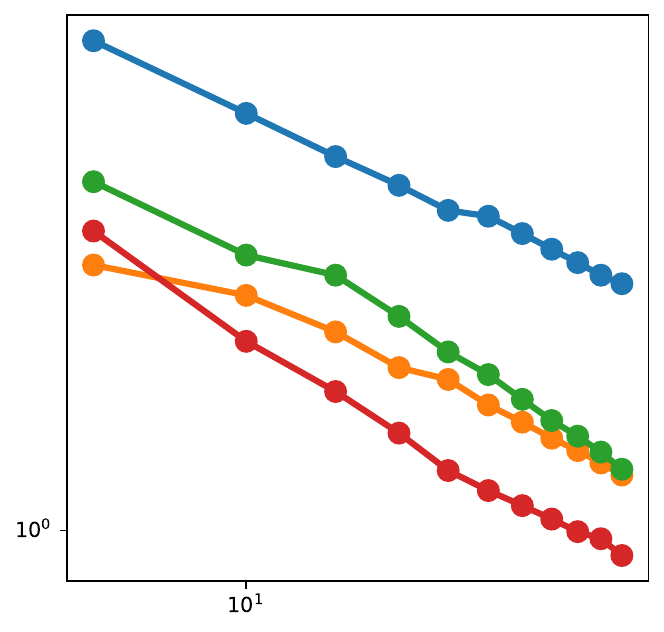}
        \\
        \multirow{1}{*}[1cm]{\rotatebox[origin=c]{90}{Lens}} &
        \includegraphics[width=\linewidth]{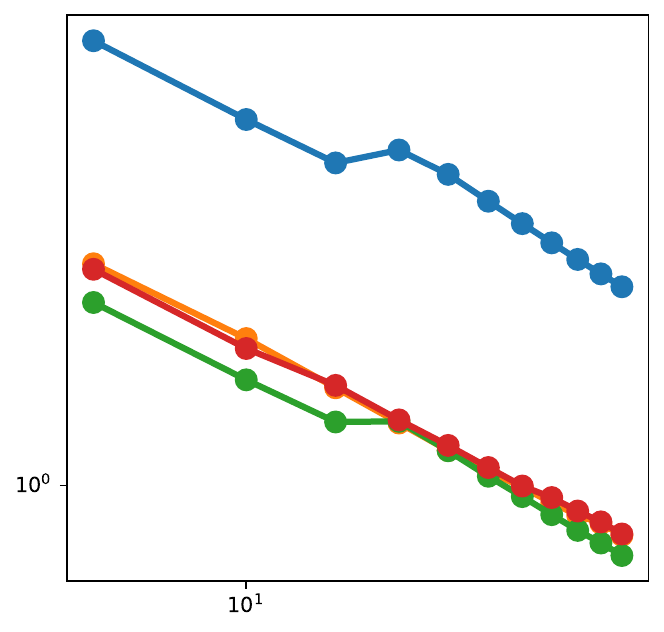} &
        \includegraphics[width=\linewidth]{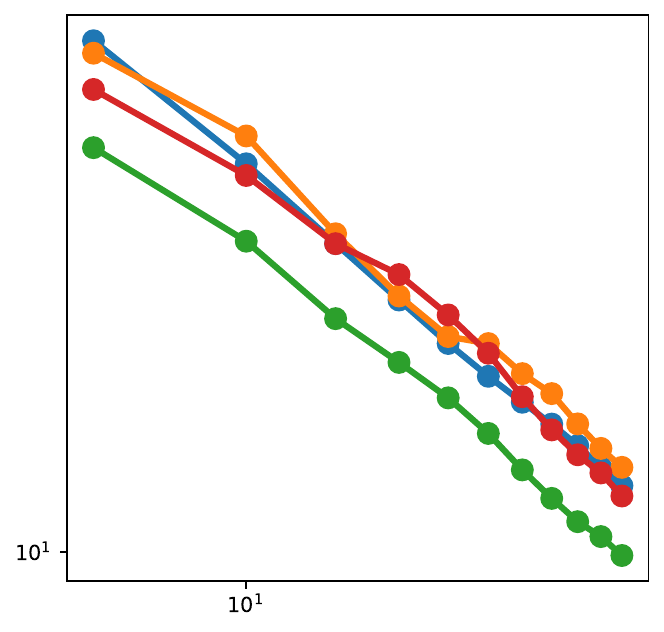} &
        \includegraphics[width=\linewidth]{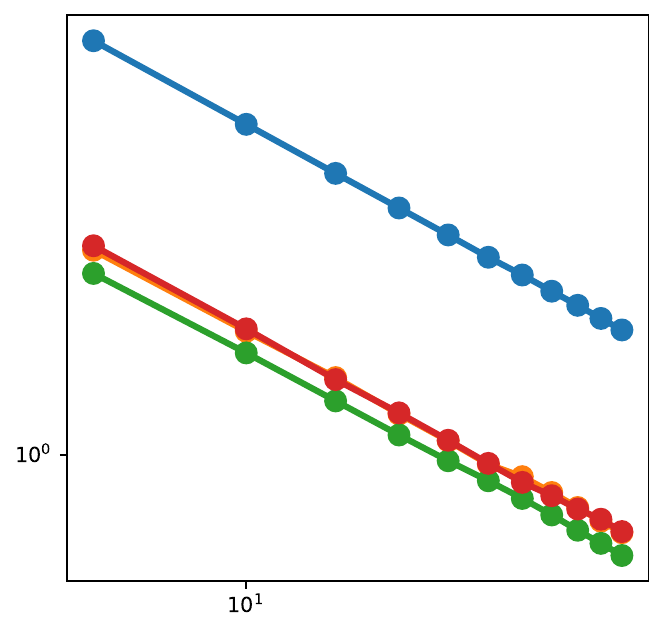} &
        \includegraphics[width=\linewidth]{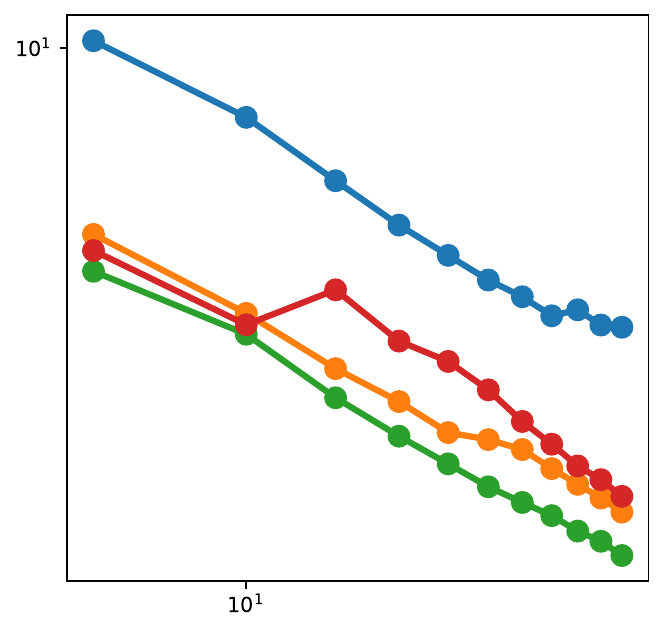}
        \\
    \end{tabular}
    \caption{
        Error convergence: the horizontal represents the render time up to 60 seconds, 
        the vertical axis is the rRMSE. Both axes are shown in log scale.
				The colors are the approaches, Blue: \textsc{Fixed}, Orange: \textsc{Global}, Green: \textsc{RA-Grid}, and Red: \textsc{RA-Quadtree}.}
    \label{fig:error_progress}
\end{figure}

{
\cref{fig:error_progress} shows the convergence of the rRMSE with respect to render time (up to 60 seconds) in log-log plots. The plot shows a similar tendency, while the regional adaptive approaches exhibit a better performance compared to approaches with fixed/non-regional parameters.
}

\begin{figure*}[t]
    \centering 
    \setlength{\tabcolsep}{0.3pt}
    \renewcommand{\arraystretch}{.1}
    \begin{tabular}{m{0.3cm} *{4}{m{.23\linewidth}}m{.3cm}}
\rotatebox{90}{Fireplace Room}
&\imgtext{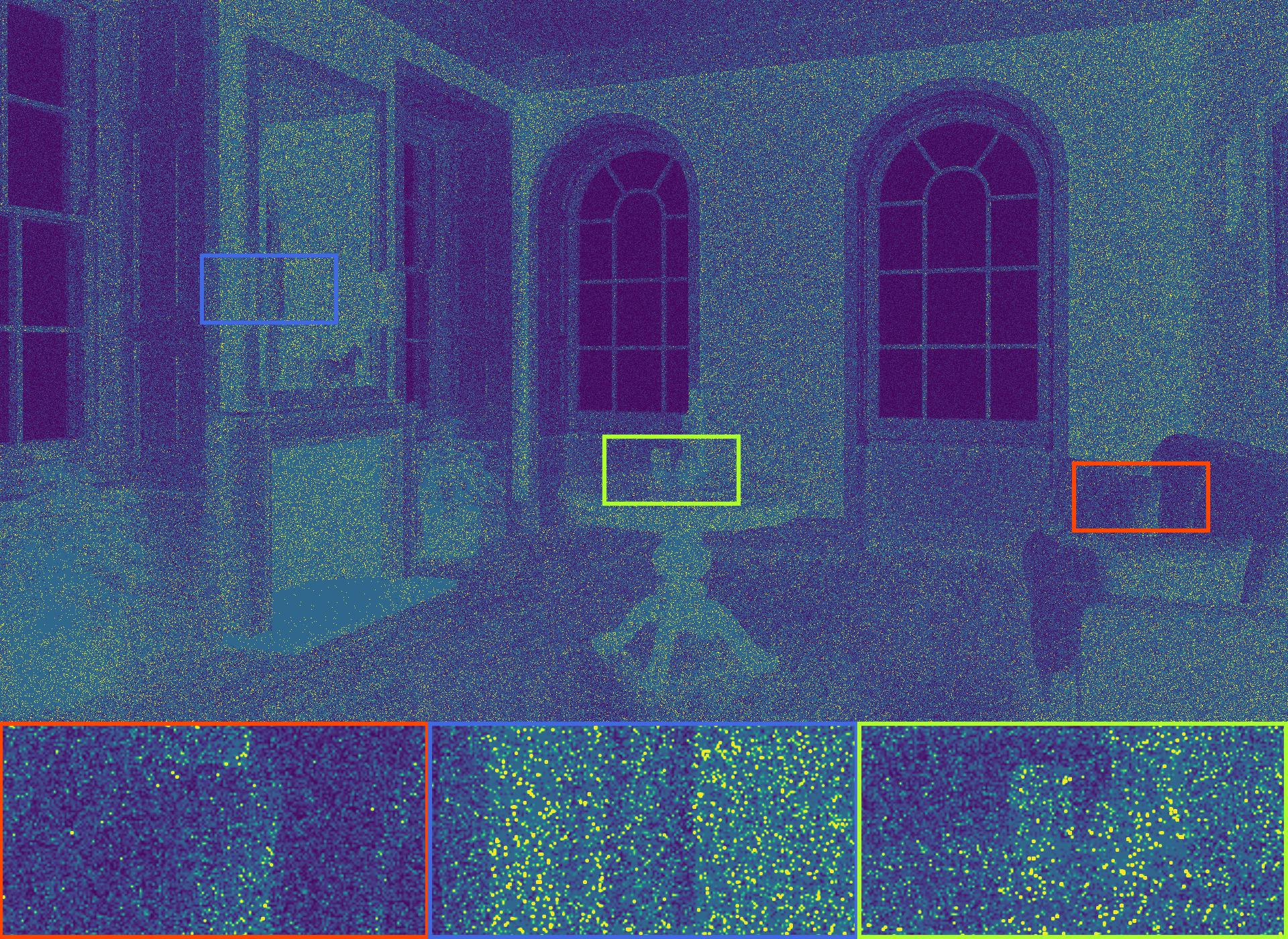}{\ltext{\textsc{Fixed}}}{\rtext{rRMSE 0.6318}}
&\imgtext{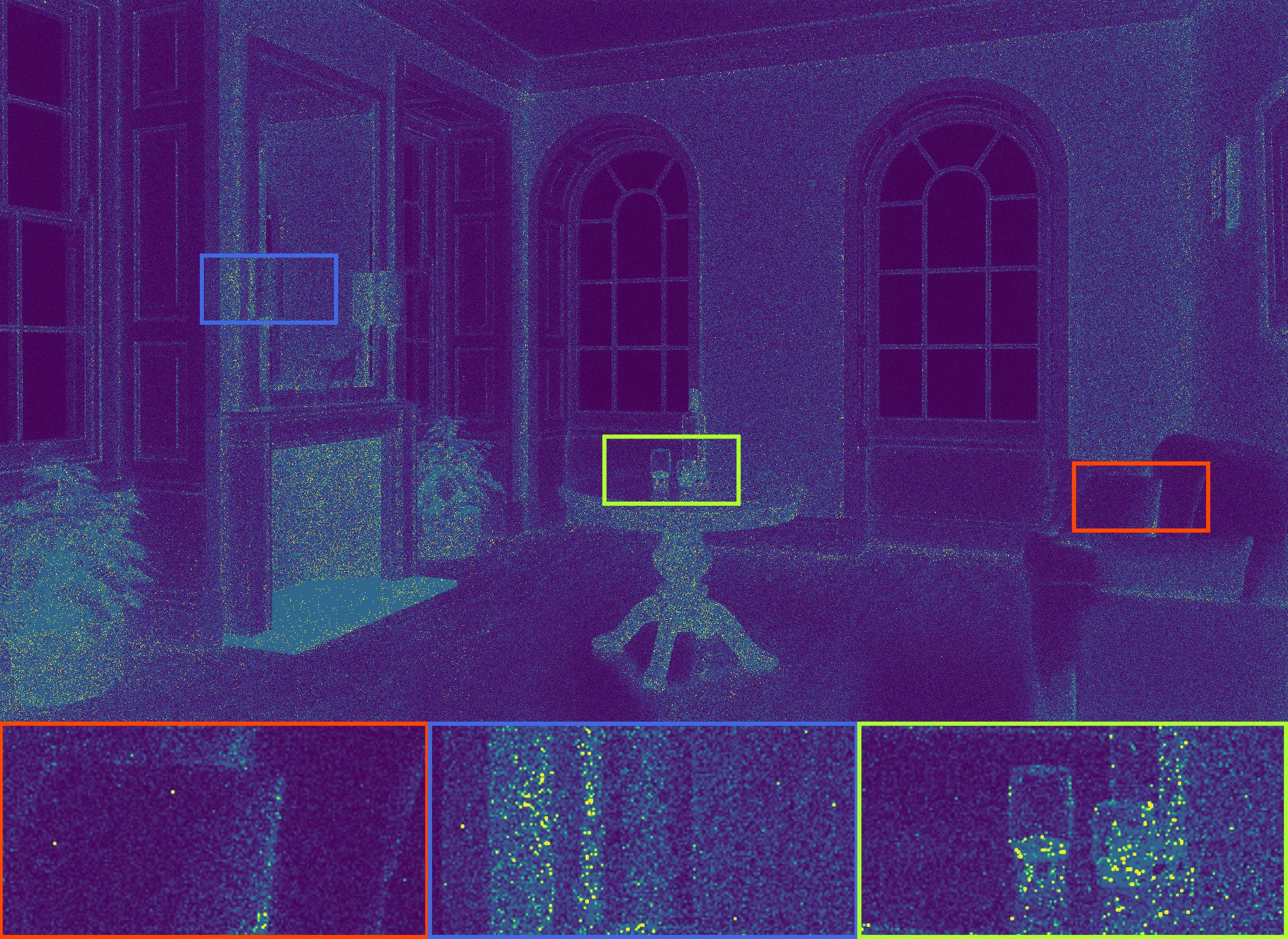}{\ltext{\textsc{Global}}}{\rtext{rRMSE 0.2340}}
&\imgtext{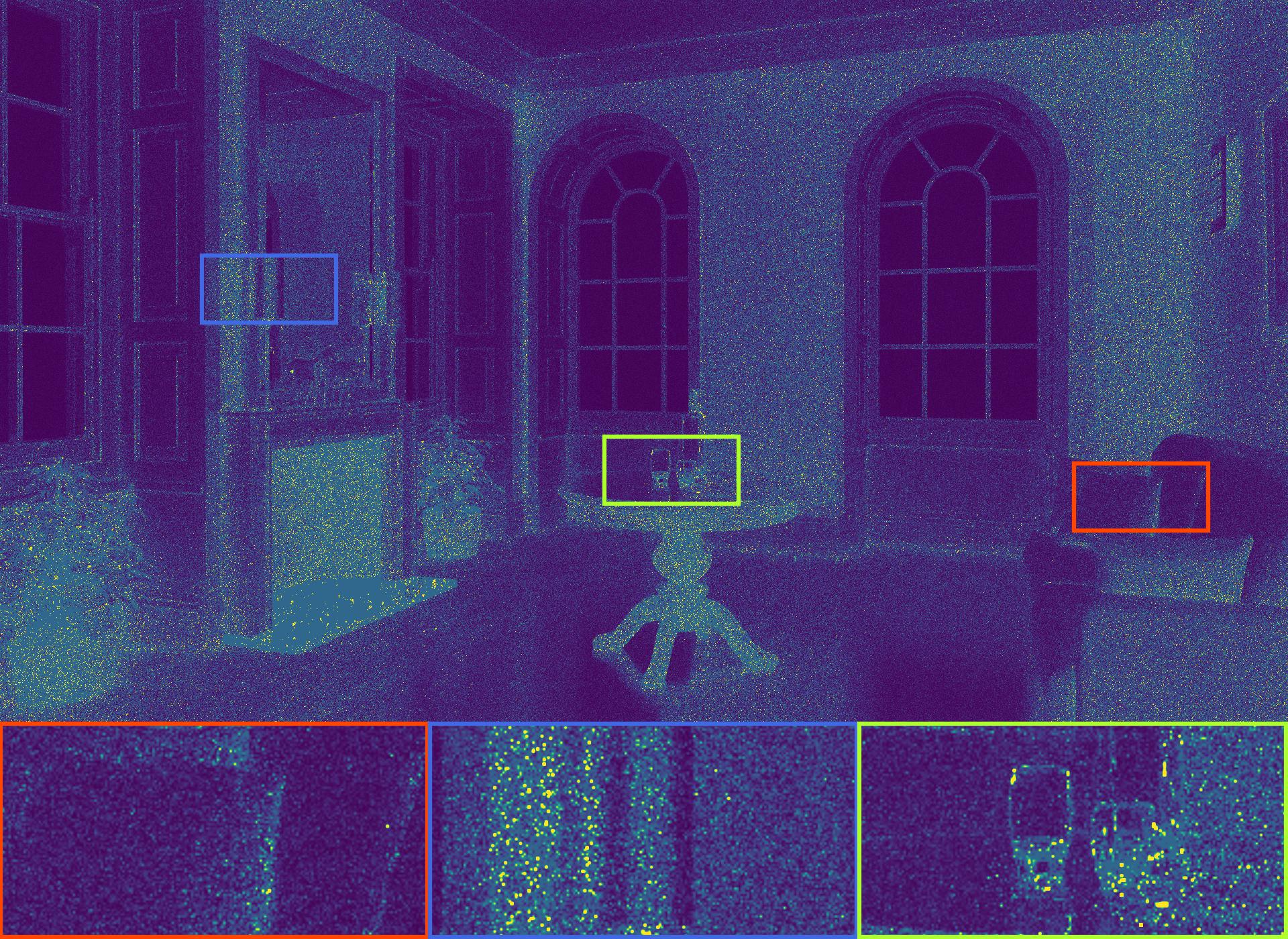}{\ltext{\textsc{RA-Grid}}}{\rtext{rRMSE 0.2501}}
&\imgtext{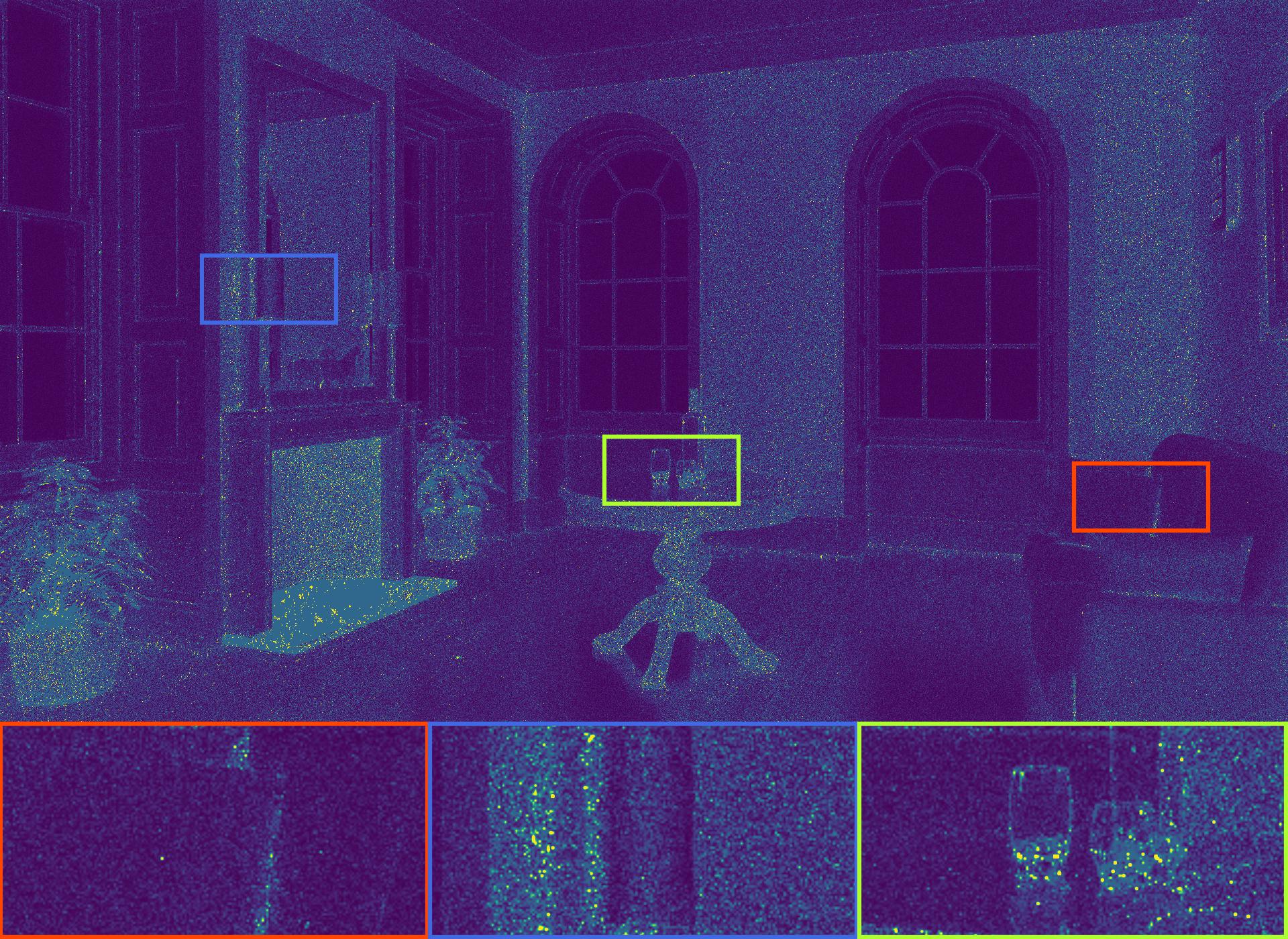}{\ltext{\textsc{RA-Quadtree}}}{\rtext{rRMSE 0.1989}}
&\includegraphics[height=3cm]{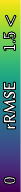}\\
\rotatebox{90}{Necklace}
&\imgtext{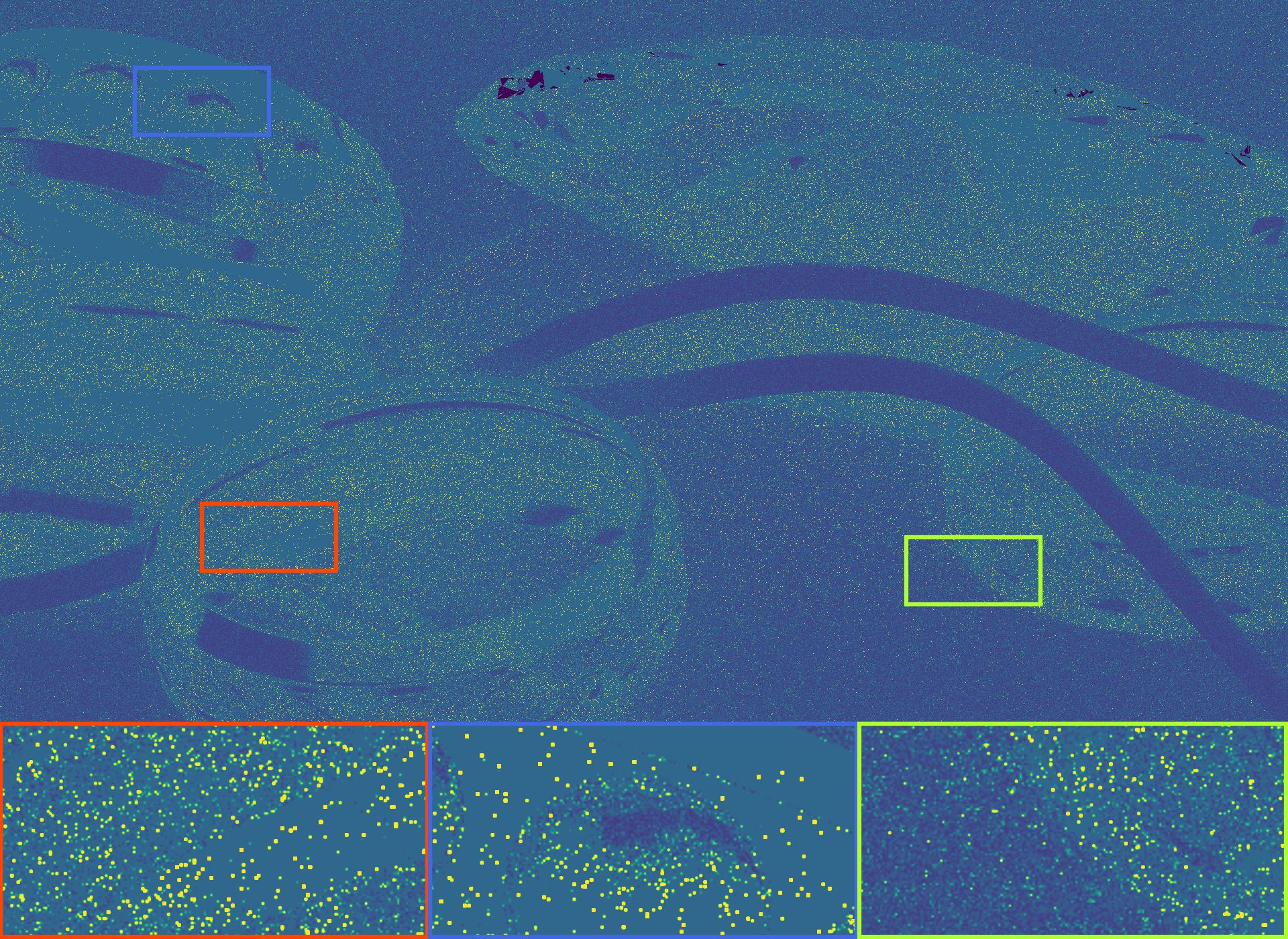}{\ltext{\textsc{Fixed}}}{\rtext{rRMSE 2.1921}}
&\imgtext{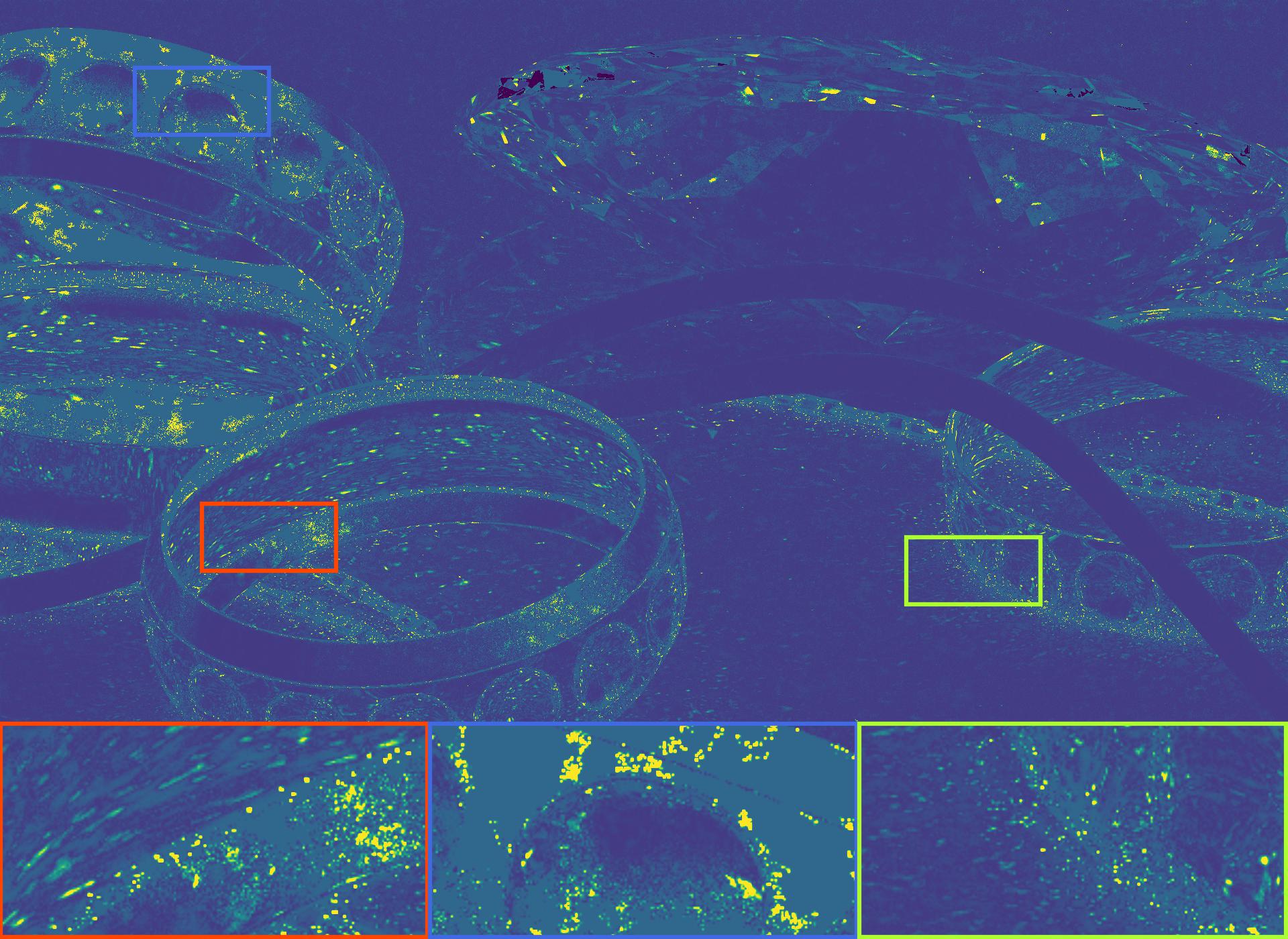}{\ltext{\textsc{Global}}}{\rtext{rRMSE 0.8403}}
&\imgtext{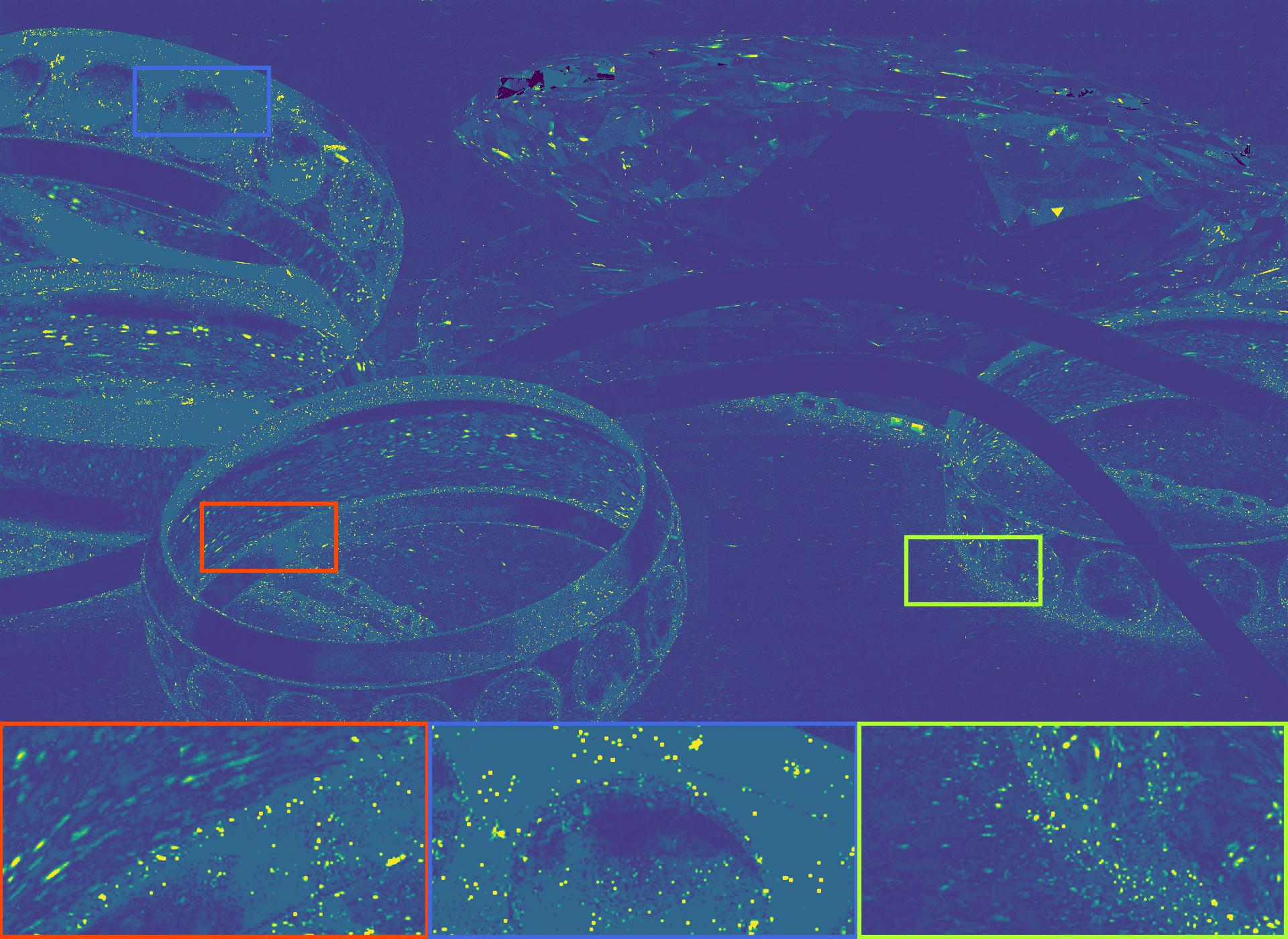}{\ltext{\textsc{RA-Grid}}}{\rtext{rRMSE 0.8805}}
&\imgtext{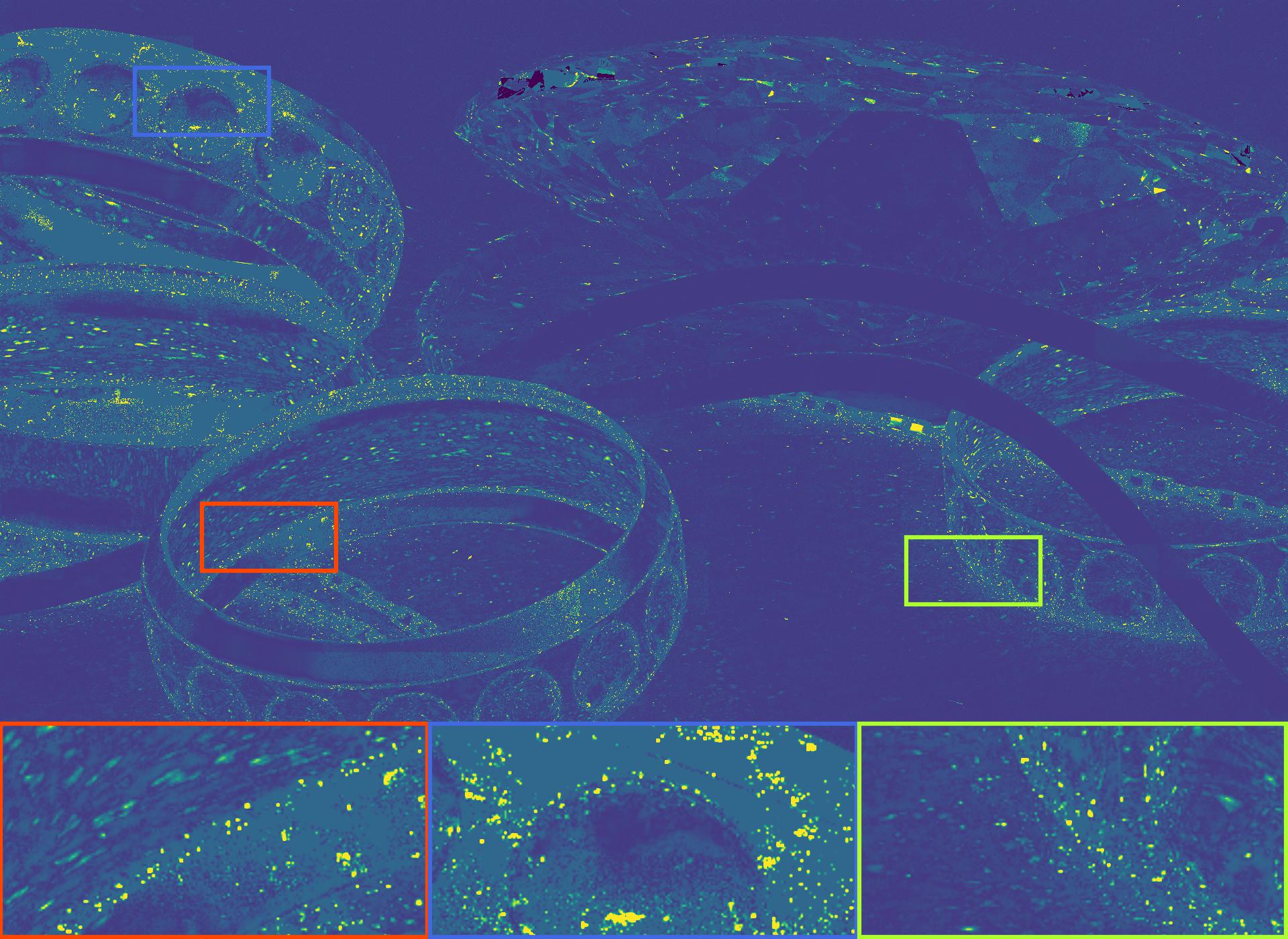}{\ltext{\textsc{RA-Quadtree}}}{\rtext{rRMSE 0.9537}}
&\includegraphics[height=3cm]{cbar.pdf}\\
\rotatebox{90}{Living Room}
&\imgtext{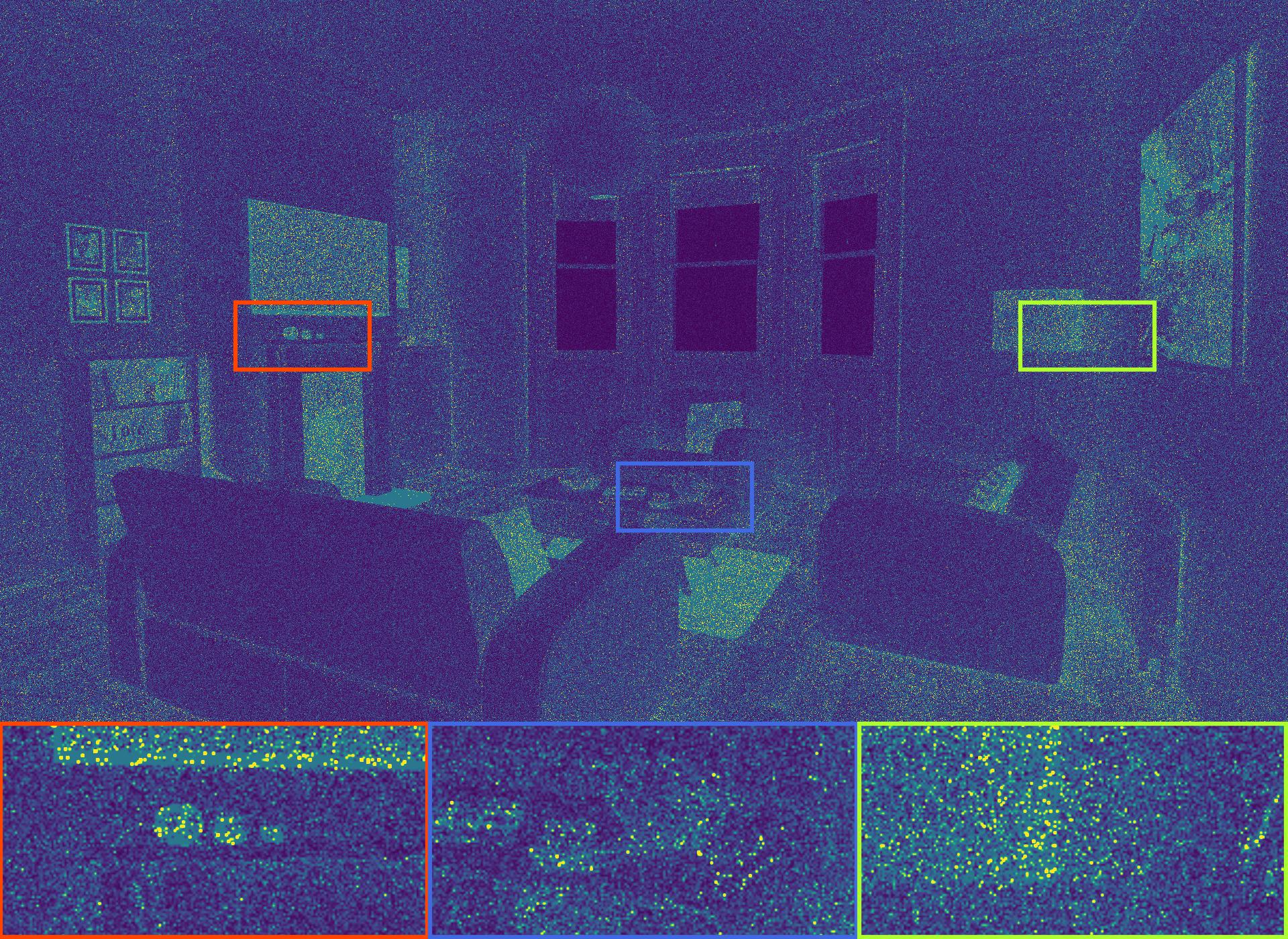}{\ltext{\textsc{Fixed}}}{\rtext{rRMSE 0.4933}}
&\imgtext{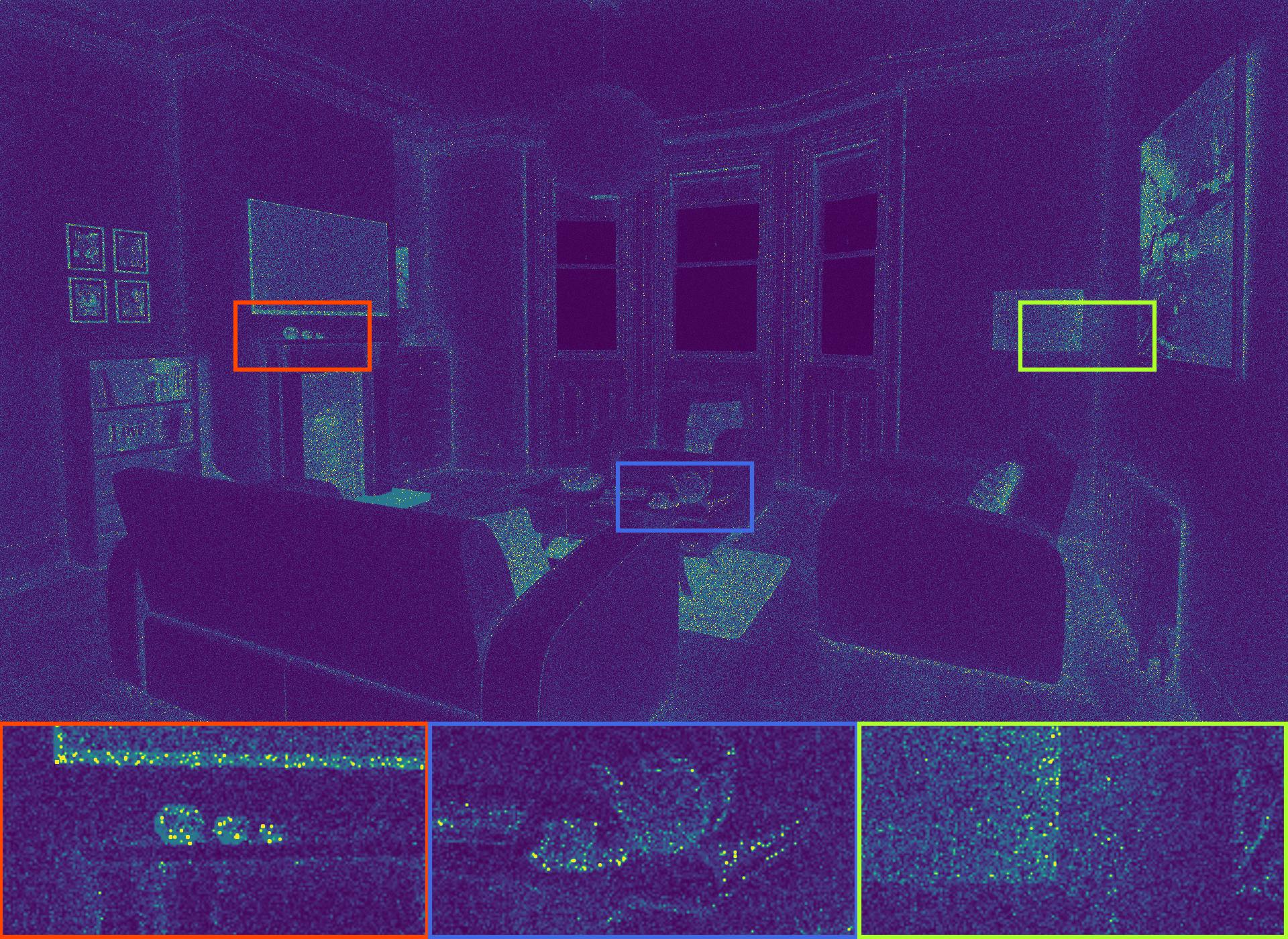}{\ltext{\textsc{Global}}}{\rtext{rRMSE 0.2314}}
&\imgtext{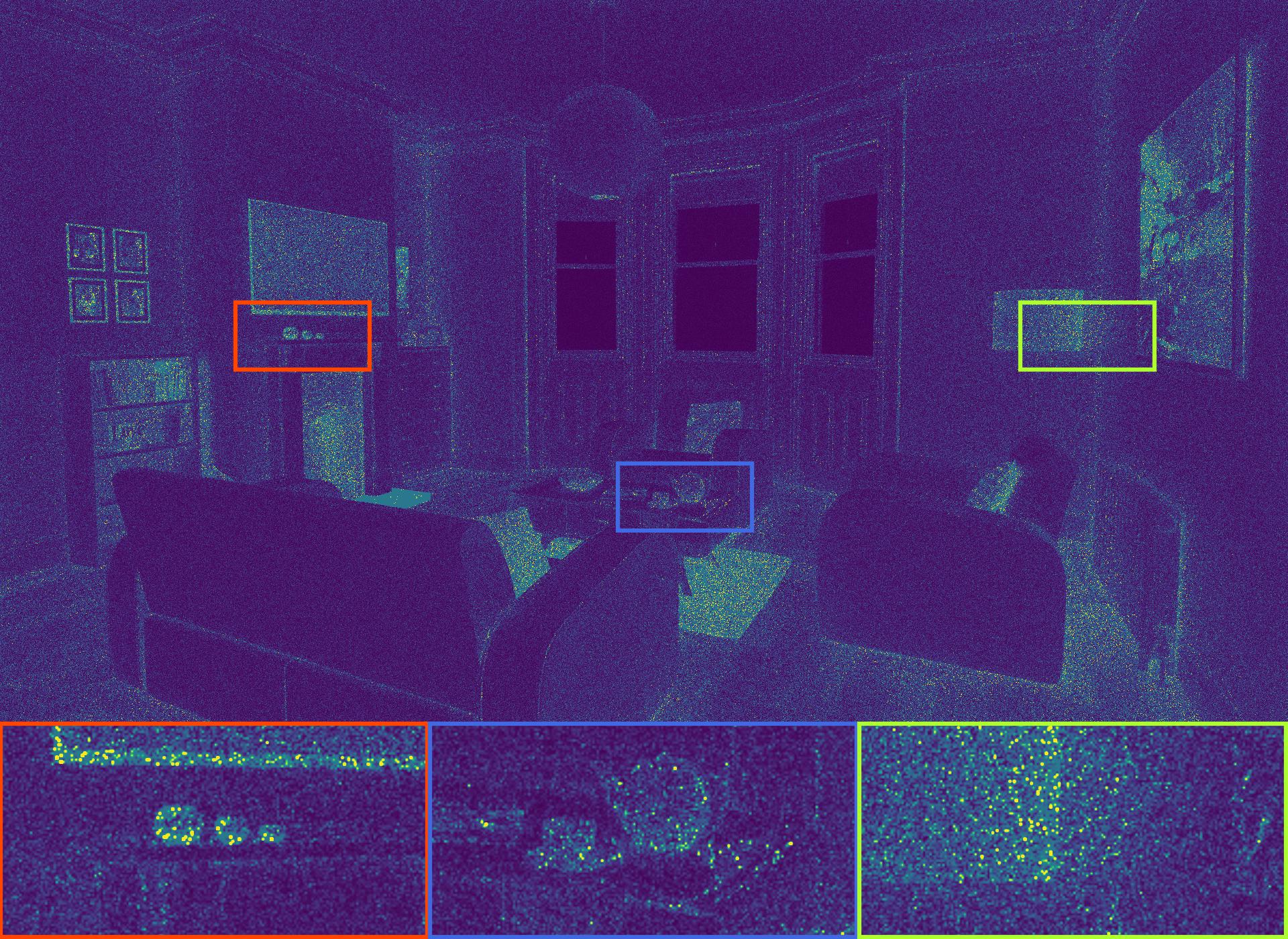}{\ltext{\textsc{RA-Grid}}}{\rtext{rRMSE 0.2375}}
&\imgtext{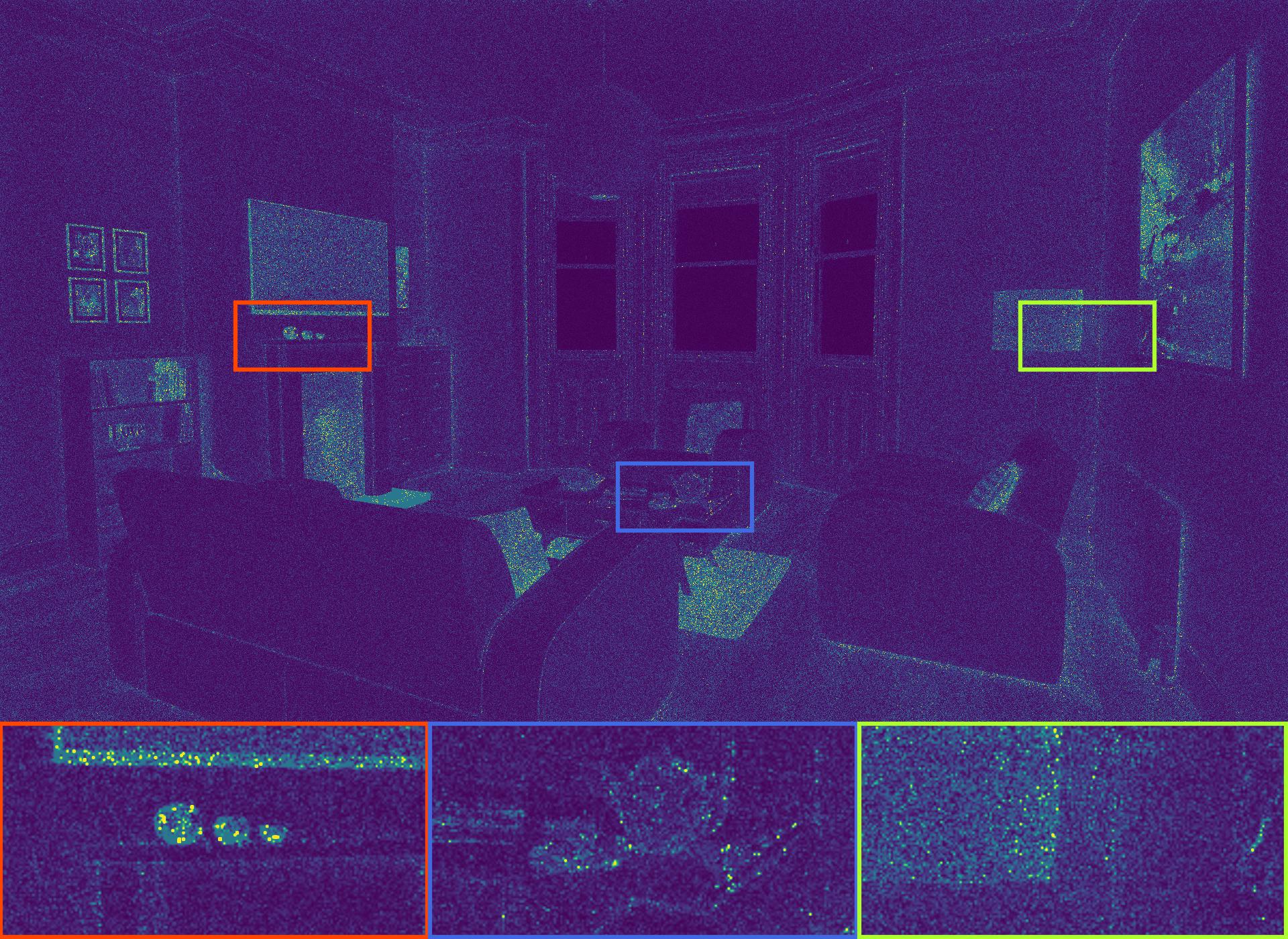}{\ltext{\textsc{RA-Quadtree}}}{\rtext{rRMSE 0.2098}}
&\includegraphics[height=3cm]{cbar.pdf}\\
\rotatebox{90}{Ajar Door}
&\imgtext{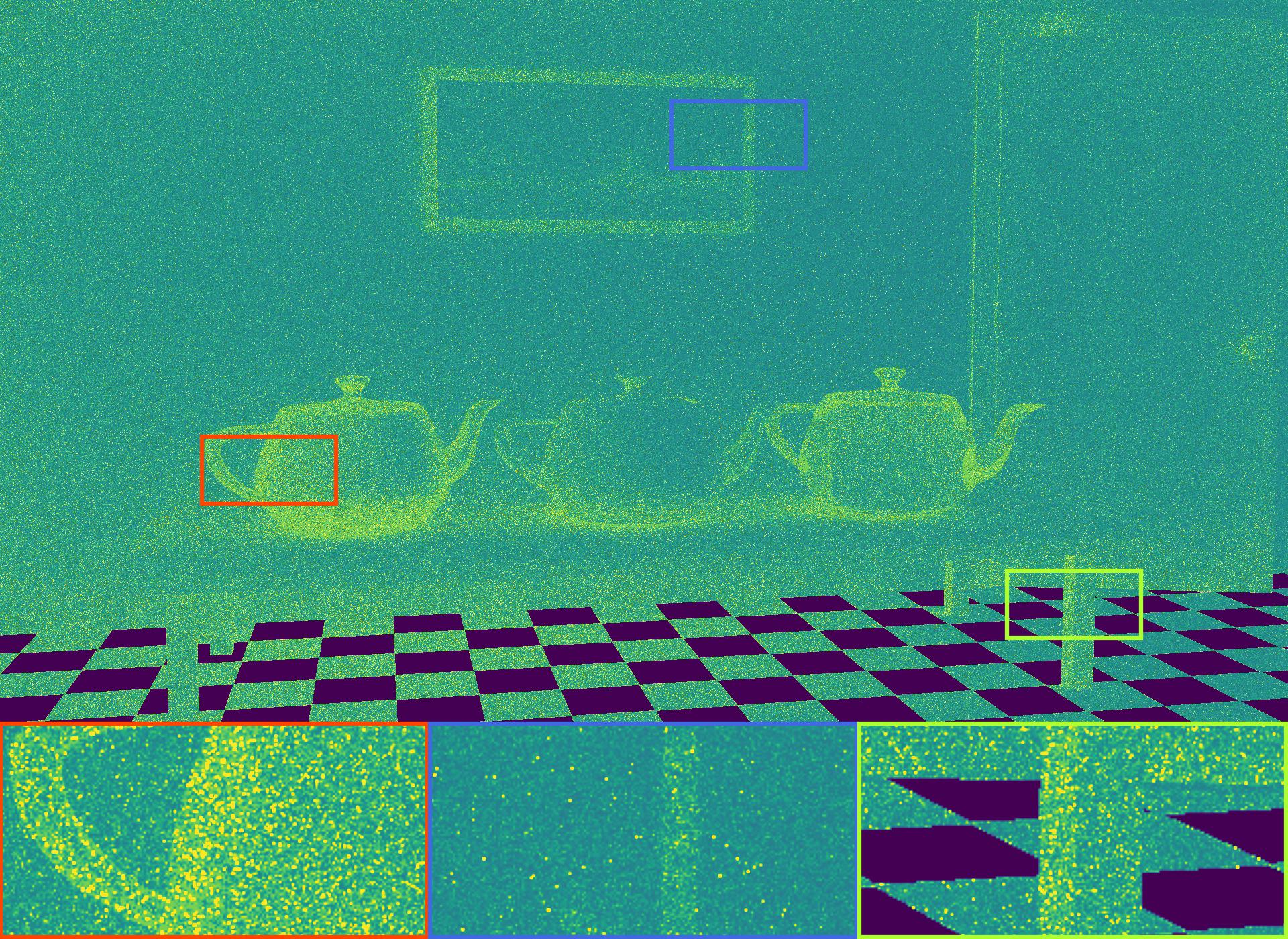}{\ltext{\textsc{Fixed}}}{\rtext{rRMSE 0.9388}}
&\imgtext{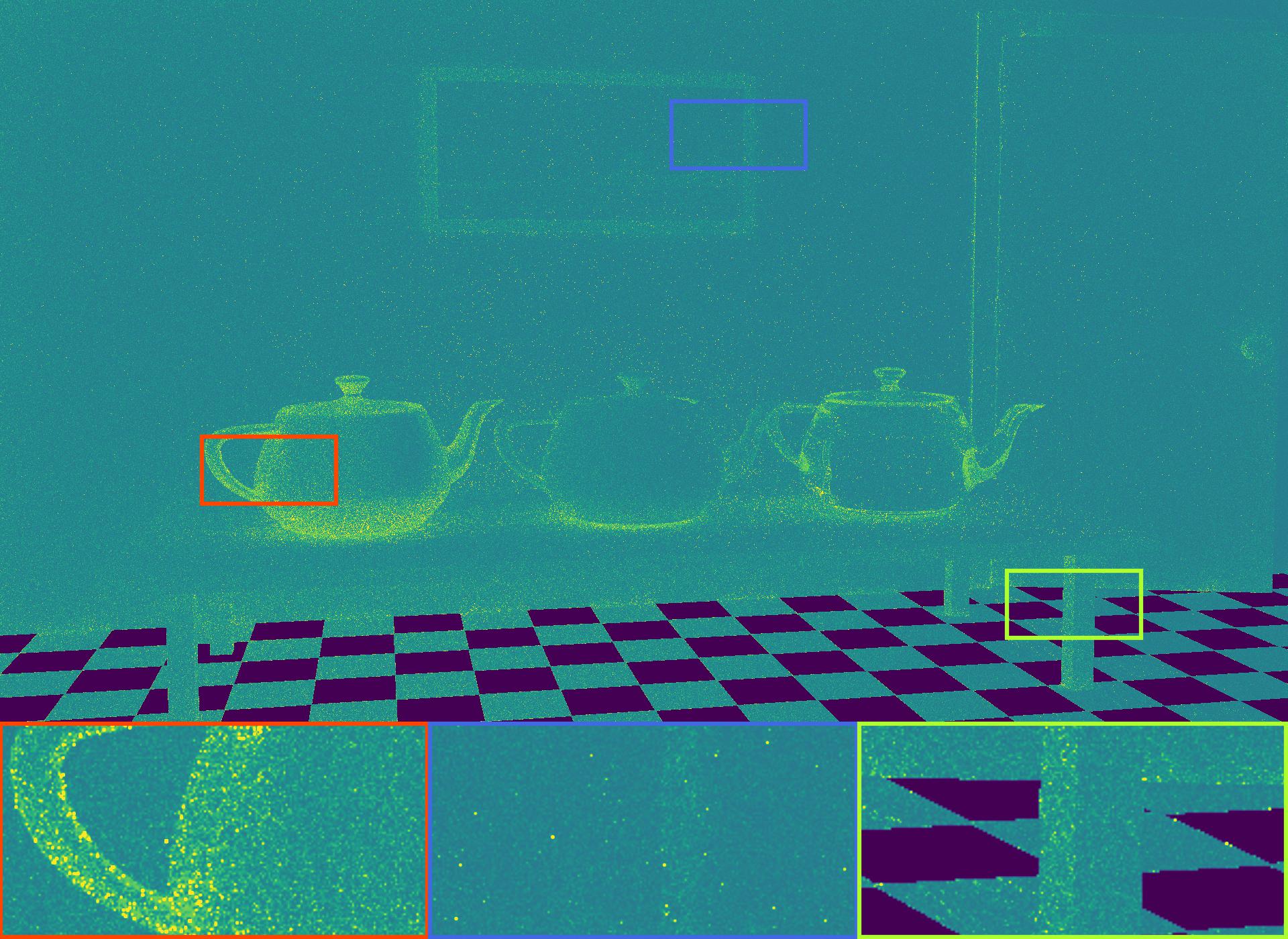}{\ltext{\textsc{Global}}}{\rtext{rRMSE 0.4232}}
&\imgtext{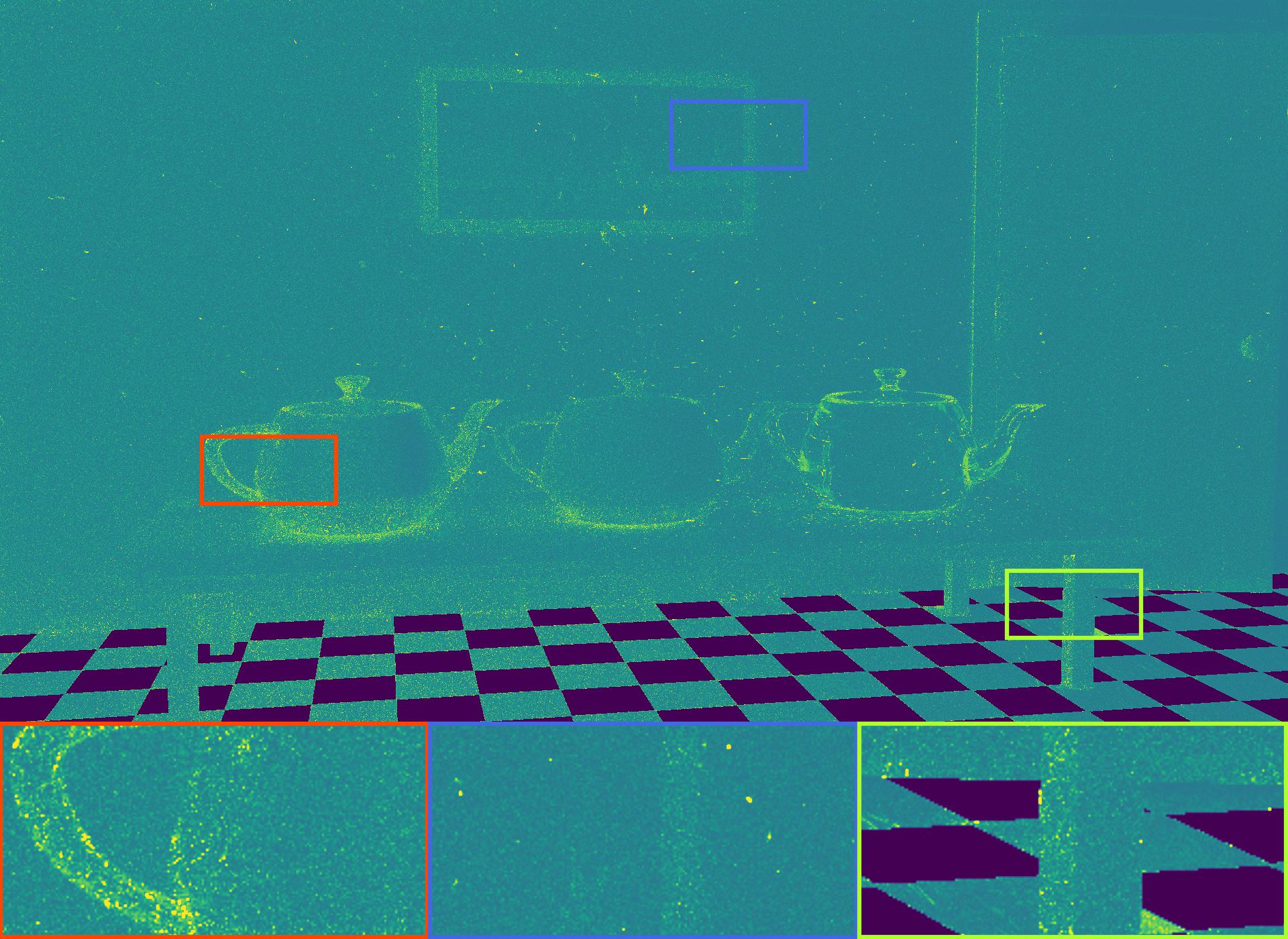}{\ltext{\textsc{RA-Grid}}}{\rtext{rRMSE 0.3650}}
&\imgtext{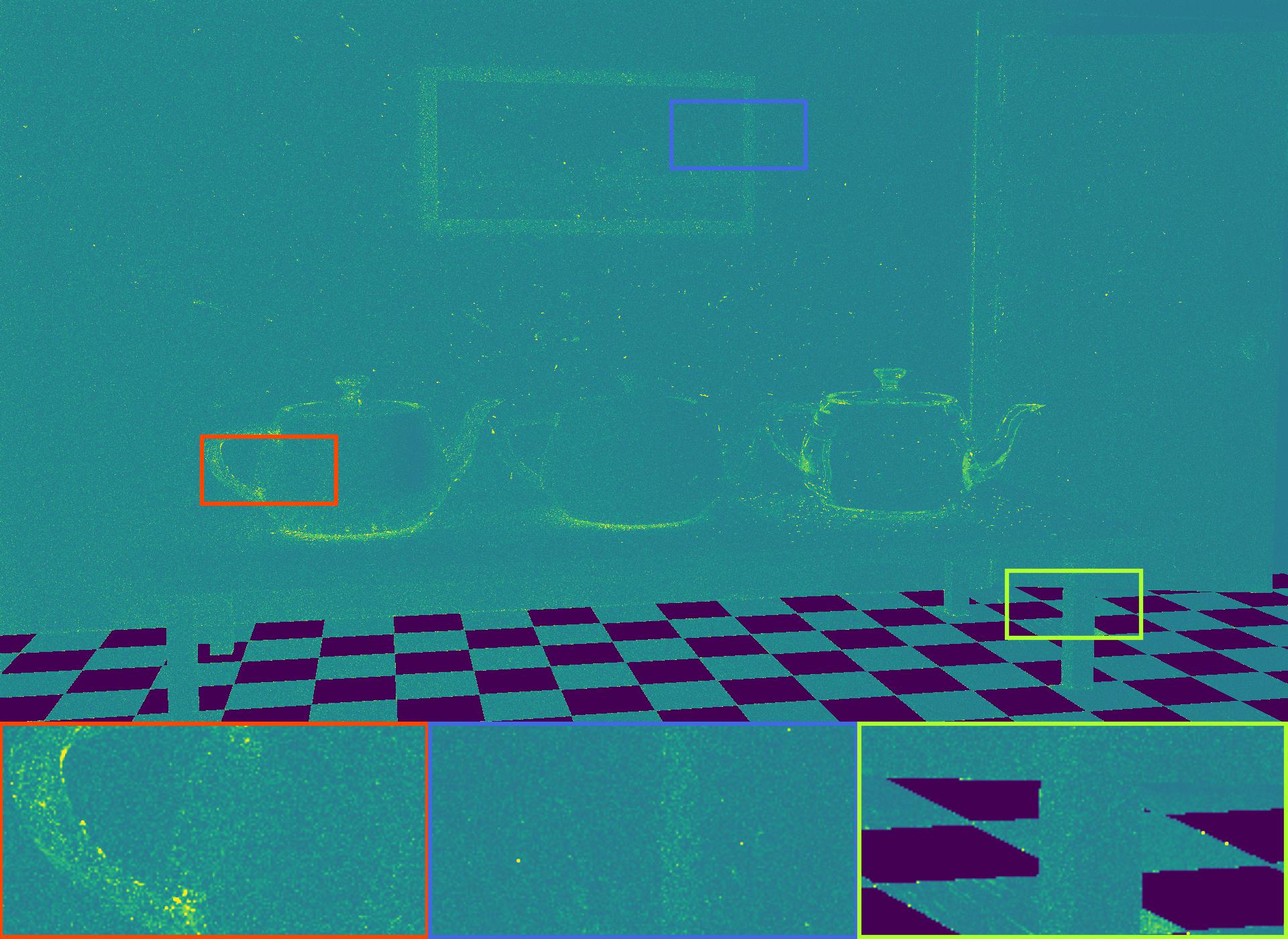}{\ltext{\textsc{RA-Quadtree}}}{\rtext{rRMSE 0.3176}}
&\includegraphics[height=3cm]{cbar.pdf}\\
\end{tabular}

    \caption{Pixel-wised error distribution of the four variants of multi-chain perturbations (\textsc{Fixed}, \textsc{Global}, \textsc{RA-Grid}, \textsc{RA-Quadtree}) rendered with the four scenes (Fireplace Room, Necklace, Living Room, Ajar Door).}
    \label{fig:error_multichain}
\end{figure*}

\cref{fig:error_multichain} shows the pixel-wise error distribution using rRMSE. We also display the rRMSE for the entire image. A similar behavior as discussed above can be observed. Interestingly, some of the results, e.g. \textsc{Global} for the Necklace scene, exhibit better performance in terms of rRMSE even though there are obvious correlation artifacts in the image.

\begin{figure*}[t]
    \centering 
    \setlength{\tabcolsep}{0.3pt}
    \renewcommand{\arraystretch}{.1}
    \begin{tabular}{m{0.3cm} *{4}{m{.23\linewidth}}m{.3cm}}
\rotatebox{90}{Fireplace Room}
&\imgtext{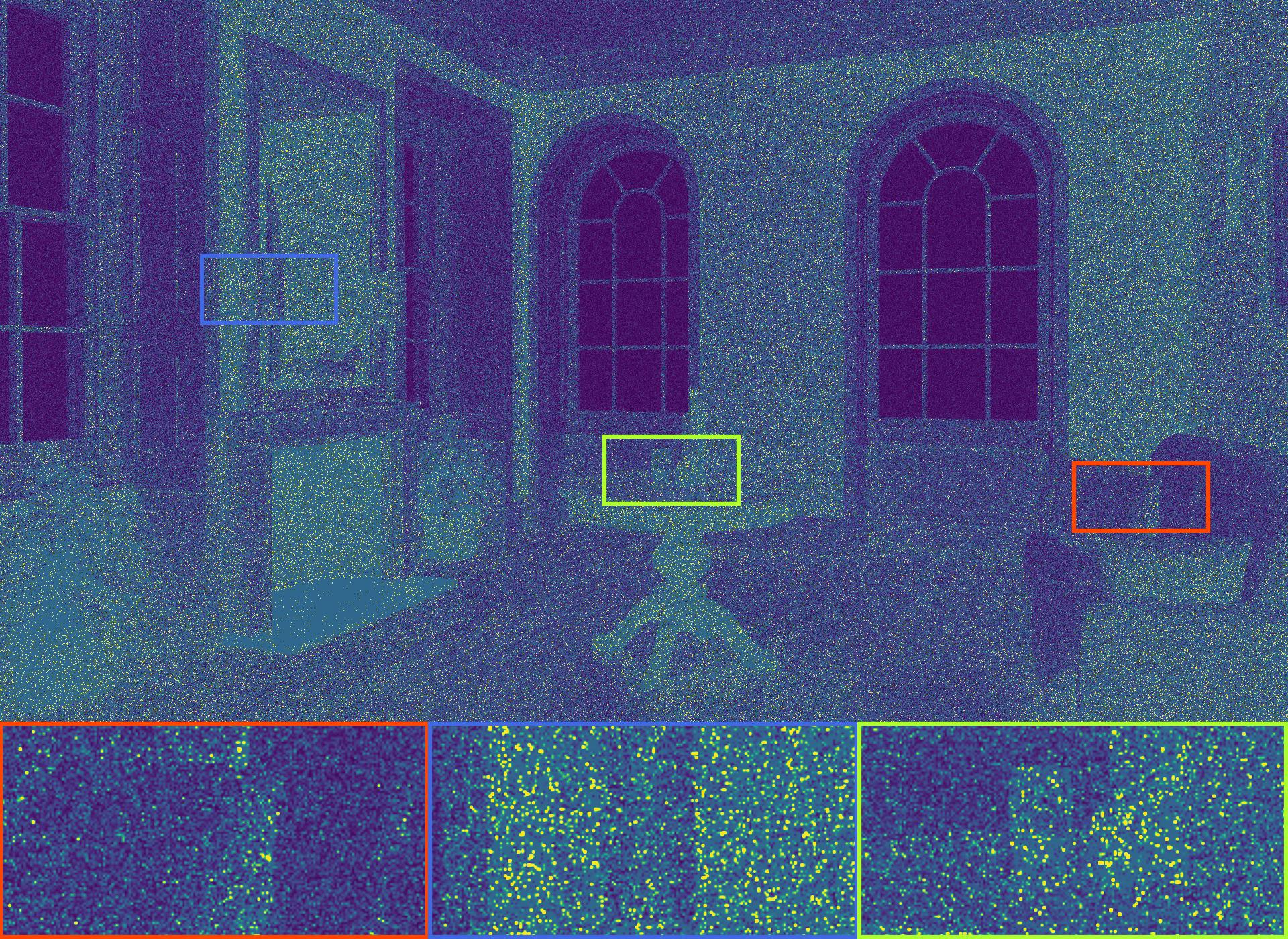}{\ltext{\textsc{Fixed}}}{\rtext{rRMSE 0.6329}}
&\imgtext{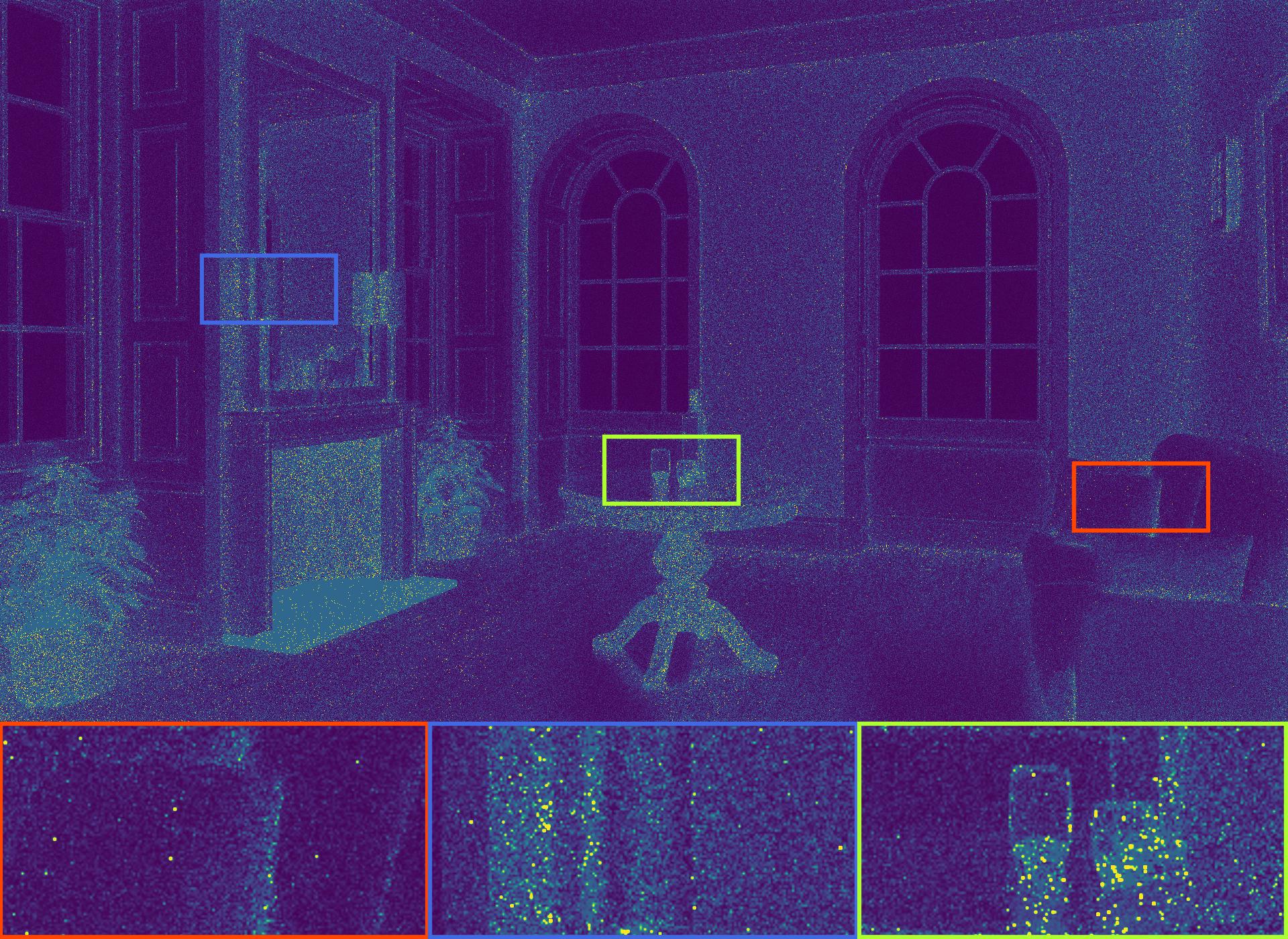}{\ltext{\textsc{Global}}}{\rtext{rRMSE 0.2458}}
&\imgtext{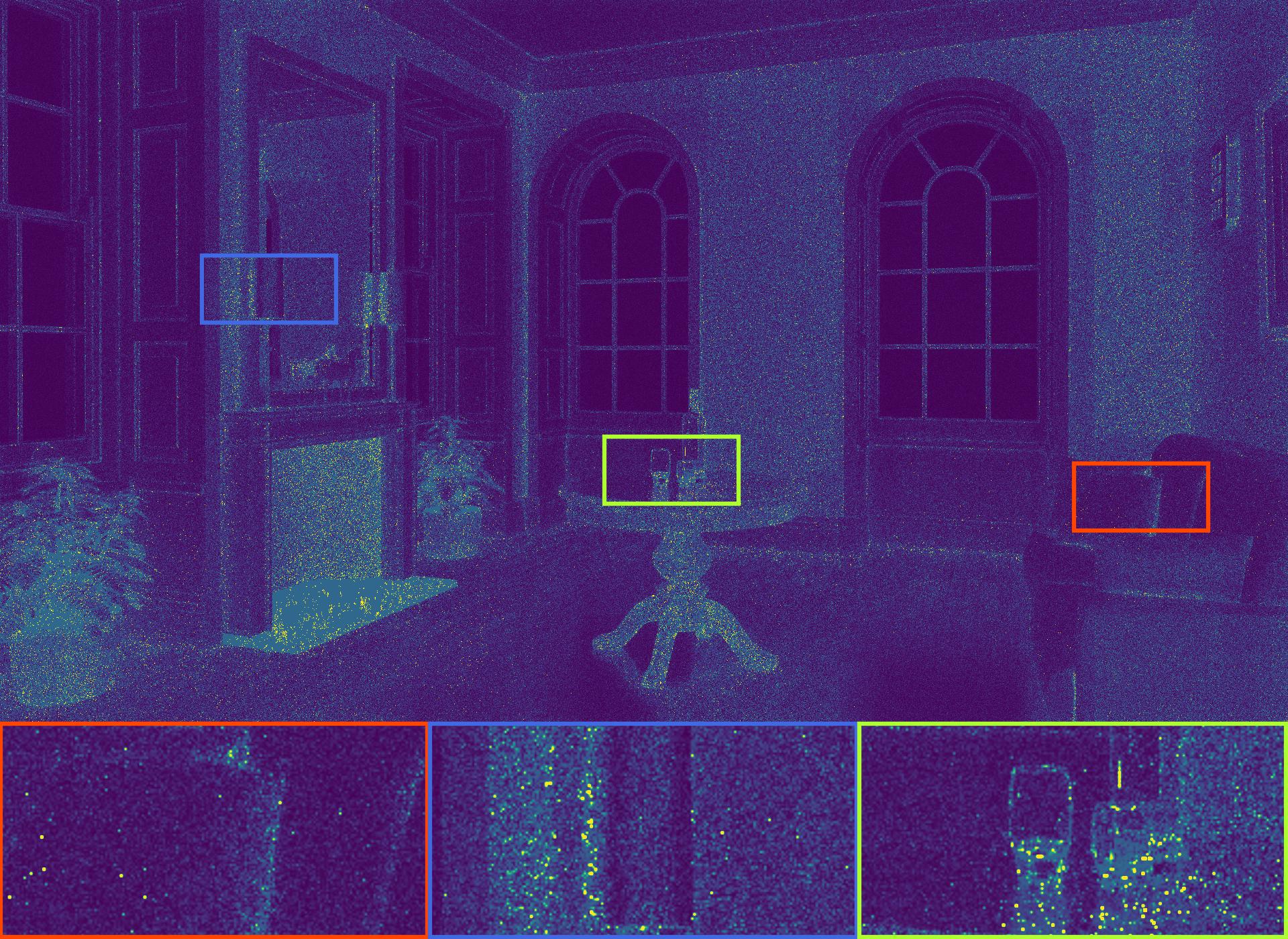}{\ltext{\textsc{RA-Grid}}}{\rtext{rRMSE 0.2179}}
&\imgtext{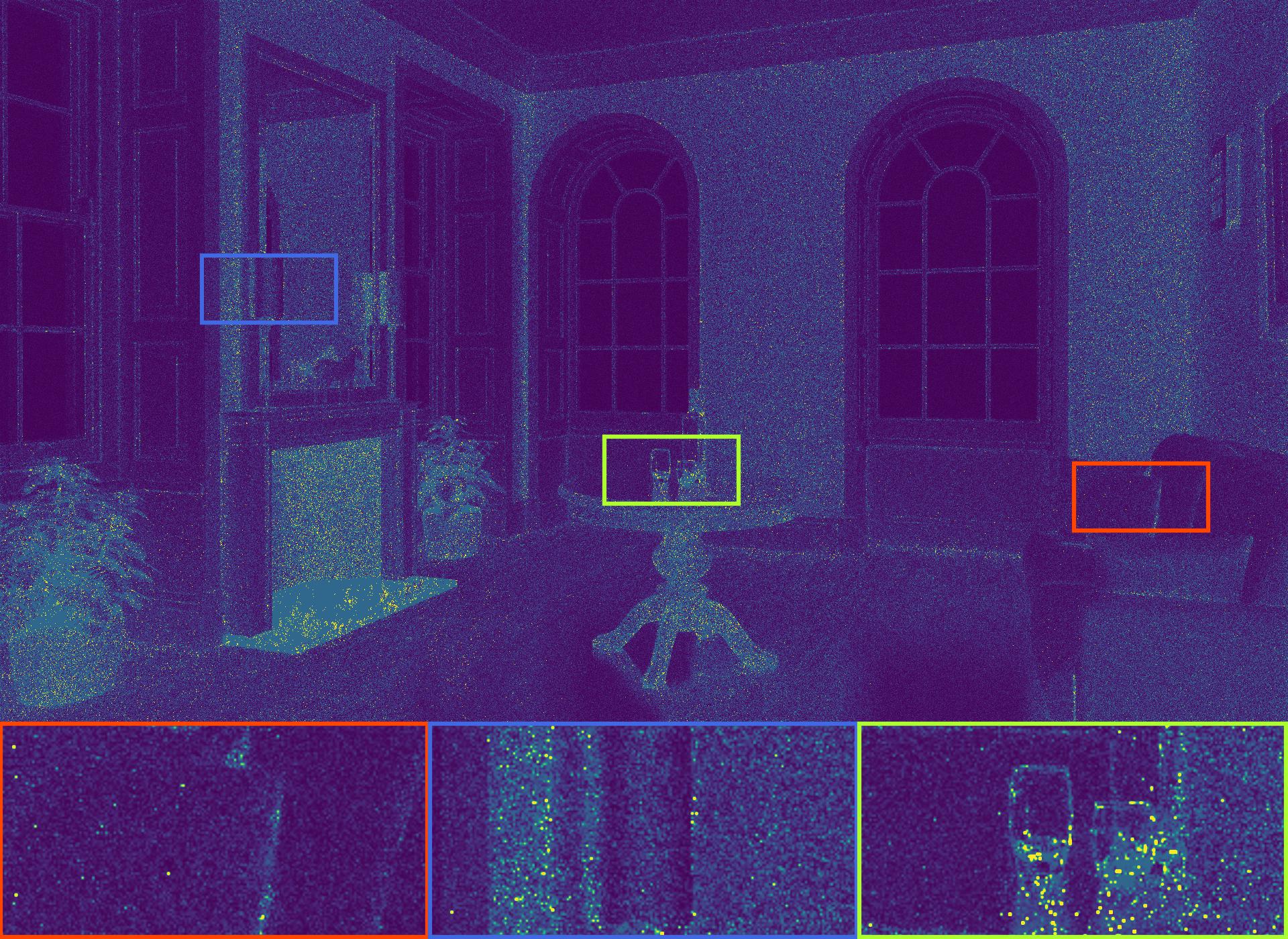}{\ltext{\textsc{RA-Quadtree}}}{\rtext{rRMSE 0.2142}}
&\includegraphics[height=3cm]{cbar.pdf}\\
\rotatebox{90}{Necklace}
&\imgtext{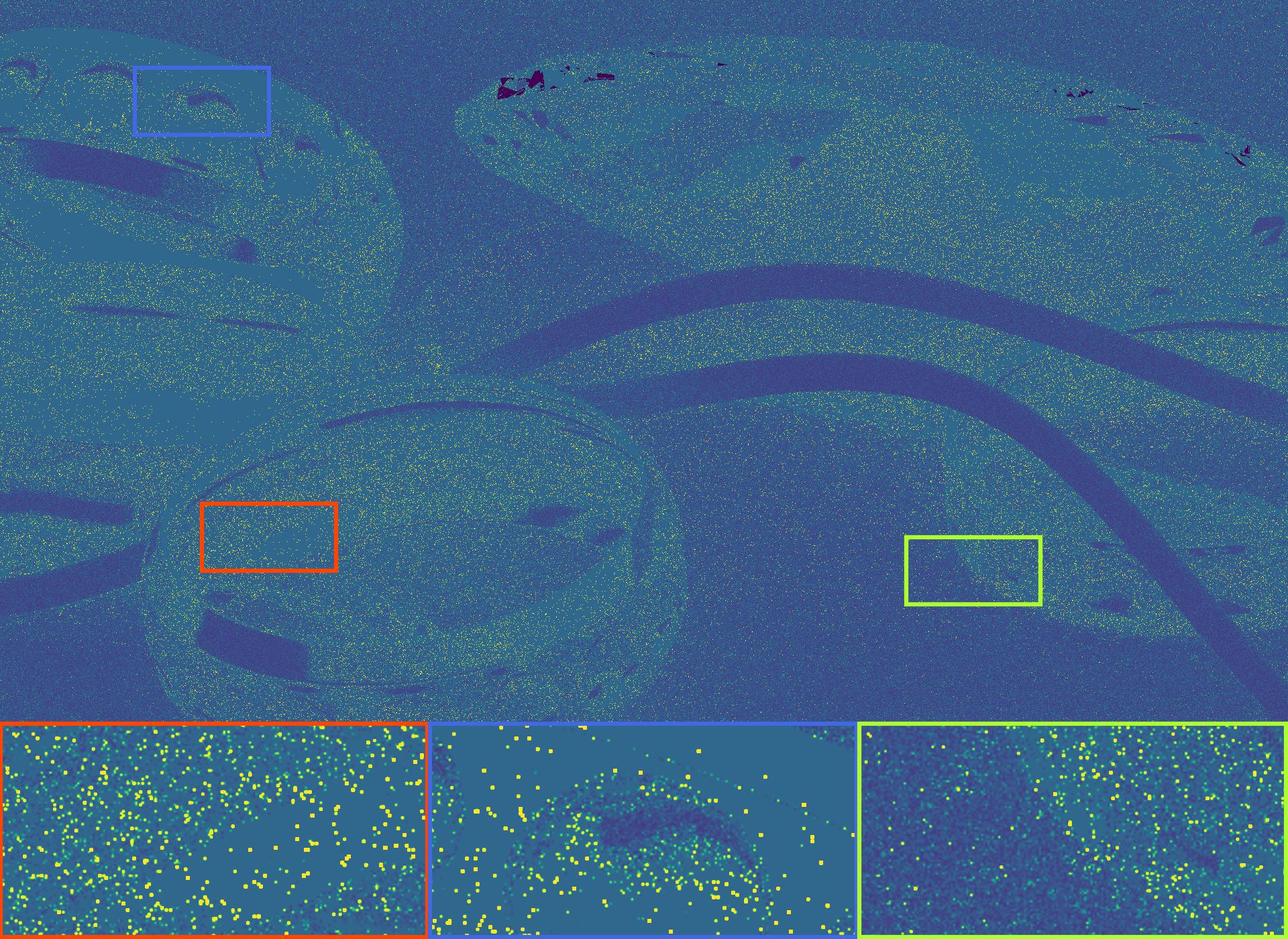}{\ltext{\textsc{Fixed}}}{\rtext{rRMSE 2.5103}}
&\imgtext{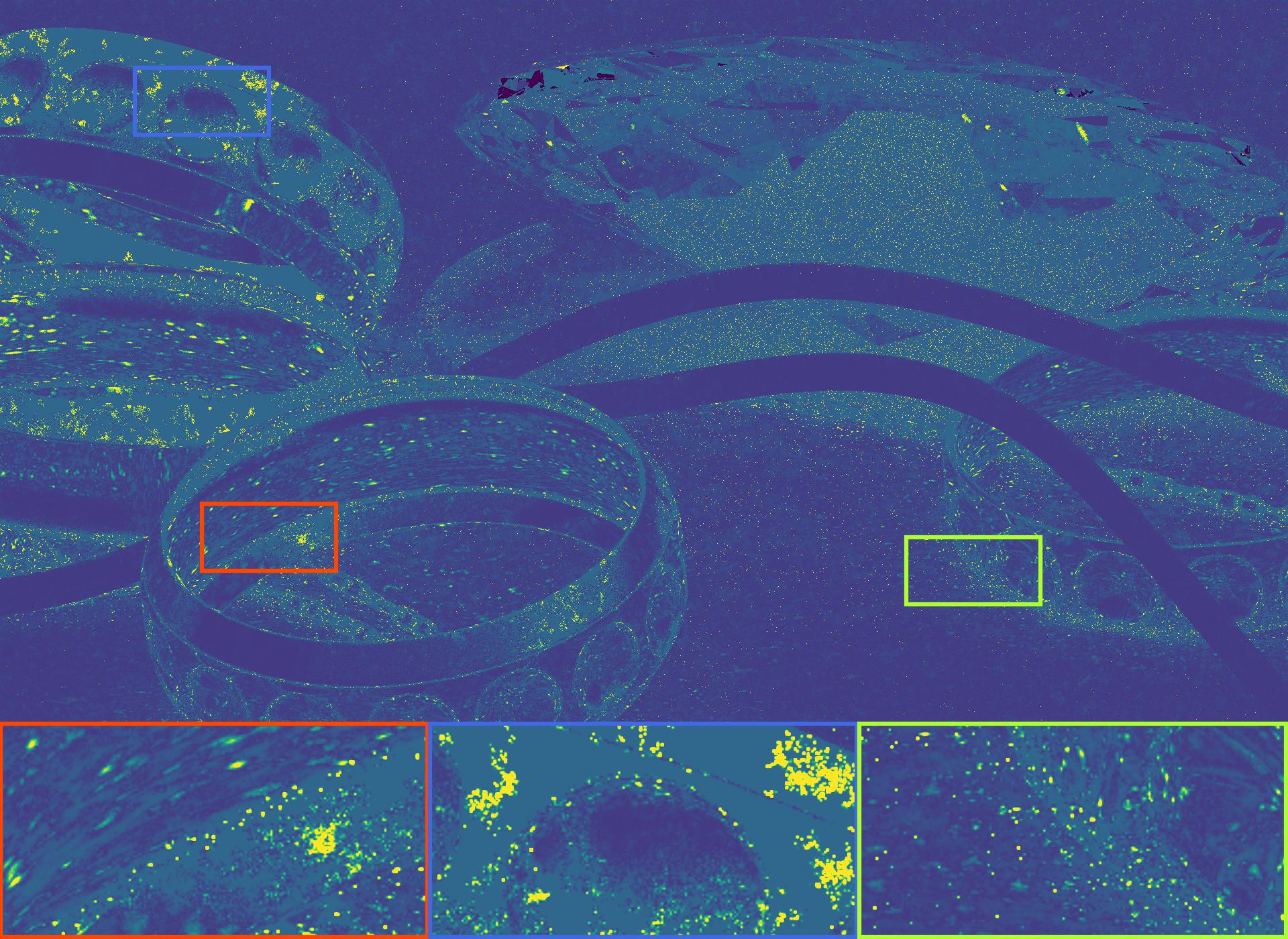}{\ltext{\textsc{Global}}}{\rtext{rRMSE 2.6005}}
&\imgtext{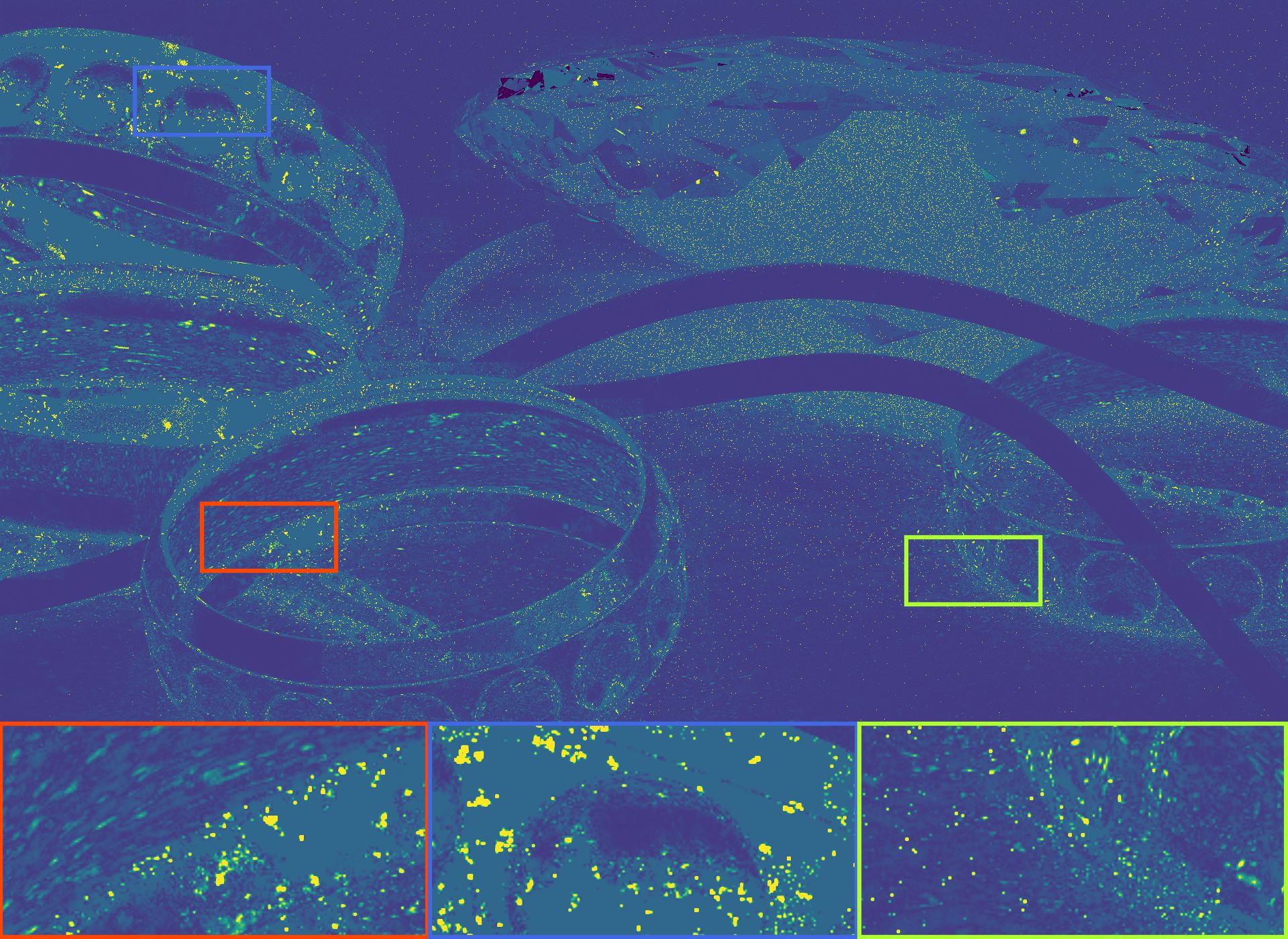}{\ltext{\textsc{RA-Grid}}}{\rtext{rRMSE 2.0825}}
&\imgtext{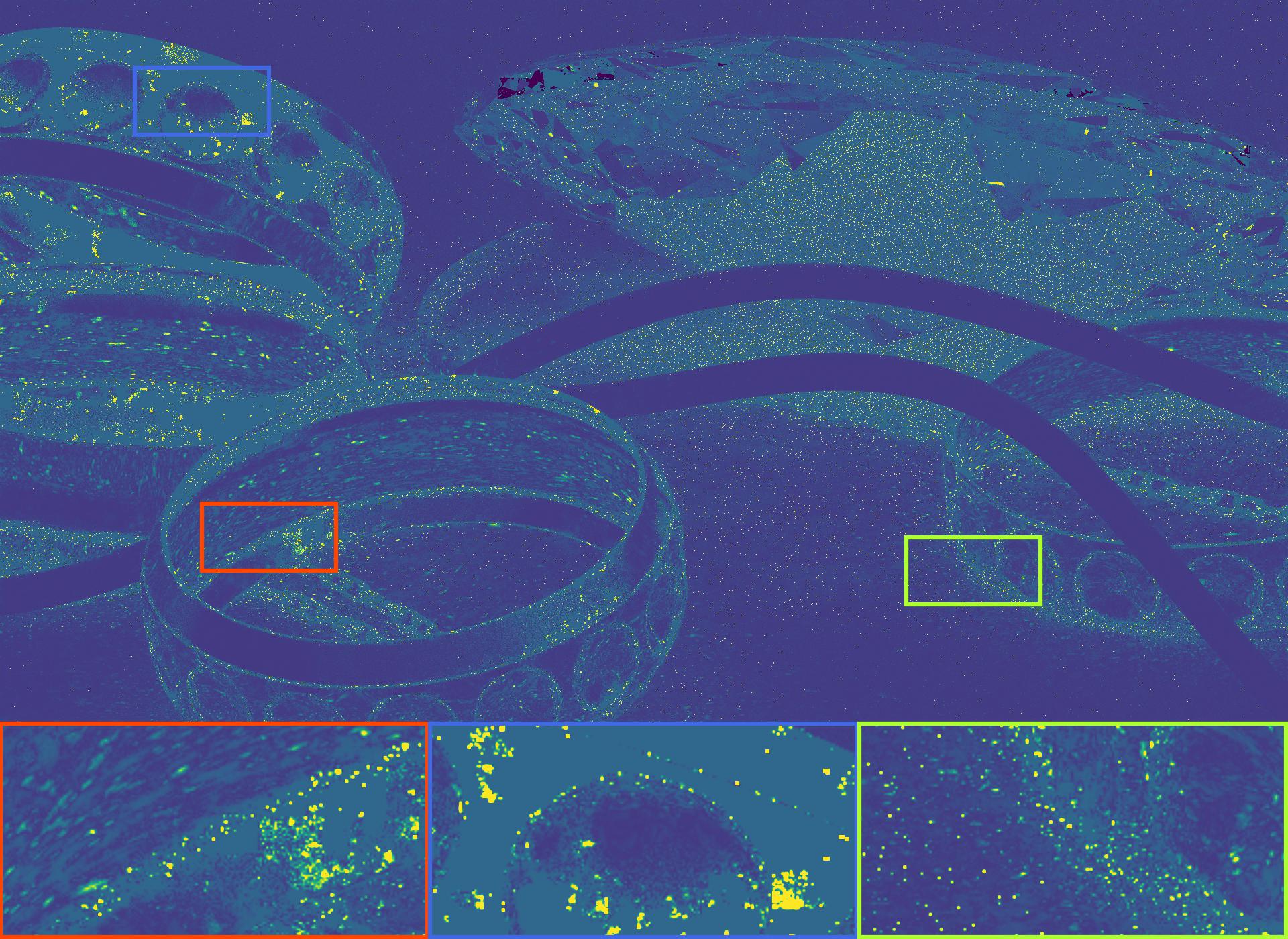}{\ltext{\textsc{RA-Quadtree}}}{\rtext{rRMSE 2.1058}}
&\includegraphics[height=3cm]{cbar.pdf}\\
\rotatebox{90}{Living Room}
&\imgtext{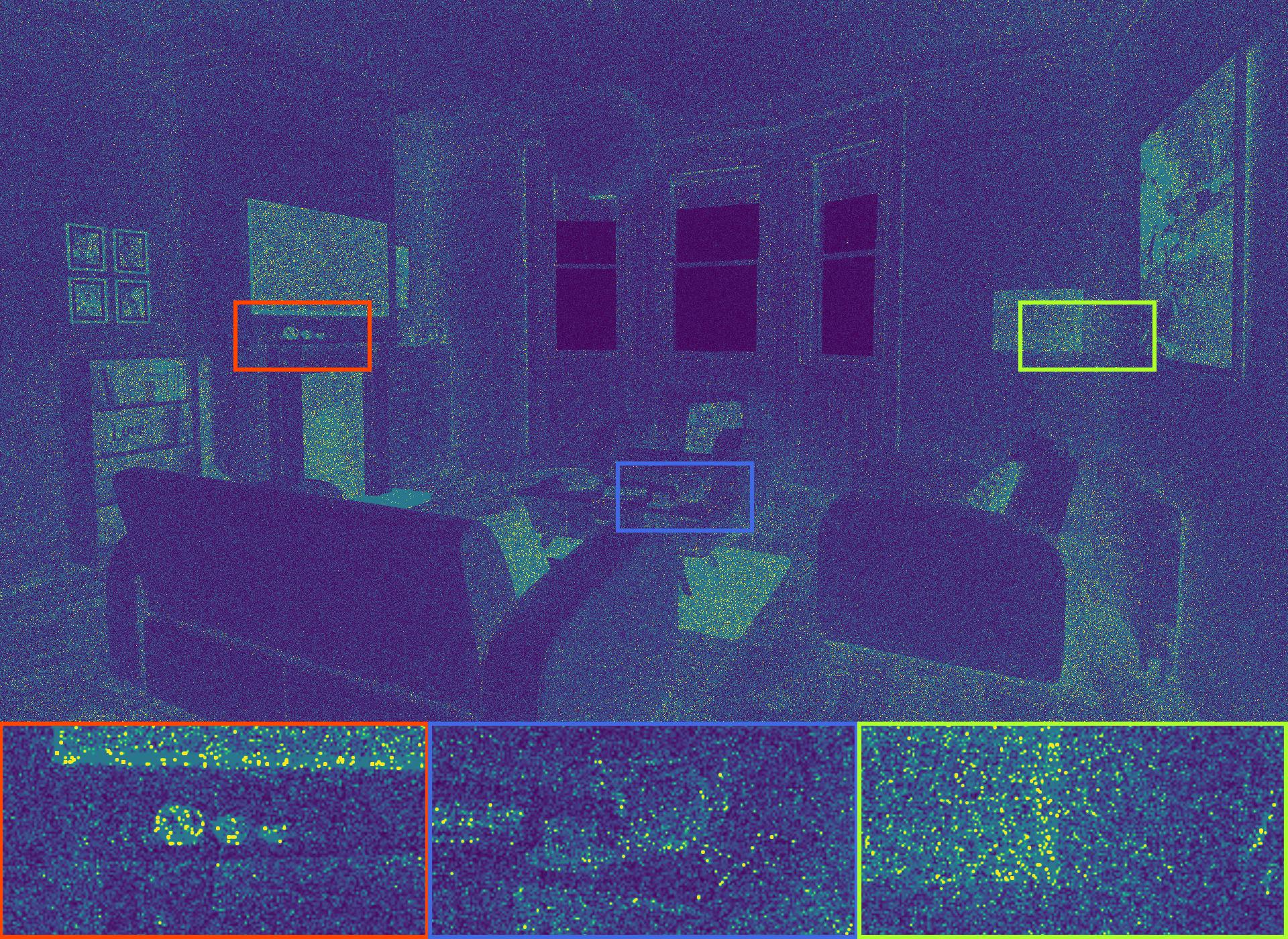}{\ltext{\textsc{Fixed}}}{\rtext{rRMSE 0.4993}}
&\imgtext{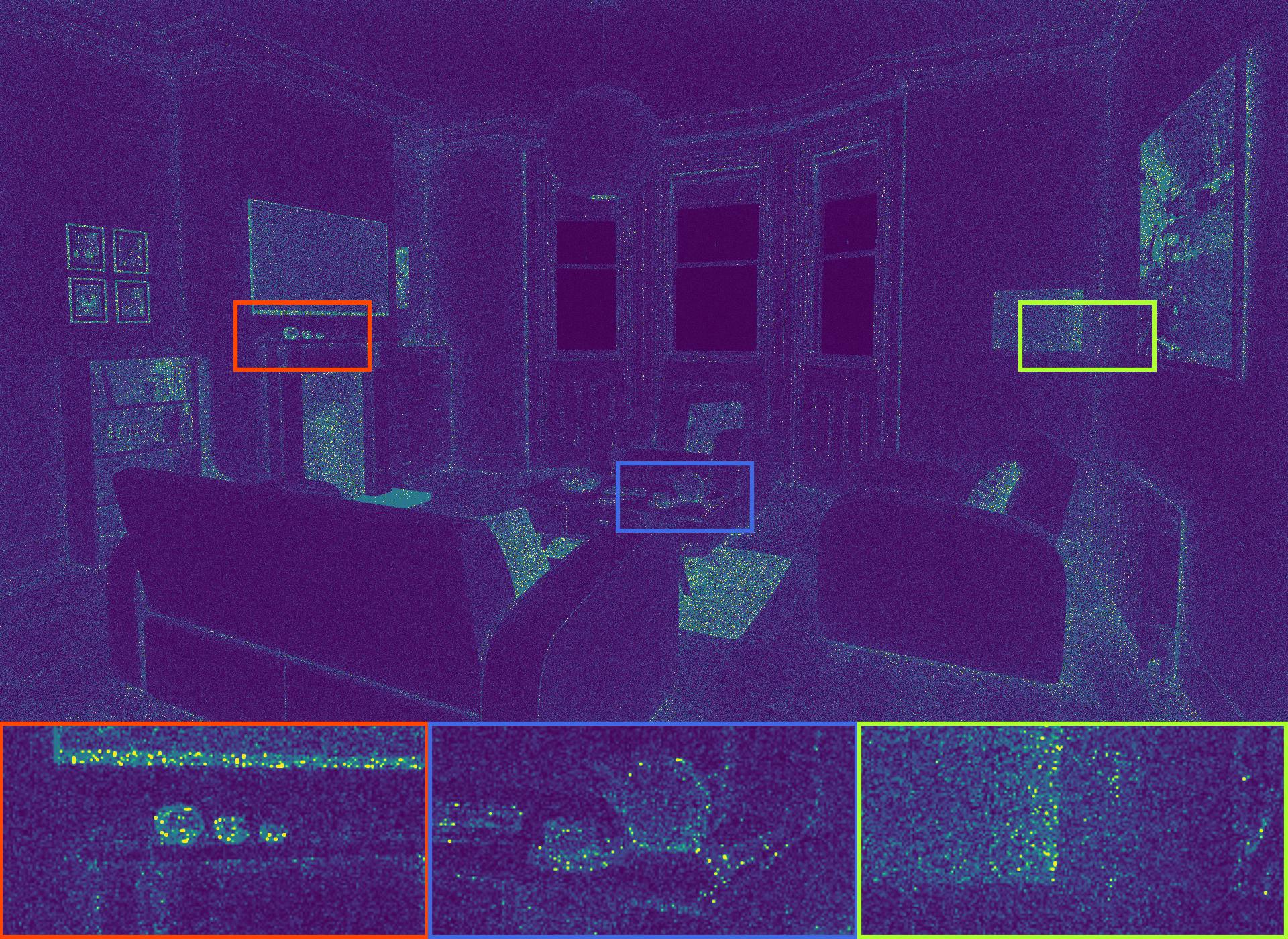}{\ltext{\textsc{Global}}}{\rtext{rRMSE 0.2315}}
&\imgtext{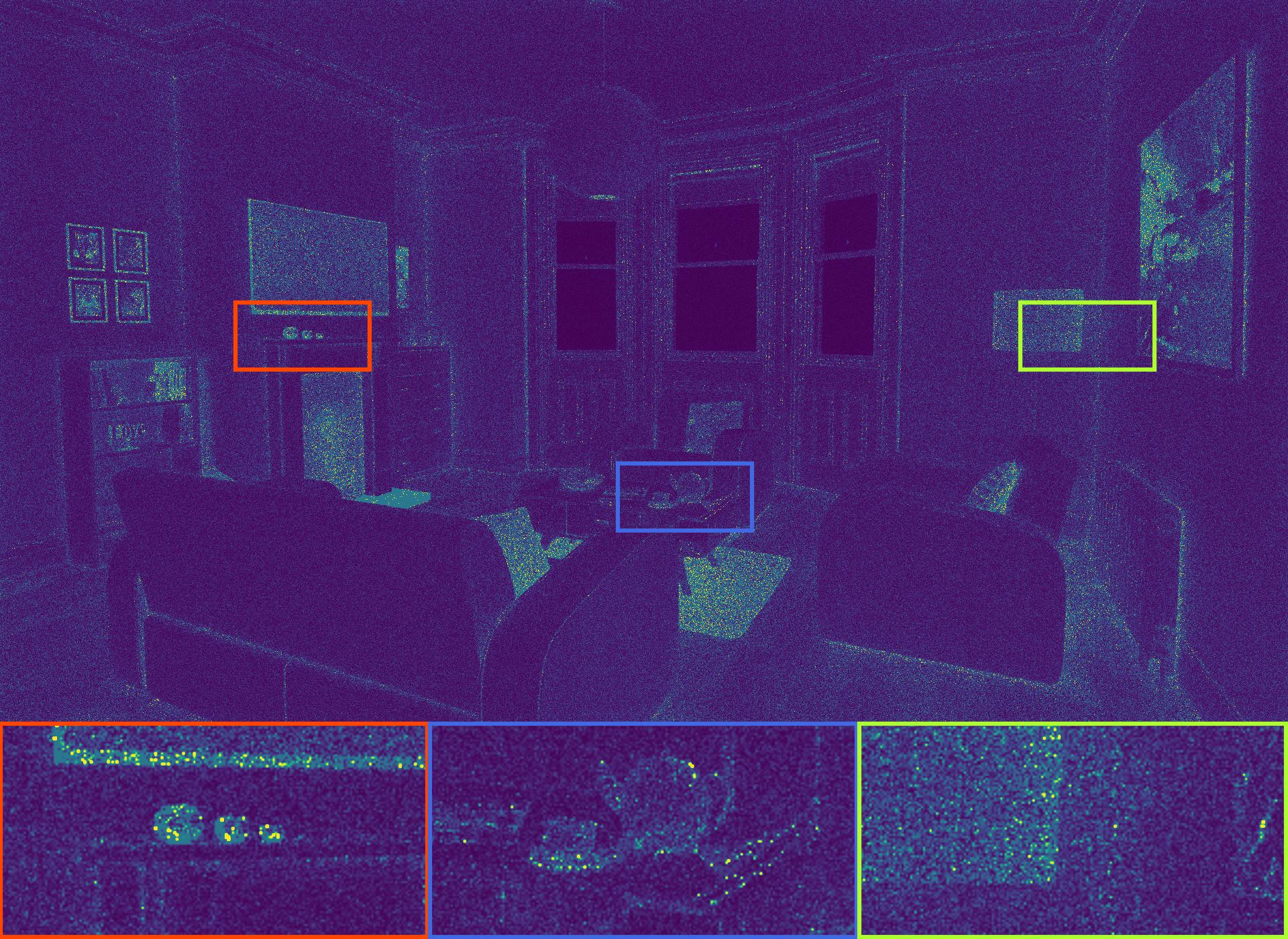}{\ltext{\textsc{RA-Grid}}}{\rtext{rRMSE 0.2209}}
&\imgtext{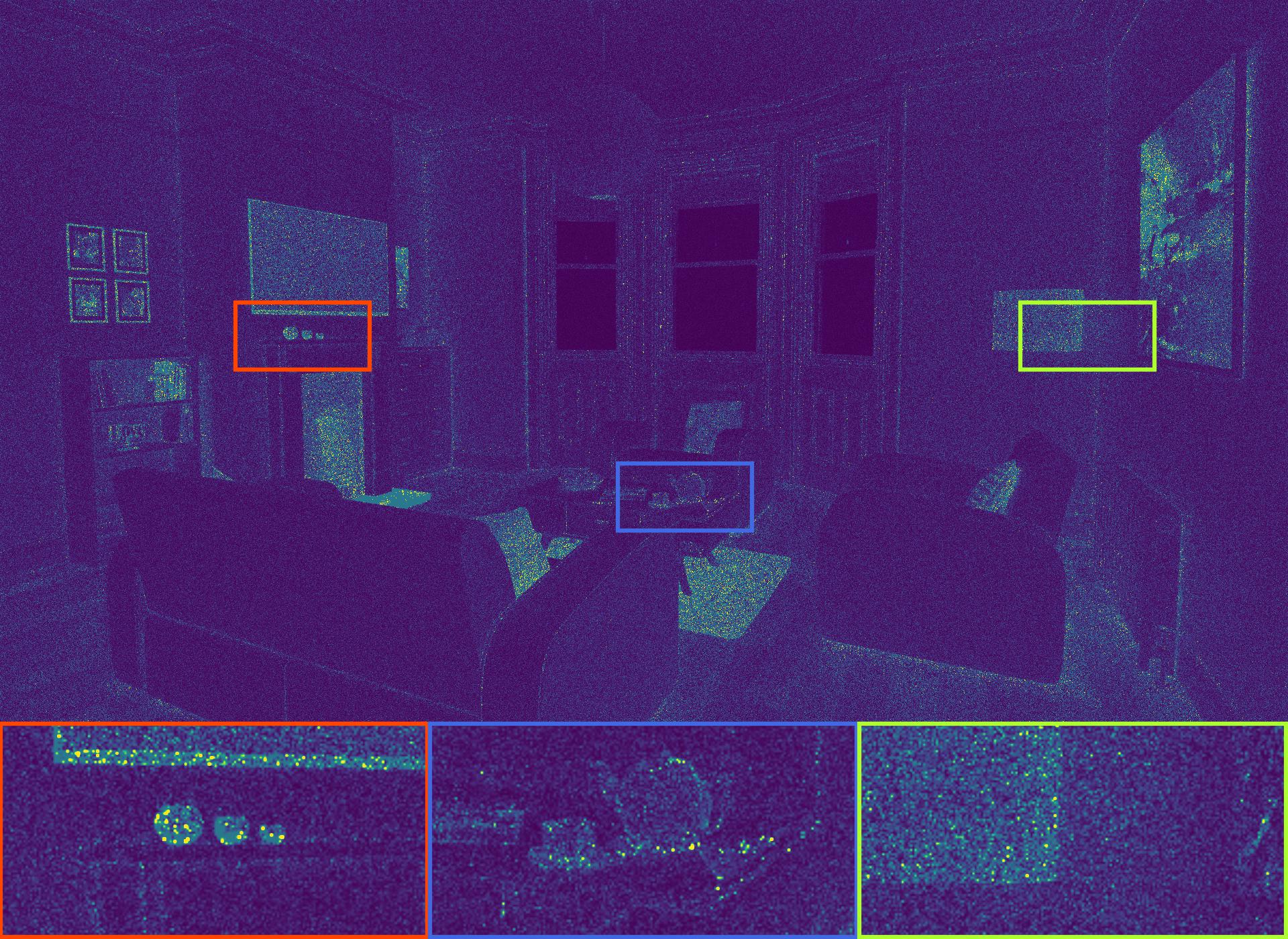}{\ltext{\textsc{RA-Quadtree}}}{\rtext{rRMSE 0.2173}}
&\includegraphics[height=3cm]{cbar.pdf}\\
\rotatebox{90}{Ajar Door}
&\imgtext{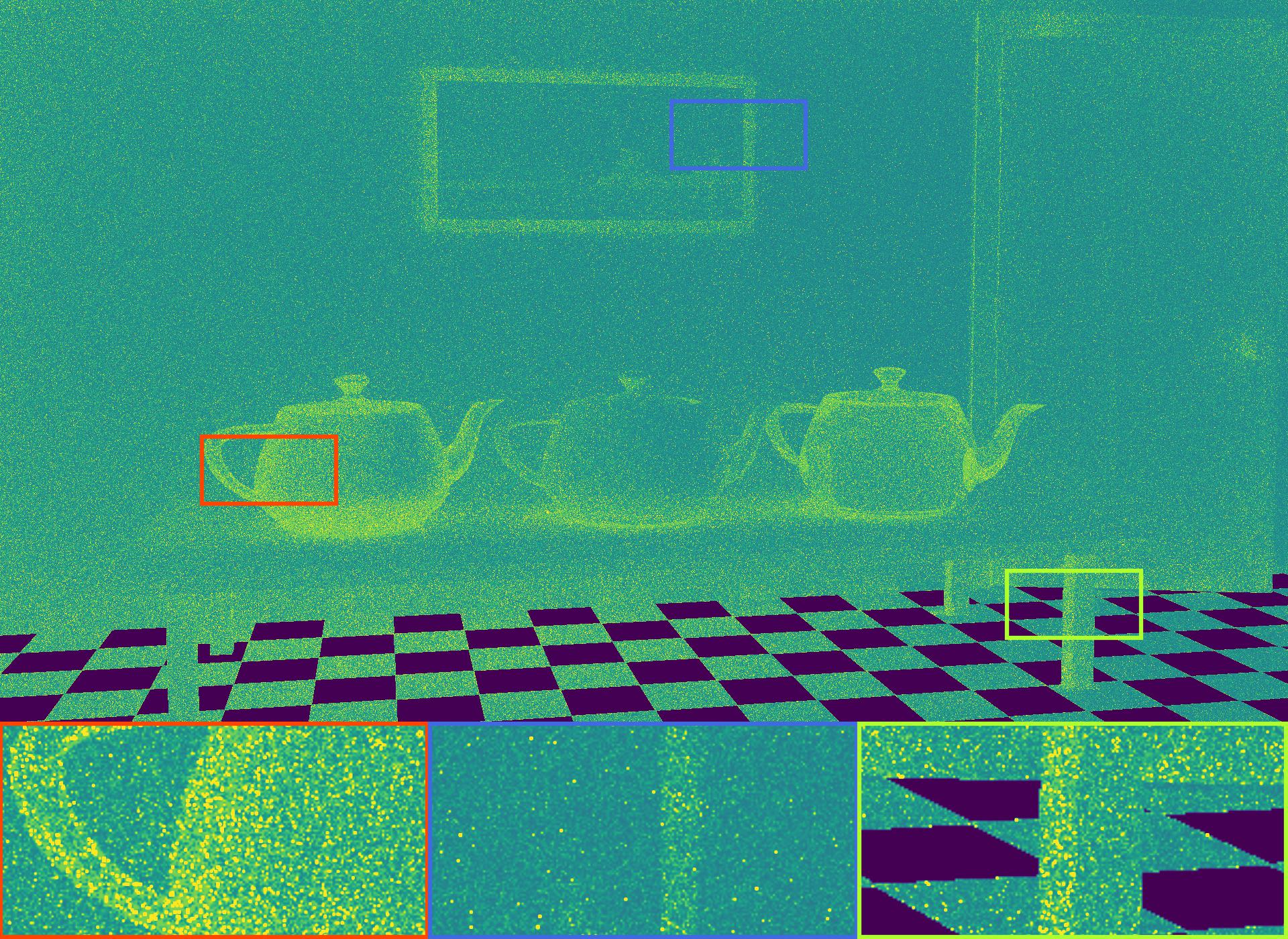}{\ltext{\textsc{Fixed}}}{\rtext{rRMSE 0.8745}}
&\imgtext{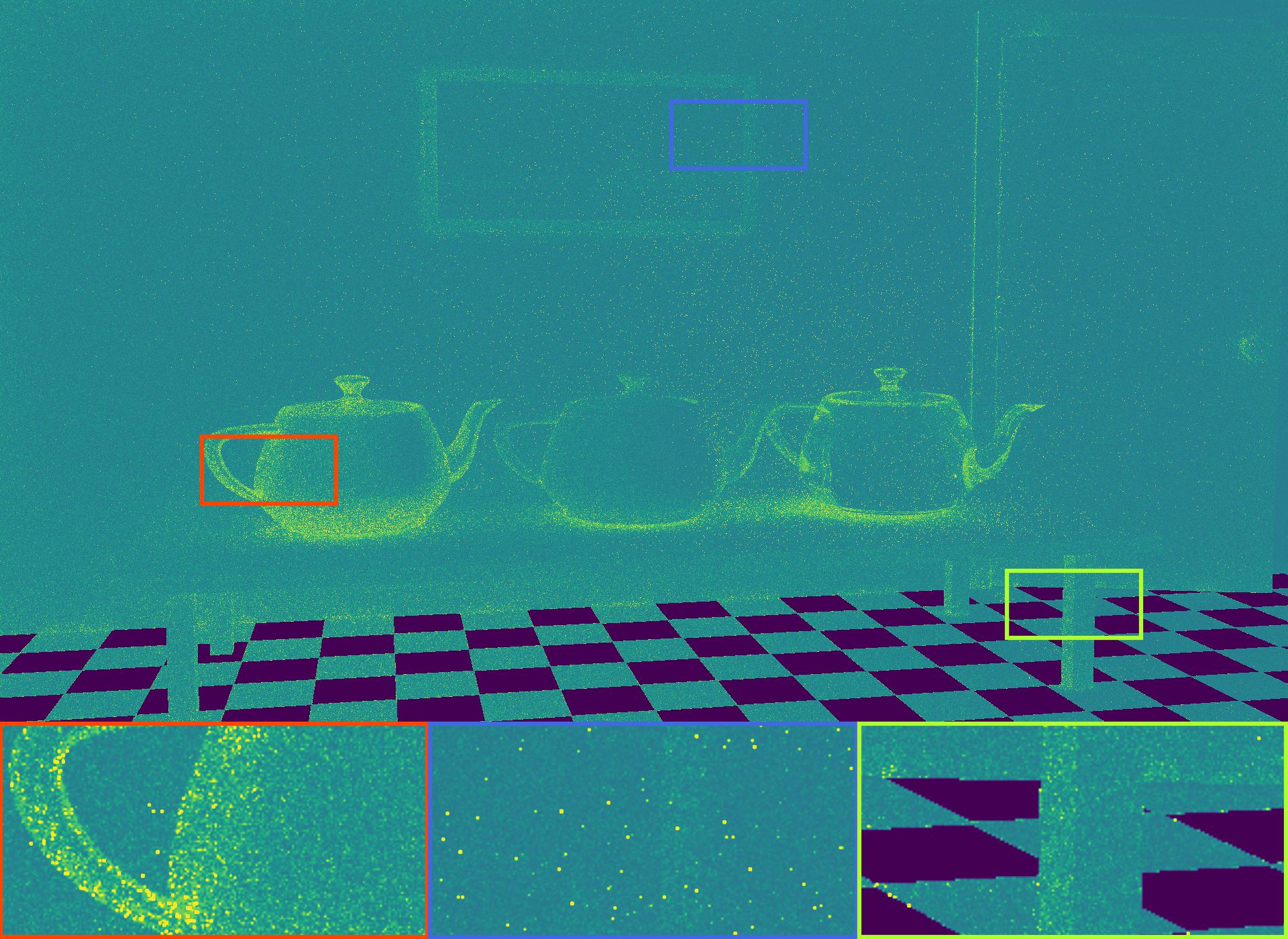}{\ltext{\textsc{Global}}}{\rtext{rRMSE 0.5408}}
&\imgtext{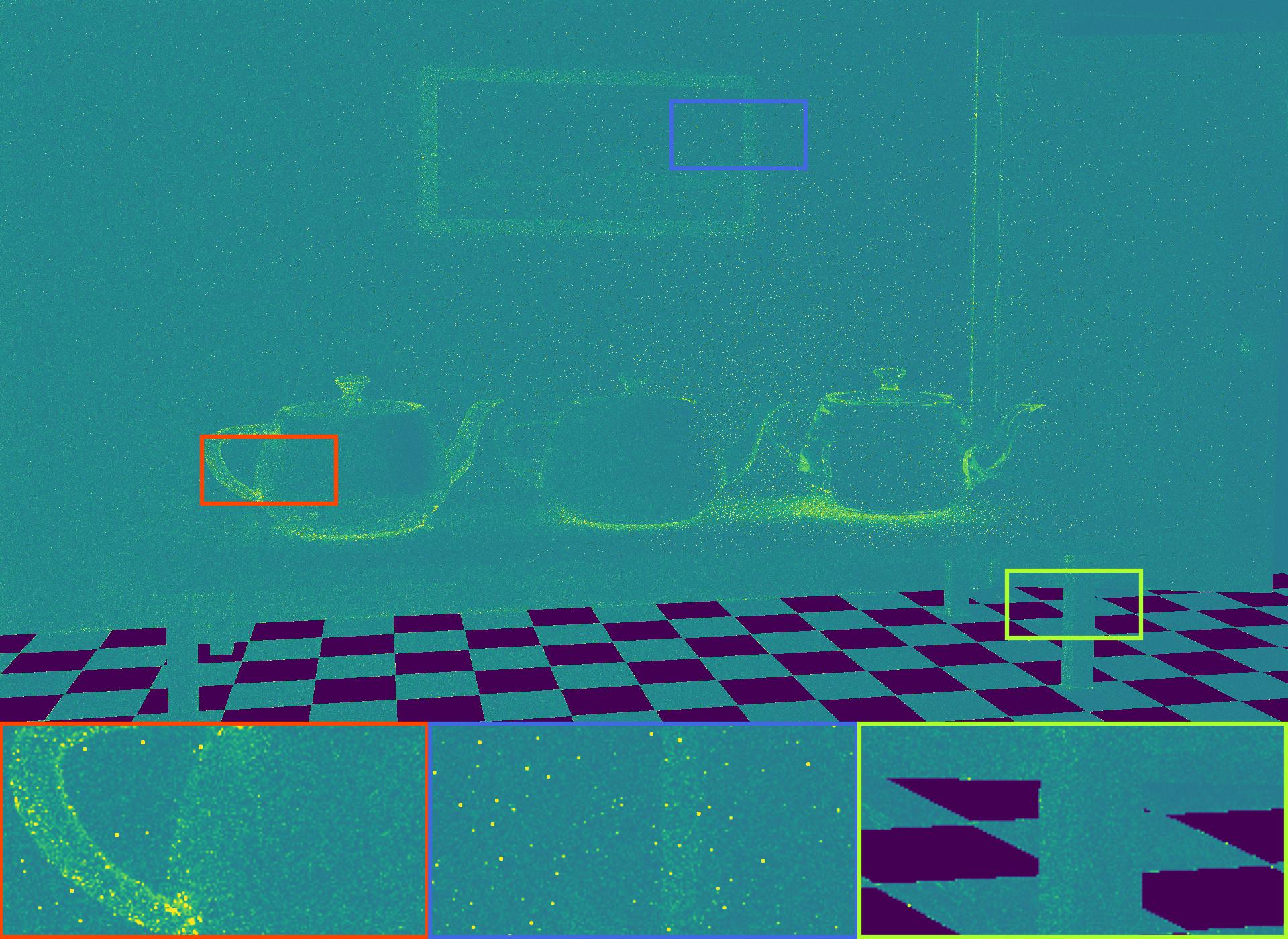}{\ltext{\textsc{RA-Grid}}}{\rtext{rRMSE 0.4612}}
&\imgtext{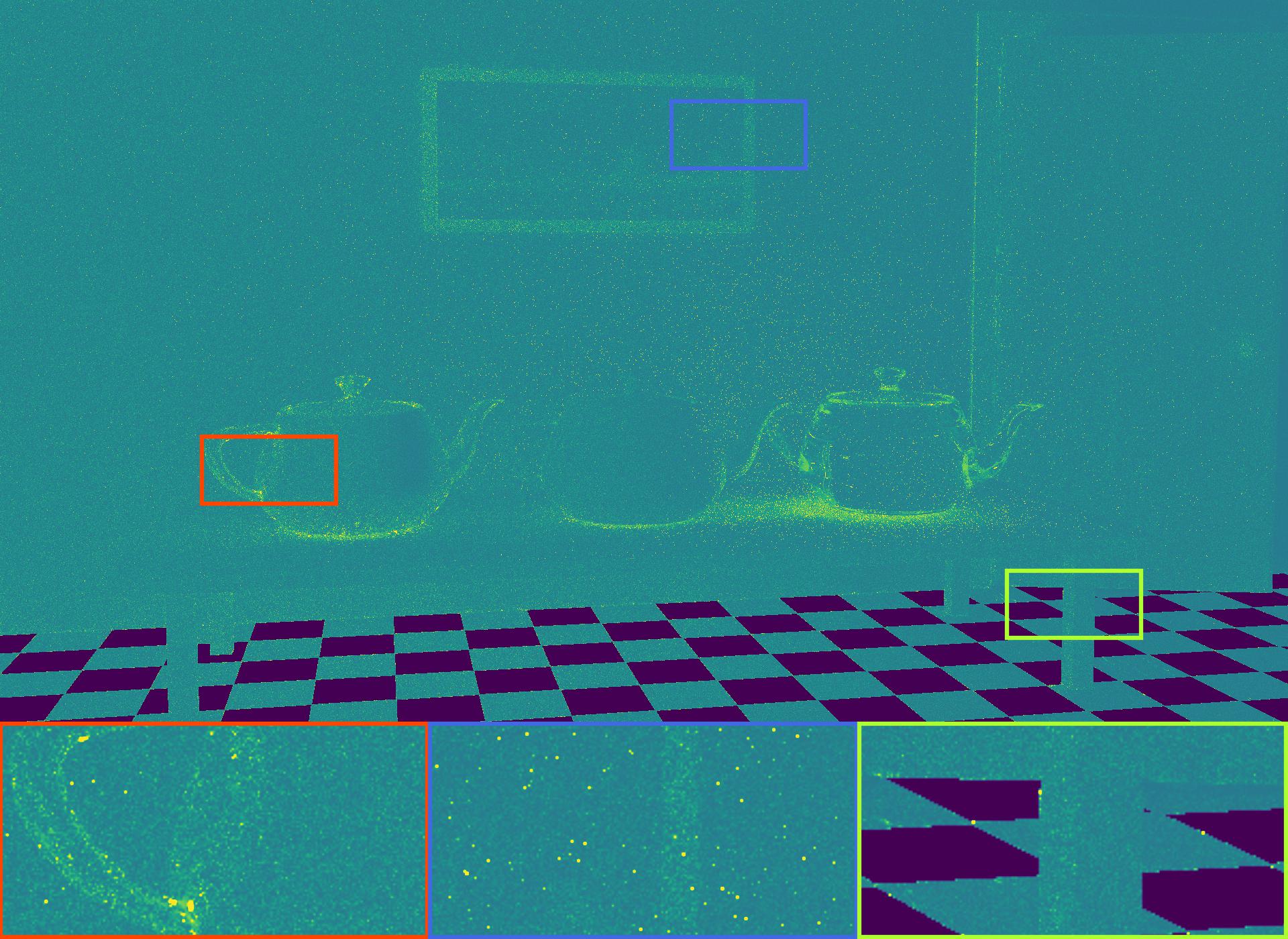}{\ltext{\textsc{RA-Quadtree}}}{\rtext{rRMSE 0.4706}}
&\includegraphics[height=3cm]{cbar.pdf}\\
\end{tabular}

    \caption{Pixel-wised error distribution of the four variants of lens perturbations (\textsc{Fixed}, \textsc{Global}, \textsc{RA-Grid}, \textsc{RA-Quadtree}) rendered with the four scenes (Fireplace Room, Necklace, Living Room, Ajar Door).}
    \label{fig:error_lens}
\end{figure*}

\paragraph*{Lens Perturbation.}
We also conducted experiments for the variants of the lens perturbation
(\cref{fig:rendering_lens} and \ref{fig:error_lens}). We can observe that the
regional adaptive approaches (\textsc{RA-Grid} or \textsc{RA-Quadtree}) consistently perform
better than global adaptation (\textsc{Global}), in terms of both visual quality
and rRMSE. 
Unlike in the multi-chain experiments, however, the differences between \textsc{RA-Grid}
and \textsc{RA-Quadtree} are less noticeable. In fact the grid often shows better
performance. 
This is because the lens perturbation uses a simple 2d canonical space and thus
each region receives enough samples to accurately estimate the adaptation.
In this case, the quadtree produces unnecessary overhead.

\section{Discussion}
\label{sec:discussion}

\paragraph*{Target Acceptance Probability.}
As we discussed in \cref{subsec:optimal_scaling}, the target acceptance
probability 0.234 is obtained from the strong assumption imposed on the shape
of the target function and the proposal distribution, which is obviously not
applicable in the rendering context. 
Thus, we consider $\bar{\alpha}^*$ a configurable
hyperparameter that controls a correlation of the chain. 
During our experiments we observed that $\bar{\alpha}^*=0.234$ performs moderately worse
than higher values of, e.g., $\bar{\alpha}^*=0.5$ or $0.8$, across many scenes.
A theoretical argument how to choose the optimal target acceptance thus remains
an interesting problem for future work.

\paragraph*{State Space.}
Our approach is built on path space MLT, not primary sample
space~\cite{kelemen2002simple}. 
This decision was made because in path space we can adapt the low-dimensional partitions exactly to
the type of perturbation. Primary sample space MLT always mutates all dimensions at once which makes it hard
to decide which dimensions to partition if only a few can fit into memory.

\paragraph*{Capturing Correlation Between Dimensions.}

Currently, a single scalar parameter $\lambda$ is used to control the perturbation of a high dimensional kernel. 
This is equivalent to applying a perturbation to each dimension independently using the same scaling parameter. 
Although this simplifies the adaptive process, it also does not capture correlation between the dimensions. 
It would be an interesting future work to use the AM-style adaptation scheme
that learns a full covariance matrix for the high dimensional (truncated) normal
distribution.

\paragraph*{Difficult Specular Transport.}

Since our regional adaptive mutation techniques are based on existing path
perturbations, the general performance characteristics are limited by the
underlying techniques.
For instance, as shown in the \emph{Necklace} scene in
\cref{fig:rendering_multichain}, the caustics illuminated by a small area light
are hard to render with the multi-chain perturbation. This is
inherited by the regional adaptive version of the technique.

\paragraph*{{Differences to Two-Stage MLT}}
In his PhD thesis, Veach proposed a variation of MLT called two-stage
MLT~\cite{veach1998rmc} which stores information in screen space similar to our
grid partitioning.
In the first stage, two-stage MLT renders a low-resolution image, possibly with low sampling density. 
In the second stage, the image is rendered using 
the target function divided by the intensity obtained from the first stage.
This approach reduces the relative error over the image
by moving samples from brighter parts of the image to darker parts.
Although both approaches use screen space information, 
two-stage MLT and our approach are still fundamentally different: 
our approach updates the mutation kernel using the gathered information, while 
two-stage MLT, in contrast, modifies the target distribution, but not the mutation kernel.

\section{Conclusion}

We presented a novel path perturbation technique for Metropolis Light Transport, which adaptively updates state-dependent parameters of the proposal distribution, based on the history of the Markov chain. 
Our approach partitions the path space along a few dimensions, resulting in low-dimensional canonical spaces.
A set of parameters is associated with each region in the partition, and the adaptation of these parameters is carried out independently, based on the statistics collected when a path sample visits the region. 
To improve the efficiency of the adaptation, we introduced a quadtree-based adaptive structure. This partitioning is adaptively refined throughout the process, by occasionally splitting the leaf nodes based on counting-based statistics. 
We demonstrated that our approach can generate a Markov chain with less correlation, and thus results in better exploration performance compared to fixed-sized kernels or global adaptation. 

\section*{Acknowledgements}
The \emph{Ajar Door} scene was originally modeled by Miika Aittala, Samuli Laine, and Jaakko Lehtinen. The \emph{Necklace} scene is made by Alex Telford.
We thank blendswap.com artist Wig42 for the \emph{Fireplace Room} scene and Jay-Artist for the \emph{Living Room} scene.
This work was funded by DFG grant DA 1200/8-1, project 405788923.

\printbibliography

\appendix

\section{Correlation and Convergence\label{sec:appendix_correlation_and_convergence}}

Let $g\left(x_{i}\right)=\frac{f\left(x_{i}\right)}{\pi\left(x_{i}\right)/b}$.
Under certain assumptions~\cite{brooks2011handbook}, Markov chain
CLT states that
\begin{align}
\sqrt{n}\left(\hat{I}-I\right) & \overset{D}{\to}N\left(0,\sigma^{2}\right),\nonumber \\
\sigma^{2} & =\var\left(g\left(x_{1}\right)\right)+2\sum_{k=1}^{\infty}\cov\left(g\left(x_{1}\right),g\left(x_{1+k}\right)\right)\label{eq:variance}
\end{align}
where $\overset{D}{\to}$ means convergence in distribution and $\sigma^{2}$
is the asymptotic variance. Alternatively, $\sigma^{2}$ can be expressed
by the \emph{autocorrelation time} $\tau$:
\begin{align*}
\sigma^{2} =\frac{\tau}{N}\var\left(g\left(x_{1}\right)\right),\;
\tau =1+2\sum_{k=1}^{\infty}\corr\left(g\left(x_{1}\right),g\left(x_{1+k}\right)\right),
\end{align*}
where $\corr\left(x,y\right):=\cov\left(x,y\right)/\var\left(x\right)$.
Also, we can write
\[
\sigma^{2}=\frac{1}{N_{\mathrm{eff}}}\var\left(g\left(x_{1}\right)\right),
\]
where $N_{\mathrm{eff}}:=N/\tau$ is the \emph{effective sample size}.
We note that the similarity of the variance formula for i.i.d. case
with the same number of samples $N$, where $N_{\mathrm{eff}}$ is
simply replaced by $N$. This formula implies that the asymptotic
variance of an estimate with Markov chain depends on the correlation
of the chain.

\section{Region Independence\label{sec:appendix_region_independence}}

Given the current state $\bar{x}\in\mathcal{U}$, then the expected
acceptance ratio $\bar{\alpha}$ can be written by
\begin{align*}
\bar{\alpha}:=\mathbb{E}_{\pi\otimes T}\left[\alpha\left(\bar{x},\bar{y}\right)\right] & =\int_{\mathcal{P}^{2}}\alpha\left(\bar{x},\bar{y}\right)\frac{\pi\left(\bar{x}\right)}{b}T\left(\bar{y}\mid\bar{x}\right)d\bar{x}d\bar{y},
\end{align*}
where $b$ is the normalization factor $b:=\int_{\mathcal{P}}\pi\left(\bar{x}\right)d\bar{x}$.
Substituting $T\left(\bar{y}\mid\bar{x}\right)$ with the definition
(\cref{eq:regional_proposal_mixture}), we have
\begin{align*}
\bar{\alpha} & =\int_{\mathcal{P}^{2}}\alpha\left(\bar{x},\bar{y}\right)\frac{\pi\left(\bar{x}\right)}{b}\left[\sum_{k=1}^{m}\mathbf{1}_{\mathcal{P}_{k}}\left(\bar{x}\right)T_{k}\left(\bar{y}\mid\bar{x}\right)\right]d\bar{x}d\bar{y}\\
 & =\sum_{k=1}^{m}\int_{\mathcal{P}^{2}}\frac{\pi\left(\bar{x}\right)}{b}\mathbf{1}_{\mathcal{P}_{k}}\left(\bar{x}\right)T_{k}\left(\bar{y}\mid\bar{x}\right)d\bar{x}d\bar{y}.
\end{align*}
Let $b_{k}:=\int_{\mathcal{P}}\mathbf{1}_{\mathcal{P}_{k}}\left(\bar{x}\right)\frac{\pi\left(\bar{x}\right)}{b}d\bar{u}=\frac{1}{b}\int_{\mathcal{P}_{k}}\pi\left(\bar{x}\right)d\bar{x}$.
Note that $\mathbb{P}\left(\bar{x}\in\mathcal{P}_{k}\right)=\mathbb{E}_{\pi}\left[\mathbf{1}_{P_{k}}\left(\bar{x}\right)\right]=b_{k}$.
Then using conditional expectation, $\bar{\alpha}$ can be further
written as
\begin{align*}
\bar{\alpha} & =\sum_{k=1}^{m}\int_{\mathcal{P}^{2}}\alpha\left(\bar{x},\bar{y}\right)\frac{\pi\left(\bar{x}\right)}{b}\mathbf{1}_{\mathcal{P}_{k}}\left(\bar{x}\right)T_{k}\left(\bar{y}\mid\bar{x}\right)d\bar{x}d\bar{y}\\
 & =\sum_{k=1}^{m}b_{k}\frac{\mathbb{E}_{\pi\otimes T_{k}}\left[\alpha\left(\bar{x},\bar{y}\right)\mathbf{1}_{\mathcal{P}_{k}}\left(\bar{x}\right)\right]}{\mathbb{P}\left(\bar{x}\in\mathcal{P}_{k}\right)}\\
 & =\sum_{k=1}^{m}b_{k}\mathbb{E}_{\pi\otimes T_{k}}\left[\alpha\left(\bar{x},\bar{y}\right)\mid\bar{x}\in\mathcal{P}_{k}\right].
\end{align*}
Since $\left\{ \mathcal{P}_{k}\right\} $ is a partition, $\sum_{k=1}^{m}b_{k}=1$.
Also, since $\bar{\alpha}^{*}$ is constant, $\bar{\alpha}^{*}=\sum_{k=1}^{m}b_{k}\bar{\alpha}^{*}$.
Therefore, the difference between the expected and the optimal acceptance
ratio can be written as
\[
\bar{\alpha}-\bar{\alpha}^{*}=\sum_{k=1}^{m}b_{k}\left(\mathbb{E}_{\pi\otimes T_{k}}\left[\alpha\left(\bar{x},\bar{y}\right)\mid\bar{x}\in\mathcal{P}_{k}\right]-\bar{\alpha}^{*}\right).
\]
Let $\bar{\alpha}_{k}=\mathbb{E}_{\pi\otimes q_{k}}\left[\alpha\left(\bar{u},\bar{v}\right)\mid\bar{u}\in\mathcal{U}_{k}\right]$.
This equation implies that if for every $k$, $\bar{\alpha}_{k}\approx\bar{\alpha}^{*}$,
then we can also say $\bar{\alpha}\approx\bar{\alpha}^{*}$. This
means we can configure the parameter $\lambda_{k}$ based only on
the information collected within a region.

\end{document}